%% file: main.tex
\documentclass[a4paper,11pt]{article}
\usepackage{float}
\usepackage{relsize}
\usepackage{graphicx}
\usepackage{caption}
\usepackage{subcaption}
\usepackage{comment}

\newcommand{\req}[1]{(\ref{#1})}

\def\bet{\beta}
\def\fc#1#2{\frac{#1}{#2}}

\newcommand{\nwc}{\newcommand}
\nwc{\ba}  {\begin{array}}
\nwc{\ea}  {\end{array}}
\nwc{\bdm} {\begin{displaymath}}
\nwc{\edm} {\end{displaymath}}

\nwc{\bea} {\begin{equation}\ba{lcl}}
\nwc{\eea} {\ea\end{equation}}

\nwc{\be} {\begin{equation}}
\nwc{\ee} {\end{equation}}

\nwc{\bda} {\bdm\ba{lcl}}
\nwc{\eda} {\ea\edm}

\nwc{\bc}  {\begin{center}}
\nwc{\ec}  {\end{center}}

\nwc{\ds}  {\displaystyle}

\nwc{\nn} {\nonumber}
\nwc{\nnn} {\nonumber \vspace{.2cm} \\ }
\nwc{\ra}{\rightarrow}
\nwc{\lra}{\longrightarrow}
\def\lf{\left}\def\ri{\right}

\nwc{\p} {\partial}

\def\Kc{{\cal K}}\def\Jc{{\cal J}}
\def\ap{\alpha'}
\def\lng{\langle}

\def\Pc{{\cal P}}\def\Mc{{\cal M}}\def\Cc{{\cal C}}

\def\ov{\overline}

\def\al{\alpha}
\def\bet{\beta}

\def\Oc{{\cal O}}

\def\Lc{{\cal L}}

\def\Sc{{\cal S}}

\def\al{\alpha}

\def\si{\sigma}
\def\om{\omega}
\def\bet{\beta}\def\bet{\beta}

\input{packages.tex}
\numberwithin{equation}{section} 

\title{\boldmath Resurgence of high-energy string amplitudes}
\author[]{Xavier Kervyn}
\author[]{and Stephan Stieberger}

\affiliation[]{Max-Planck-Institut für Physik, Werner-Heisenberg-Institut, \\ Boltzmannstra\ss e 8, 85748 Garching bei München, Germany}

\emailAdd{xavier.kervyn@mpp.mpg.de}
\emailAdd{stephan.stieberger@mpp.mpg.de}

\abstract{  
We analyze the fixed-angle high-energy ($\alpha' \to \infty$) structure of $n$-point tree-level string amplitudes from complementary perspectives: locally via saddle-point expansions, algebraically via difference equations and their asymptotic structure, analytically via Aomoto-Gauss-Ma\-nin connection and Mellin-Barnes representation, and geometrically via twisted intersection theory and Lefschetz thimbles. Using, in turn, saddle-point analysis and finite-difference equations in the kinematic variables, we show that the perturbative coefficients in the resulting asymptotic series in $1/\alpha'$ are organized by Bernoulli-number data, rather than by the multiple zeta values characteristic of the low-energy $\alpha' \to 0$ regime.  Resurgence theory allows upgrading these divergent series to transseries whose Stokes data capture the analytic continuation between unphysical and physical kinematic regions in the form of non-perturbative monodromy contributions. 
We derive the transseries for four-point open string amplitudes explicitly. We also construct a differential and Mellin formulation which place their low- and high-energy expansions in a common analytic framework and unifies them as asymptotic sectors of the same underlying object. We extend the 
difference-equation analysis to $n \geq 5$, where it yields perturbative
high-energy asymptotic expansions and leads naturally to a higher-rank connection problem.
Finally, translating our asymptotic analysis into the language of twisted de Rham theory, we derive an alternative double-copy representation of the high-energy limit of closed-string amplitudes in terms of Lefschetz thimbles for any $n$.

}

\begin{document}
\maketitle
\flushbottom

\setlength{\parskip}{5pt}


\section{Introduction}

Divergent perturbative expansions are a defining feature of quantum field theory and string theory. While perturbation theory often provides the only practical handle on physical observables, the associated series coefficients typically grow factorially, making them asymptotic and often non-Borel-summable; see, e.g.,~\cite{Dyson:1952tj, tHooft:1977xjm} and 
\cite{Mende:1989wt,Shenker:1990uf} for early accounts of this phenomenon respectively in the context of field and string theory.\footnote{String perturbation theory is an expansion in the genus (sum over worldsheet topologies). It is widely expected that this expansion is divergent (asymptotic), with zero radius of convergence and coefficients growing like $(2g)!$ at genus $g$.} Resurgence theory, initiated by Jean \'Ecalle \cite{EcalleLesFonctionsResurgentes}, provides a canonical way to endow such expansions with analytic meaning by relating their large-order behaviour to non-perturbative effects through \textit{transseries} and \textit{Stokes phenomena}. A transseries \cite{Edgar:2009} is a generalized expansion that goes beyond ordinary power series by including, in addition to perturbative terms, non-perturbative contributions such as exponentials of inverse coupling and logarithms.
As such, it provides a unified framework that extends the original, ill-defined asymptotic series to a well-defined analytic function across \textit{sectors} in the complex plane of the formal expansion parameter (such as the coupling constant). Stokes phenomena \cite{Stokes:1864, Costin:2008}, on the other hand, describe how the relative importance of these different contributions can change abruptly when one analytically continues the latter from one sector to the other, crossing so-called \textit{Stokes lines} in the complex plane. Together, transseries and Stokes phenomena allow one to consistently relate perturbative and non-perturbative physics in a unified framework.

From a physics viewpoint, resurgence turns the `breakdown' of perturbation theory into a feature: the singularities of the Borel transform of an asymptotic series dictate both the form of non-perturbative effects and the location of Stokes discontinuities in the complex plane, while the associated residues encode their relative weights. Altogether, these contributions conspire to cancel the ambiguities between expansions in different sectors, thereby offering a sharp and stringent notion of consistency for resummations \cite{Dorigoni:2014hea,Aniceto:2018bis}.
A particularly well-studied arena is planar AdS/CFT, where the strong-coupling expansion of the cusp anomalous dimension derived from the Beisert-Eden-Staudacher (BES) equation \cite{Beisert:2006ez} exhibits a resurgent transseries structure tied to the $O(6)$ sigma-model mass gap and to analyticity constraints \cite{Aniceto:2015rua,Dorigoni:2015dha}.
In parallel, resurgence has become central to the non-perturbative analysis of topological strings, where the divergent genus expansion can be systematically related to instanton sectors and Stokes data, spurring conjectural non-perturbative completions \cite{Marino:2012zq,Aniceto:2018bis}. 

Motivated by these developments, we investigate the resurgent structure underlying high-energy string scattering amplitudes, with the goal of extracting the `non-perturbative' information encoded in their large-order asymptotics and clarifying which transseries sectors control the high-energy regime at a given order in the genus expansion. Here `non-perturbative' refers to the transseries sectors that are non-analytic in the Gross-Mende asymptotic parameter $1/\alpha'$, i.e.,~exponentially small/oscillatory contributions beyond the single saddle-point expansion.\footnote{This is distinct from non-perturbative effects in the string coupling $g_s$, which arise from the genus expansion and are not at issue at tree level. More precisely, these terms are non-perturbative in the semiclassical parameter controlling the steepest-descent analysis of the Koba-Nielsen integral (large $\alpha'$ at fixed kinematics), thereby avoiding any ambiguity associated with coupling.} We focus on the fixed-angle high-energy (`hard scattering') regime of tree-level string amplitudes initiated by Gross and Mende \cite{Veneziano:1968yb,Gross:1987kza}. In this limit the amplitudes admit a formal expansion in small string tension $T$ (equivalently in $1/\alpha'$) around the leading classical contribution to the path integral \cite{Kervyn:2025wsb}, but the resulting subleading series is generically nonconvergent and should therefore be viewed as a formal asymptotic expansion. Working at four points, we analyze this $1/\alpha'$-expansion through the lens of resurgence, identify the relevant Borel singularities and Stokes data, construct the associated transseries completion, and thereby extract the `non-perturbative' information encoded in its large-order growth. 
By viewing the aforementioned transseries as solutions to finite difference (shift) equations in the Mandelstam variables, we then obtain an alternative route to accessing the high-energy perturbative behaviour of string amplitudes that bypasses both the explicit evaluation and the regularization of the underlying string integrals, for an arbitrary number of external states.

It is natural to ask whether the fixed-angle high-energy $1/\alpha'$-expansion can be formulated in a language that treats the low- and high-energy regimes on the same footing. A particularly convenient route is to recast the 
$\alpha'$-dependence into a first-order differential system (of Gauss-Manin/Aomoto type) and, equivalently, into a complex Mellin/Laplace representation. In such a framework the various asymptotic limits are expected to arise as solutions of this differential equation or by standard contour manipulations, and provide an analytic handle on the emergence of Stokes phenomena. This question is sharpened by the fact that the $\alpha'$- and $1/\alpha'$-expansions of string amplitudes are of markedly different nature: the low-energy $\alpha'\to 0$ expansion is tied to localization on the string worldsheet and the coefficients in the expansion typically involve Riemann zeta values of positive weight, $\zeta(n>1)$, whereas the fixed-angle 
high-energy $\alpha'\to\infty$ regime is governed by a saddle-point expansion in $1/\alpha'$, whose coefficients naturally involve zeta values at non-positive integers $\zeta(n\le 0)$ (equivalently Bernoulli-number data) \cite{Kervyn:2025wsb}. A unified differential/Mellin formulation should therefore not only connect two limits, but also clarify how these two arithmetic structures---positive-weight versus non-positive-weight zeta data---arise from a common analytic mechanism.

The high-energy limit of string amplitudes has recently garnered renewed attention from multiple directions. High-energy string theory naturally connects to the so-called \textit{tensionless limit} ($\alpha' \to \infty$ or $T \to 0$), which describes a corner in which strings become long and floppy. In this regime, extremely extended strings probe very high energies and correspond to small observers, i.e.,~sub-Planckian length scales. As a theory of quantum gravity, this is precisely the context where genuinely stringy effects and the extended nature of strings are expected to dominate, in sharp contrast to the field-theory limit ($\alpha' \to 0$), where strings effectively become point-like and quantum-gravitational effects are suppressed. We refer to the recent review \cite{Bagchi:2026wcu} on tensionless strings for details.

Exact computations of tensionless string amplitudes require the formulation of worldsheet techniques that directly probe this ultra-high-energy regime. Such a description has only been proposed recently in terms of null-string dynamics and  the tensionless string amplitudes are related to  the tensile string amplitudes in the high energy limit \cite{Bagchi:2026iyu}. These results support the earlier suggestion \cite{Kervyn:2025wsb} that the functional dependence of tensile string amplitudes in the high-energy limit, together with their subleading corrections, captures the structure of tensionless string scattering after an appropriate identification of parameters.
Therefore  to probe the ultra-high-energy regime it is fruitful  to study the (asymptotic) behaviour of tensile string amplitudes in the $\alpha' \to \infty$ limit, which is the topic of this work. 

Although indirect, this strategy has already proven highly fruitful. As mentioned above, string amplitudes exhibit a striking simplicity in this limit: they are exponentially suppressed at tree level \cite{Veneziano:1968yb}, and the worldsheet path integral localises onto dominant classical configurations of vertex operators, or \textit{saddles} \cite{Gross:1987kza}. This simplification readily renders several salient features of the theory manifest, most notably the emergence of \textit{higher-spin symmetry}, which appears in the form of an infinite set of linear relations among scattering amplitudes of different string states, valid order by order in perturbation theory in the $\alpha' \to \infty$ limit \cite{Gross:1988ue}. As a framework with very few intrinsic parameters, string theory is expected to possess a large symmetry structure that is not manifest in its conventional formulation.\footnote{By analogy with field theory, where low-energy dynamics obscure the full unbroken gauge symmetry, it is natural to expect that high-energy scattering amplitudes probe a larger, unbroken symmetry. The same reasoning is believed to apply in the $\alpha' \to \infty$ limit of string theory, where the tensionless sector is expected to be an UV fixed point and encode a vast, yet poorly understood symmetry algebra; see \cite{Lee:2025akf} for a review.} The study of tensionless strings has therefore long been intertwined with the search for emergent higher-spin theories, such as Vasiliev’s \cite{Sundborg:2000wp, Vasiliev:2003cph}, with the expectation that the spontaneous breaking of higher-spin symmetries gives rise to the infinite tower of massive states of the tensile string. Related ideas have also been invoked in discussions of the high-temperature Hagedorn phase \cite{Atick:1988si}, where the exponential growth of the single-string density of states may signal a transition to a phase governed by degrees of freedom distinct from those of conventional (tensile) string theory.

\paragraph{Outline.} 
This work is organized as follows. In \S\ref{sec: sec2}, we formulate the problem and describe the structure of tree-level open superstring amplitudes at low and high energies, for arbitrary multiplicity, highlighting their number-theoretic features. We also review the asymptotic structure of the four-point amplitude, obtained via steepest descent, and its analytic continuation.  In \S\ref{sec: sec3}, we apply resurgence theory to derive high-energy transseries for general kinematics, working explicitly at four points where the full resurgent structure is accessible. We then show that these transseries follow from difference equations in the Mandelstam invariants satisfied by the worldsheet integrals, without requiring their exact analytic form. This allows us to derive asymptotic $1/\alpha'$-expansions for arbitrary multiplicity. In \S\ref{MellinStokes}, we propose for four points a relation between low- and high-energy expansions via an Aomoto-Gauss-Manin connection and a Mellin-Barnes representation providing a unified description of both expansions.
Finally, in \S\ref{sec: sec5}, we reinterpret our results in terms of twisted intersection theory. In particular, we formulate a high-energy double-copy relation in terms of Lefschetz thimbles. We conclude with future prospects in \S\ref{sec: conclusion}.


\input{resurgenceString.tex}
\input{complexMellinSpace.tex}
\input{resurgenceClosedKLT.tex}

\section{Concluding remarks}
\label{sec: conclusion}

In this work, we have investigated the high-energy structure of tree-level string amplitudes from complementary perspectives: locally via saddle-point expansions (\S\ref{sec: sec2}), algebraically via difference equations and their asymptotic structure (\S\ref{sec: sec3}), analytically via the Gauss-Manin connection and Mellin-Barnes representations (\S\ref{MellinStokes}), and geometrically via twisted intersection theory and Lefschetz thimbles (\S\ref{sec: sec5}). In particular, we systematically explored the high-energy structure of tree-level string amplitudes using resurgence theory, and analyzed how monodromies arising from analytic continuation in the Mandelstam invariants are encoded in the transseries completion of the asymptotic expansion. 

By reformulating string amplitudes as solutions to difference equations in the kinematic invariants, we derived asymptotic expansions for arbitrary multiplicity and kinematics in \S\ref{sec: sec3}. A key observation is that, in the high-energy limit, these expansions involve only rational (Bernoulli) coefficients, in stark contrast with the low-energy regime where intricate patterns of multiple zeta values appear. Thus the ring of periods appearing in the power series expansions \req{low-energy expansion} and \req{Morse} at low and high $\ap$, respectively changes. We further verified that the algebraic spectral curve associated with the difference connection is in precise one-to-one correspondence with the scattering equation curve. To our knowledge, this is the first study of high-energy string amplitudes from the perspective of their difference equations. A complete understanding of the high-energy structure of string amplitudes will require solving the connection problem outlined in \S\ref{sec: physical kinematics and Stokes}. In light of the discussion in \S\ref{sec: sec5}, this data should not only reproduce the results of \cite{Eberhardt:2024twy}, but also describe intersection numbers of thimbles, without requiring their explicit construction.

As mentioned in the introduction, the high-energy limit of string theory is closely related to the tensionless limit. Together with the recent work \cite{Kervyn:2025wsb}, the present work builds on and extends the seminal results of \cite{Gross:1987kza, Gross:1989ge}. In this picture, the subleading terms in the $1/\alpha'$ expansion arise naturally as quantum fluctuations around the leading classical saddle. Our discussion of Stokes phenomena and connection problems in \S\ref{sec: sec3} shows that a careful treatment of `non-perturbative' contributions is essential to consistently connect different kinematic regimes, in particular those involving massive string resonances. The role of these effects may directly be anticipated at the level of tensionless string   vertex operators  recently constructed in \cite{Bagchi:2026iyu}. 
 
The reformulation in \S \ref{MellinStokes} reveals that the apparent dichotomy between the $\ap$-expansion and the fixed-angle $1/\ap$ saddle-point regime is in fact an artifact of asymptotic truncation. At the level of the differential system \req{DGL} and Mellin integral \req{eq:MB-bridge} they correspond to distinct Stokes sectors of a single resurgent object. Although this construction has only been established for $n=4$, it would be interesting to extend this description to $n > 4$~\cite{Progress1}.

Most of the machinery underlying \S\ref{sec: sec3} and \ref{sec: sec5} is currently restricted to genus-zero amplitudes, where the framework of twisted de Rham cohomology and its  antiholomorphic
counterpart are well established. 
It remains an open and interesting question to investigate difference equations and thimbles in connection with twisted intersection theory 
at genus one superstring amplitudes.
Recent progress has instead focused on developing new genus-one contour prescriptions to regulate divergences \cite{Manschot:2024prc, Eberhardt:2023xck, Baccianti:2025gll}. These methods enabled the numerical evaluation of one-loop amplitudes in physical kinematics at finite $\alpha'$ \cite{Baccianti:2025whd}, and were further refined in \cite{Baccianti:2026lpc} to incorporate the infinite families of complex saddles contributing in the $\alpha' \to \infty$ limit. Performing a steepest descent analysis in this setting requires splitting the integration over the modular parameter into independent integrations over $\tau$ and $\bar{\tau}$. This effectively enlarges the space of saddle points, introducing additional contributions beyond those associated with monodromies of the punctures, present at tree level. The multiplicities of these saddles are pinned down in \cite{Baccianti:2026lpc} through a bootstrap procedure combined with numerical matching.
From our perspective, these multiplicities should be naturally encoded in the Stokes data of the problem. It would therefore be very interesting to investigate whether resurgence techniques can reproduce the results of \cite{Baccianti:2026lpc}, or complement the bootstrap method. In light of our discussion in \S\ref{sec: sec5}, solving the connection problem may also provide new insight into the structure of twisted intersection numbers at genus one. Finally, a systematic derivation of asymptotic expansions in this context 
would shed light on their period structure. 
We plan to report on a generalization of our results to higher loops in future work.

In \S \ref{sec: sec5} in eq.~\req{KLTa} we reformulated the KLT representation of closed string amplitudes
in terms of Lefschetz thimbles associated with the critical points of the
Koba-Nielsen potential.
In this basis, each integral localizes onto a single saddle configuration,
and KLT admits a geometric interpretation as a bilinear pairing of these
saddle contributions \req{Morsei} via the intersection matrix.
The leading high-energy limit is governed by the scattering equations \req{SQE},
while subleading corrections \req{WefindKir} are systematically controlled by higher derivatives of the
Koba-Nielsen potential, providing a geometric origin of string-theoretic
$\alpha'$-corrections beyond the leading high-energy order.

The high-energy limit of string amplitudes has also gained 
recent interest in the study of celestial holography. The string worldsheet is
identified with the celestial sphere in the zero tension limit of tree-level string amplitudes \cite{Stieberger:2018edy,Castiblanco:2024hnq,CanazasGaray:2025xlh}.
This suggests an interesting link between celestial conformal field theories and tensionless string theories \cite{Kervyn:2025wsb} and illustrates how boundary conformal data at null infinity may arise from a fundamentally geometric and worldsheet-like
description  \cite{Dong:2025qiv}.
With the resurgent features of the open string amplitude at hand exhibiting  Stokes phenomena on the celestial sphere we perform an in-depth analysis of celestial string amplitude in the complex $\beta$-space in \cite{progress}, and thereby confirm and expand on the results of \cite{Kervyn:2025wsb}.

\acknowledgments

We wish to thank Murad Alim, Zoltán Bajnok, Johannes Broedel, Cl\'ement Dupont, 
Maximilian Haensch, Saiei-Jaeyeong Matsubara-Heo, Bernd Sturmfels, Don Zagier, and Shun-Qing Zhang for many valuable discussions.
This work is supported by the DFG grant 508889767 {\it 
Forschungsgruppe `Modern foundations of scattering amplitudes'}.

\appendix
\input{ODEcoefs.tex}
\input{resurgenceGamma.tex}
\input{contiguitymatrices}

\bibliography{biblio.bib}
\bibliographystyle{JHEP}

\end{document}

%% file: packages.tex
\usepackage{jheppub} 
\usepackage{lineno}

\usepackage{empheq}

\usepackage{xcolor}
\usepackage[dvipsnames]{xcolor}

\usepackage{tikz}
\usetikzlibrary{decorations.markings,calc,arrows.meta,decorations.pathmorphing}

\usepackage{caption}
\usepackage{subcaption}

\usepackage[most]{tcolorbox}
\newtcolorbox{todo}[1][]{
  colback=red!5!white,   
  colframe=red!50!black, 
  fonttitle=\bfseries,
  boxrule=0.8pt,
  arc=3pt,                
  left=6pt,right=6pt,top=6pt,bottom=6pt
}
\newtcolorbox{remark}[1][]{
  colback=blue!5!white,   
  colframe=blue!50!black, 
  fonttitle=\bfseries,
  boxrule=0.8pt,
  arc=3pt,                
  left=6pt,right=6pt,top=6pt,bottom=6pt
}

\usepackage{physics}

\usepackage[table]{xcolor} 
\usepackage{array}         

\usepackage{bm} 

\newcommand{\TODO}{\textcolor{red}{TODO} }

\newcommand{\deriv}{\mathrm{d}}
\newcommand{\red}{\textcolor{red}}

\newcommand{\stokes}{\mathrm{S}}
\newcommand{\Stokes}{\bm{\mathrm{S}}}
\newcommand{\expterms}{\mathcal{E}} 
\newcommand{\npsaddles}{\mathcal{P}} 
\newcommand{\shift}{\sigma} 
\newcommand{\KN}{\mathrm{KN}}

\usepackage{fancyvrb}
\usepackage{listings}  

\usepackage{enumitem}

\usepackage{comment}

\usepackage[makeroom]{cancel}

\usepackage{import}

\usepackage{import}
\usepackage{xifthen}
\usepackage{pdfpages}
\usepackage{transparent}


%% file: resurgenceString.tex
\section{Low- and high-energy structure of tree-level  string  amplitudes}
\label{sec: sec2}

Physical quantities (such as prepotentials, coupling constants, cusp anomalous dimensions, etc.)~can be expanded at different points in their moduli space. Comparing the number-theoretic content and  underlying quantum effects 
at theses points often leads to new physical insights. String amplitudes depend on the inverse string tension $\ap$ and may thus also be expanded at different values in  $\ap$. The expansion around $\ap=0$ corresponds to the low-energy expansion, while the expansion at $\ap=\infty$ is tied to high-energy physics. Comparing 
these two expansions should give new insights into high- and low-energy physics and possible connections between these two regimes.

Clearly, the inverse string tension $\ap$ is a positive real parameter, but
from a mathematical viewpoint it may be considered as complex number, i.e.: 
\be\label{Spacealpha}
\ap\in \mathbb{C}.
\ee
The low-energy expansion  of string amplitudes relates to localization on the string worldsheet (through moduli-space integrals of punctured
Riemann surfaces) and is by now well understood. On the other hand, comparatively little is known about the high-energy behaviour of string amplitudes, driven by saddle-point approximations.

\subsection{Periods in low- and high-energy string expansions}

\label{sec: periods low high amplitudes}

Tree-level $n$-point open string amplitudes are described by a basis of $(n-3)!$  worldsheet disk integrals on the  configuration space of $n$ marked points on the boundary of the disk  ${\cal M}_{0,n}(\mathbb{R})=\{(z_2,z_3,\ldots,z_{n-2})\in (\mathbb{R}\mathbb{P}^1)^{n-3}\;|\;z_i\neq z_j\ {\rm for}\ i\neq j\}$ \cite{Mafra:2011nv,Mafra:2011nw,Broedel:2013tta}
\begin{align}\label{Zint}
Z_{\pi\rho}(\ap)&\coloneqq Z_\pi(1,\rho(2,\ldots,n-2),n,n-1)\cr
&\coloneqq \lf(\prod_{j=2}^{N-2}
\int\limits_{C_\pi} \deriv z_j\ri) \   \fc{\prod_{i<j}^{n-1} \abs{z_{ij}}^{-\ap\hat{s}_{ij}}}{  z_{1 2_\rho} z_{2_\rho 3_\rho} \cdots z_{(n-3)_\rho (n-2)_\rho}},
\end{align}
with permutations $\pi,\rho\in S_{n-3}$, $j_\rho \coloneqq \rho(j)$ and $z_{ij} \coloneqq z_i-z_j$. The real (iterated) integrals \req{Zint} refer to a particular (colour) ordering $\pi$ and are integrated over the $(n-3)$-cycle parameterized by the domain
\be\label{domain}
C_\pi\coloneqq\{z_j\in \mathbb{R}\ |\ 0<z_{\pi(2)}<\ldots<z_{\pi(n-2)}<1\}.
\ee
Furthermore, the functions \req{Zint} are parametrized by the kinematic invariants
\begin{equation}\label{Mandelstam}
    \hat{s}_{ij} \coloneqq (k_i + k_j)^2, \quad  s_{ij}=\ap\;\hat{s}_{ij}, 
\end{equation} 
subject to on-shell (massless) condition $k_i^2=0$ and momentum conservation $\sum_{i=1}^n k_i=0$. Convergence of \req{Zint} is ensured by imposing $\Re (s_{i,i+1}) <0$ and $\Re(s_{ij}) <1, \, j\!\neq\! i+1$. In addition, in \req{Zint} we have fixed  the PSL$(2, \mathbb{R})$ symmetry by choosing $z_1 = 0$, $z_{n-1} =1$ and $z_n = \infty$.   After parametrizing the boundary integration $C_\pi$  the integrals \req{Zint}
 reduce to generalized Euler integrals \cite{Oprisa:2005wu,Stieberger:2006te}.
The dependence on the inverse string tension $\ap$ in \req{Zint} enters 
solely through the  $\tfrac{n}{2}(n-3)$ independent Mandelstam variables \req{Mandelstam}.

\paragraph{Low-energy regime.}
For the low-energy expansion  at $\ap=0$ one typically  obtains a power series of the form
\be \label{low-energy expansion}
Z_{\pi\rho}(\ap)=(-\ap)^{3-n}\ S^{-1}_{\pi\rho}+\sum_{k=2}^\infty (-\ap)^{3-n+k}\ b_k.
\ee
The first term of this series can be completely specified by the field-theory limit of the momentum kernel $S$  \cite{Kawai:1985xq,Bern:1998sv,Bjerrum-Bohr:2010pnr}, a $(n-3)!
\times (n-3)!$ homogeneous matrix\footnote{The $(n-3)!\times (n-3)!$ matrix $S$ has entries $S_{\rho,\si}=S[\rho|\si]$, with the rows and columns corresponding to the orderings $\rho \equiv \{\rho(2),\ldots,\rho(n-2)\}$ and
$\si \equiv \{\si(2),\ldots,\si(n-2)\}$, respectively. We have~$S^\mathrm{t}=S$.} of degree $(n-3)$ in the
Mandelstam variables $\hat{s}_{ij}$,
\be\label{KLTkernel}
S[\rho|\si]:=S[\, \rho(2,\ldots,n-2) \, | \, \si(2,\ldots,n-2) \, ] = \prod_{j=2}^{n-2} \Big( \, \hat{s}_{1j_\rho} \ + \ \sum_{k=2}^{j-1} \theta(j_\rho, k_\rho) \, \hat{s}_{j_\rho,k_\rho} \, \Big),
\ee
with $\theta(j_\rho,k_\rho)=1$ if the ordering of the legs $j_\rho,k_\rho$ is the same in both $\rho(2,\ldots,n-2)$ and $\si(2,\ldots,n-2)$, and zero otherwise.
The expansion coefficients $b_k$ in \eqref{low-energy expansion} are polynomials of weight $3-n+k$ in the Mandelstam invariants \req{Mandelstam} multiplied by some 
 (Euler-Zagier) multiple zeta values (MZVs) of weight $k$ \cite{Brown:2009qja,Stieberger:2009rr,Schlotterer:2012ny,Broedel:2013aza}. The latter are defined as \cite{Zagier1994}
\be\label{mzv}
\zeta(n_1,n_2,\ldots,n_r)=\sum_{0<k_1<k_2<\leq\ldots<k_r}\fc{1}{k_1^{n_1}k_2^{n_2}\cdots k_r^{n_r}}
\ee
at depth $r$ with positive integers $n_i, i=1,\ldots,r$ and $n_r\geq 2$ with weight $\sum_{i=1}^r n_i$. 

\paragraph{High-energy regime.}
To access the high-energy regime, on the other hand, one rewrites the generalized Euler integrals \req{Zint} as
\be\label{KLaplace}
Z_{\pi\rho}(\ap)=\oint_{\Cc} \deriv^{n-3}\vec z\ g(\vec z)\; e^{-\ap S}
\ee
with $\Cc$ a suitable integration cycle (to be specified later) and the Morse function 
\be
\label{eq: morse action}
S(\{z_l\};\hat s)=\sum_{i<j}\hat{s}_{ij}\;\ln(z_i-z_j)
\ee
on ${\cal M}_{0,n}$. The generic structure of the large-$\ap$ expansion then takes the form
\be\label{Morse}
Z_{\pi\rho}(\ap)=\lf(\fc{2\pi}{-\ap}\ri)^{\tfrac12 (n-3)}\ \sum_{i=1}^{(n-3)!}Z^{(i)}_{\rho}(\ap),
\ee
which can be anticipated by a Laplace-type approximation\footnote{It is convenient \cite[\S 6.6]{Bender:1999box} to write the integrand in exponential form \req{KLaplace}. For $\ap\in \mathbb{R}_+$ we have: 
$$e^{-\ap S(z)}=e^{-\ap\Re S(z)-i\ap\Im S(z)},$$
with the globally single-valued Morse function \req{eq: morse action}. 
In the steepest-descent analysis the dominant contributions arise from critical
points of the logarithmic potential \req{eq: morse action},
\be\label{cond1}
\nabla \Re S(z)=0,
\ee
which determine the saddle points $z^{(i)}$ of the magnitude of the integrand. For Euler-type integrals  \req{Zint}  one can choose the integration cycle $\Cc$ so that the
contour passes through saddles $z^{(i)}$ where the phase of the integrand is constant along this cycle:
\be\label{cond2}
\Im S(z)=\lf.\Im S(z)\ri|_{z=z^{(i)}}.
\ee
At such points the integrand is real (multiplied by a total constant phase factor $e^{-i\ap\Im S(z)}|_{z=z^{(i)}}$) and positive, so the local expansion reduces to a Laplace-type integral with quadratic leading behaviour.
This choice simplifies the stationary phase analysis and ensures that the
steepest-descent contour is aligned with directions along which
$\Re S$ decreases most rapidly away from the saddle. 

Furthermore, the conditions \req{cond1} and \req{cond2} also admit a natural geometric interpretation in terms of
steepest-descent cycles. Splitting $S$ into its real and imaginary parts, the gradient flow of $\Re S(z)$ determines the  directions along which the magnitude of the integrand decreases most rapidly. The corresponding (real) 
($n-3$)-dimensional integration cycles, called steepest-descent paths or \textit{Lefschetz thimbles},
follow the downward flow of $\Re S$. Along such cycles the branch
$e^{-i\ap\Im S(z)}$ remains constant. By choosing the cycle
so that it passes through a saddle point with $\Im S={\rm const.}$, the integrand is
real and positive at the critical point and the local expansion takes the
standard Laplace form. In this way the multidimensional stationary phase
approximation reduces locally to a Gaussian integral governed by the Hessian
of $S$ at the saddle.} (stationary phase or steepest descent approximation) of \req{KLaplace} resulting in a sum over $(n-3)!$  saddles $\{z_l^{(i)}\}$. The latter follow from the $n-3$ saddle-point equations \cite{Gross:1987ar,Gross:1987kza} (or \textit{scattering equations} \cite{Cachazo:2013hca}):
\be\label{SQE}
\fc{\p S}{\p z_i}=\sum_{j\neq i} \fc{\hat{s}_{ij}}{z_i-z_j}=0.
\ee
Then, in \req{Morse} there are the individual $(n-3)!$ contributions 
\be\label{Morsei}
Z^{(i)}_{\rho}(\ap) = \left[\det H(\{z_l^{(i)}\})\right]^{-1/2}\ e^{-\ap S(\{z_l^{(i)}\})}\ \sum_{k\geq0}\fc{c^{(i)}_k}{(-\ap)^k}
\ee
referring to one saddle-point solution $i$ of \req{SQE}. 
The Hessian $H$ is given by the matrix $H_{ij}=\lf.\p_i \p_j S \ri|_{z_l^{(i)}}$ at the saddle point $\{z_l^{(i)}\}$. 
The saddles $z^{(i)}$ are assumed to be non-degenerate, i.e., $\det H\neq 0$, which can be achieved generically by choosing appropriate kinematic invariants \req{Mandelstam}.
Note that there are $\tfrac n2 (n-3)$ dimensional subregions of the latter, giving rise to $(n-3)!$ real  solutions  of  \req{SQE}  with each solution corresponding to one ordering $\pi\in S_{n-3}$ of \req{Zint} \cite{Cachazo:2016ror}.
In this case in \req{KLaplace} the contour $\Cc$ is a real line integral from $-\infty$ to $+\infty$. Other kinematic configurations may lead to complex saddles with $\Cc$ a path of steepest descent (or \textit{Lefschetz thimble}) and a 
steepest descent approximation  is appropriate. 

\begin{figure}[H]
    \centering
    \includegraphics[scale=.6]{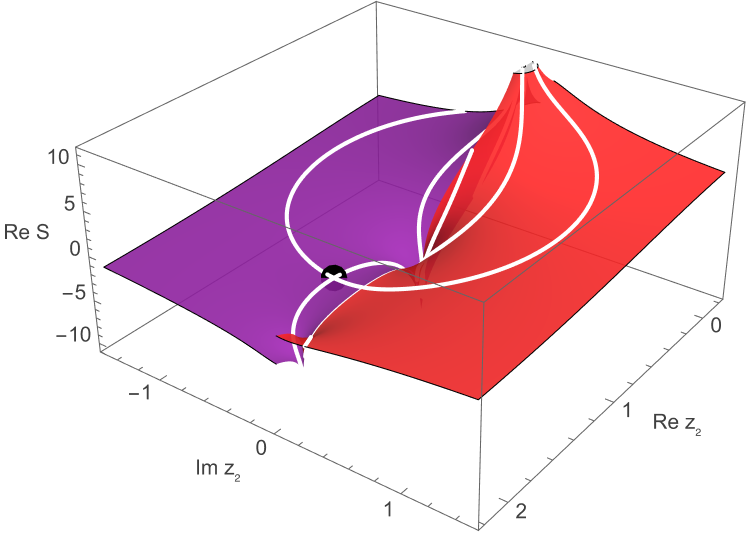}
    \caption{Morse function $\Re S$ with $(z_1,z_3,z_4)=(0,1,2)$ for $n=4$ for $\hat{s}_{12}=-3+\tfrac{i}{2},\;\hat{s}_{13}=1+\tfrac{i}{2},\;\hat{s}_{14}=2-i$ with saddle point $z_2=\tfrac32-\tfrac{i}{4}$ (black dot) and Lefschetz thimbles (white curves).}
    \label{Thimbles}
\end{figure}

As an illustrative example in Figure~\ref{Thimbles} we display the case $n=4$ for a specific choice of kinematics \req{Mandelstam}.  We draw the real part of the Morse potential
\req{eq: morse action} in the complex $z_2$-plane. There is one saddle point following from \req{SQE}. Various cycles of steepest descent and steepest ascent can be constructed starting at this saddle point and enforcing the condition \req{cond2}. These curves generically end on divisors $z_i=z_j$ where the potential \req{eq: morse action} becomes singular.

The coefficients $c_k$ in \eqref{Morse} can be obtained by Taylor expanding $S$ and $g$ and performing Gaussian integrations.  
For arbitrary $n\geq4$ we find\footnote{Our expression \req{WefindKir}  matches the formulae given in \cite{Kirwin} for $d\equiv n-3 =1,2,3$ and $H^{ij}=1_{\bf d}$, except that there are missing factors of $1/2$ in the last two lines for $d=3$.}
\begin{align}
 c_0&=1,\nonumber\\
 c_1&=\fc12\; g_{ij}\; H^{ij}-\fc12\; g_i\; S_{jkl}\;H^{ij}H^{kl}\label{WefindKir}\\
 &+g\lf(\fc{1}{8}\; S_{ijkl}\;H^{ij}H^{kl}-\fc{1}{12}\; S_{ijk}S_{lmn}\; H^{il}H^{jm}H^{kn}-\fc{1}{8}\; S_{ijk}S_{lmn}\; H^{ij}H^{kl}H^{mn}\ri), \ldots ,\nonumber
 \end{align}
with $\lf.g_i=\p_i g\ri|_{z_l^{(i)}},\ \lf.S_{ijk}=\p_i\p_j\p_k S\ri|_{z_l^{(i)}}$, etc. Thus, the coefficients $c_k$ involve only combinatorial quantities and Taylor coefficients of $S$ and $g$. 

For $n=4$ the coefficients $c_k$ involve  single zeta values $\zeta(1-2k)$ at non-positive integer and have been listed in  \cite{Kervyn:2025wsb}.
The latter are related to the Bernoulli numbers as:
\be\label{rational}
\zeta(1-2k)=-\fc{B_{2k}}{2k}=\fc{(-1)^{k}}{2^{2k-1}}\ \fc{(2k-1)!}{\pi^{2k}}\ \zeta(2k), \quad k\geq1.
\ee
Motivated by this observation, we now address the question of which types of periods are expected to appear in the large-$\ap$ expansion for $n>4$. With \req{Morse} let us refine our question into: {\it can we detect any generic periods inside the coefficients   $c^{(i)}_k$? }

Before turning to explicit calculations, note that the appearance of 
positive-weight MZVs \req{mzv} for generic $n$ can be readily excluded in \req{Morse}. 
Indeed, the perturbative WKB expansion around an isolated non-degenerate saddle $z^{(i)}$ of an Euler-type disk integral \req{Zint} is entirely controlled by local Gaussian fluctuations. Its coefficients $c^{(i)}_k$ are universal polynomials in the derivatives of the phase function \req{Morse} and in the entries of the inverse Hessian $H$ at the critical point $z^{(i)}$. At the latter the inverse Hessian and all higher derivatives $S_{ij\ldots q}$ are rational functions thereof. As a consequence, the resulting asymptotic series \req{Morse} is generated by Stirling- and Euler-Maclaurin-type structures and should involve only rational numbers and Bernoulli numbers.

From the perspective of periods and transcendental weight, this observation reflects the fact that a local saddle analysis probes only the formal neighborhood of a Morse critical point. In contrast to \req{Zint}, no iterated integration occurs in the Laplace-type approximation \req{KLaplace}, and hence no new mixed Tate periods are generated. In particular, multiple zeta values \req{mzv}---which arise from iterated integrals and non-Abelian monodromy in the small-$\ap$ regime---cannot be produced by the perturbative large-$\ap$ expansion. 

The Euler-Zagier multiple zeta function \req{mzv} admits a meromorphic continuation to $\mathbb{C}^r$. Its values at non-positive integers were determined in \cite{AkiyamaEgamiTanigawa2001}  and are given by explicit
polynomials in Bernoulli numbers. In particular, for fixed depth r one has:
\be\zeta(-n_1,\dots,-n_r)\in \mathbb{Q}.
\ee 
E.g., we have: $\zeta(-r_1,-r_2)=-\tfrac12\zeta(-r_1-r_2)$ with positive integers $r_1,r_2$ and $r_1+r_2\ \mbox{odd}$. Consequently, even if such values appear in the large $\ap$ expansion for $n\geq 5$, they do not introduce new periods beyond $\mathbb{Q}$. Furthermore, multiple zeta function with mixed-sign arguments reduce to $\mathbb{Q}$-linear combinations of positive-weight MZVs. Hence, the transcendental content of the amplitude remains governed by classical positive-weight MZVs.

To conclude, we expect that for $n\geq5$  the large $\ap$ expansion \req{Morse} cannot generate new transcendental periods beyond those already present in the small $\ap$ regime. Furthermore, even for $n\geq6$, where the integration dimension and the number of saddles increase, the perturbative sector attached to each saddle remains confined to the ring generated by rational numbers and Bernoulli numbers. One of the goals of this paper is to explicitly construct the high-energy asymptotic series describing the $\alpha' \to \infty$ regime of string amplitudes, and thereby verify this claim. Let us thus first review the four-point case.

\subsection{High-energy limit of the four-point amplitude} 
\label{sec: four point tree string}

We begin our exploration of high-energy open superstring amplitudes with the simplest non-trivial example of four-point scattering, described by two independent Mandelstam invariants. We choose\footnote{With this choice, we match the conventions of \cite{Stieberger:2018edy, Kervyn:2025wsb}.} $s \coloneqq -s_{12}$, $t \coloneqq - s_{13}$ and $u \coloneqq - s_{23}$, together with the constraint $s+t+u=0$. For $n\!=\!4$ there is only one independent canonically-colour-ordered worldsheet integral of the form \eqref{Zint} to consider, corresponding to  $\pi,\rho=1$: 
\begin{equation}
    \label{eq: integral form factor}
    F(s,u) \coloneqq s\ Z_{11}(\ap) = - s \int_0^1 \deriv x \, x^{-s-1} (1-x)^{-u}.
\end{equation} Using the integral representation of the Beta function
\begin{equation}
    \label{eq: euler integral beta}
    \int_0^1 \deriv x \, x^{\alpha -1} (1-x)^{\beta -1} \eqqcolon B(\alpha, \beta) = \frac{\Gamma(\alpha) \Gamma(\beta)}{\Gamma(\alpha+\beta)}, \quad \Re(\alpha), \Re(\beta) > 0,
\end{equation} one finds
\begin{equation}
    F(s,u) = \frac{\Gamma(1-s) \Gamma(1-u)}{\Gamma(1-s-u)}.
    \label{eq: string form factor}
\end{equation} Throughout this paper we mostly regard the Mandelstam invariants as complex variables, as is standard when discussing analytic properties of scattering amplitudes. While \eqref{eq: string form factor} can be extended to arbitrary kinematics by analytic continuation of Gamma, we stress that \eqref{eq: integral form factor} converges only if $\Re(s) < 0$ and $\Re(u) < 1$. 

Physical fixed-angle scattering in the $s$-channel instead corresponds to \textit{real} Mandelstam invariants satisfying $s>0 > t,u$. Hence, the physical region lies outside the domain of absolute convergence of \eqref{eq: integral form factor}. This situation is typical for string worldsheet integrals \req{Zint} appearing in amplitude computations, where one often derives expressions in a mathematically convenient region and subsequently analytically continues them in the kinematic invariants to reach the physical regime. Resurgence provides a systematic framework to perform this continuation directly at the level of high-energy expansions.

In \cite{Kervyn:2025wsb}, we put forward an asymptotic expansion for $F$ as $\alpha' \to \infty$ obtained from the known Stirling expansion of the Gamma function for large  (non-negative) complex values of its argument \cite{Gradshteyn:1702455},
\begin{equation}
    \Gamma(z) \sim \sqrt{\frac{2\pi}{z}} \left( \frac{z}{e} \right)^z \exp[ \sum_{k=1}^\infty \frac{B_{2k}}{2k(2k-1)} z^{1-2k}], \quad \abs{\arg(z)} < \pi.
    \label{eq: Stirling Gamma}
\end{equation} Above, $B_{2k}$ are the even Bernoulli numbers, a sequence of rational numbers which arise from the exponential generating series \cite{NIST:DLMF}
\begin{equation}
    \label{eq: generating series Bernoulli}
    \frac{t}{e^t -1} = \sum_{n=0}^\infty \frac{B_n}{n!}\ t^n, \quad \abs{t} < 2\pi, 
\end{equation} and relate to the Riemann zeta function through \eqref{rational}. With this, the open string form factor \eqref{eq: string form factor} may now be rewritten in a $1/\alpha'$ expansion. Assuming $\Re(s) < 0, \Re(u) < 1$ as well as $\Re(t) > 0$, we find
\begin{equation}
    F(s,u) \sim \sqrt{2\pi \frac{su}{t}} e^{-s \ln (s) - u \ln (u) - t \ln (t)} (-1)^{-s -u} \expterms(s,u),
    \label{eq: Stirling open string}
\end{equation} where we defined the auxiliary function
\begin{equation}
    \label{eq: def exponential factor}
    \expterms(s,u) \coloneqq \exp[- \sum_{k=1}^\infty \frac{B_{2k}}{2k(2k-1)} \left(\frac{1}{s^{2k-1}} + \frac{1}{u^{2k-1}} + \frac{1}{t^{2k-1}} \right)]
\end{equation} for notational convenience, and recall that all Mandelstams scale linearly with $\alpha'$. Owing to the restricted validity of \eqref{eq: Stirling Gamma}, \eqref{eq: Stirling open string} holds only in an unphysical region.

The asymptotic expansion of $F$ in the physical region where $s \to + \infty$, $u,t \to - \infty$ along the real axis follows from using Euler's reflection formula to recast \eqref{eq: string form factor} as
\begin{equation}
    F(s,u) = u \frac{\sin(\pi t)}{\sin(\pi s)} \frac{\Gamma(-u) \Gamma(-t)}{\Gamma(s)},
\end{equation} where the real part of the argument of each Gamma function is now strictly positive. Using \eqref{eq: Stirling Gamma} again, one obtains the asymptotic series expansion
\begin{equation}
    \label{eq: Stirling physical}
    F(s,u) \sim \sqrt{2\pi \frac{su}{t}} \frac{\sin (\pi t)}{\sin(\pi s)} e^{- s \ln(s) - t \ln (t) - u \ln (u)} (-1)^{-u -t} \expterms(s,u),
\end{equation} valid for $\Re(s) > 0 > \Re(t), \Re(u)$, which now includes the physical region. The transformation from \eqref{eq: Stirling open string} to \eqref{eq: Stirling physical} is clearly discontinuous, and shouldn't be interpreted as a discontinuity of the amplitude itself, but rather as a discontinuous change in the asymptotic representation appropriate to a given kinematic sector. Upon analytic continuation, exponentially suppressed saddle contributions switch on or off, yielding a different asymptotic expansion of the same analytic function. This is the hallmark of Stokes’ phenomenon.

\subsubsection{Asymptotic series and Padé approximant}

In \cite{Kervyn:2025wsb}, we further expanded the exponential factors $\expterms$ common to both \eqref{eq: Stirling open string} and \eqref{eq: Stirling physical} as a formal series in powers of $1/\alpha'$,
\begin{equation}
    \expterms(s, -as) = \sum_{n=0}^{\infty} \frac{c_{n}(a)}{(-s)^n}, \quad a \coloneqq - \frac{u}{s} < 0.
\end{equation} The scalar $a$ is the four-point conformal cross ratio introduced in \cite{Stieberger:2018edy}. When $s,u \in \mathbb{R}$, one has $a = \sin^2(\theta/2)$, with $\theta$ the scattering angle in the center-of-mass frame. Hence, $s,a < 0$ is unphysical, while $s > 0$, $a \in (0,1)$ describe physical $s$-channel scattering; see~Table~\ref{tab: kinematics}. 
\begin{table}[ht]
    \centering 
    \begin{tabular}{|c||c|c|c|}
    \hline
    cross ratio & $s$-channel ($12 \leftrightarrow 34$) & $u$-channel ($23 \leftrightarrow 14$) & $t$-channel ($13 \leftrightarrow 24$) \\ \hline
    $0 < a < 1$ & \cellcolor{cyan!30} $s > 0 > u,t$ & $u,t > 0> s$ & $u,t > 0 > s$ \\ \hline
    $a > 1$ & $s,t > 0 > u$ & \cellcolor{cyan!30} $u > 0 > s,t$ & $s,t > 0 > u$ \\ \hline
    $a < 0$ & $s,u > 0 > t$ & $s,u > 0 > t$ & \cellcolor{cyan!30} $t > 0 > u,s$ \\ \hline
    \end{tabular}
    \caption{Physical (blue) and unphysical kinematic regions for all channels, with $s,u\in \mathbb{R}$.}
    \label{tab: kinematics}
\end{table}

\noindent
This yields a formal asymptotic series in $1/\alpha'$ of the form
\begin{equation}
    \label{eq: asymptotic series F}
    F(s,-as) \sim \sqrt{\frac{2 \pi a s}{1-a}} (-a)^{as} (1-a)^{(1-a)s} \sum_{n=0}^{\infty} \frac{c_{n}(a)}{(-s)^n}, \quad s, a < 0,
\end{equation} for \eqref{eq: string form factor} in the unphysical region (within the convergence domain). While there is no closed form for the coefficients $c_{n}(a)$, 
one may recursively compute the latter efficiently using the first-order differential equation in $s$ satisfied by \eqref{eq: string form factor}, and thereby apply resurgence techniques to \eqref{eq: asymptotic series F} to complete it into a transseries, which extends it to an analytic function in the full complex $s$-plane for all $a \in \mathbb{R}$. We provide details about this recursion relation in Appendix~\ref{app: ODEcoefs} and delay the discussion of the transseries to \S\ref{sec: transseries string}.

\begin{figure}[h]
     \centering
    \includegraphics[width=0.8\textwidth]{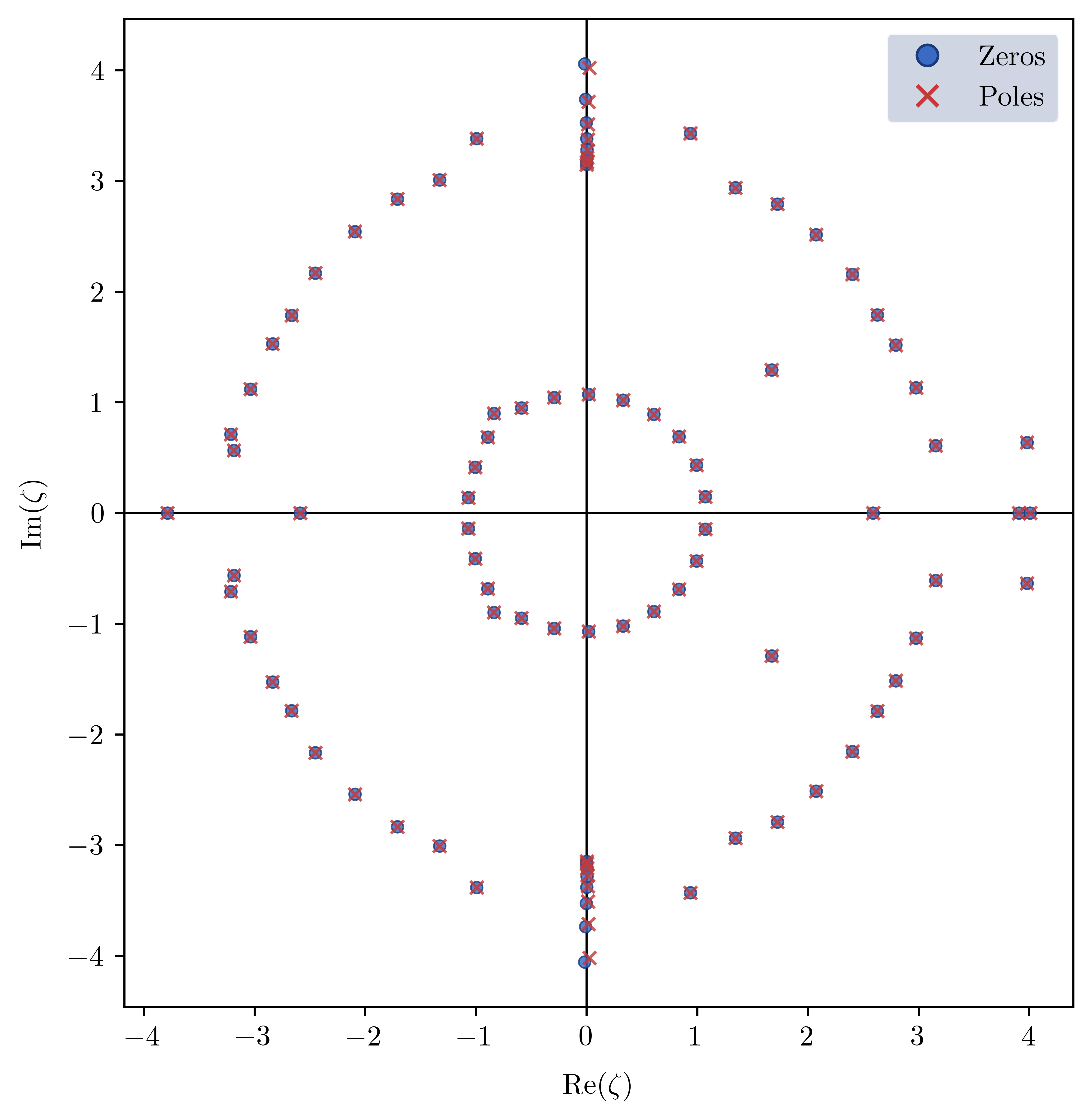}
    \caption{Zeros and singularities in the Borel plane of the diagonal $[100,100]$ Padé approximant of the Borel transform of $\sum_{n=0}^{\infty} c_n(a) / (-s)^n$, for $a=-\tfrac{1}{2}$, with coefficients up to $n_\text{max}=200$ and outliers removed for visual clarity. The rings of spurious singularities are known as \textit{Froissart doublets}, an artefact of the Padé approximation \cite{Gilewicz:1987hj, Baker:1998}. Only the `naked' poles around the imaginary axis signal a true singularity of the Borel transform.}
    \label{fig: pade plot string amp}
\end{figure}

In Figure~\ref{fig: pade plot string amp}, we display the analytic structure of the $[100,100]$ diagonal Borel-Padé approximant for the series $\sum_{n=0}^\infty c_n(a) / (-s)^n$ for $a = - \tfrac{1}{2}$, truncated at $n_\text{max}=200$. 
 The exact Borel transform is generically meromorphic (or has branch cuts) in the complex Borel plane. Padé approximants mimic branch cuts via interlacing poles and zeros \cite{Baker:1998}, so nearly coincident pole-zero pairs should be regarded as artefacts rather than genuine singularities. In contrast, the accumulation of isolated (`naked') poles along the imaginary axis in Figure~\ref{fig: pade plot string amp} indicates a sequence of true complex Borel singularities at
\begin{equation}
\zeta \in \{2 \pi i k \} \cup \{2 \pi i a k\}, \qquad k \in \mathbb{Z} \backslash \{0\}.
\end{equation} A finite-order Padé approximant does not resolve each singularity individually, but instead produces clusters of poles and zeros whose centers approximate their locations, with maximal accuracy near the expansion point $\zeta=0$. This behavior is consistent with the known Borel transform of $\ln \Gamma(z)$ (Appendix~\ref{app: transseries log Gamma}), which exhibits an infinite lattice of poles at $\zeta_k=2\pi i k$. Upon exponentiation to the Gamma function (and its ratios), this lattice is compressed by the Padé approximant into dense vertical stacks, explaining the observed pole bunching. These Borel singularities will later reappear as complex saddles or instanton actions in the corresponding transseries description.

\subsubsection{Thimble analysis and complex saddles}
\label{eq: steepest descent methods Yoda}

Our discussion above relied on evaluating \eqref{eq: integral form factor} as a ratio of Gamma functions, \eqref{eq: string form factor}, and on exploiting the known asymptotic expansion of the latter, \eqref{eq: Stirling Gamma}. This simplification is specific to four-point amplitudes, for which the integral can be evaluated in closed form. For higher-point amplitudes, however, both the \textit{exact} result and its analytic continuation become significantly more intricate. As noted in \S \ref{sec: periods low high amplitudes}, information  as $\alpha' \to \infty$ is typically extracted via a steepest-descent analysis of the string worldsheet integral around its saddle points. Within the region of convergence, the (iterated) Euler cycle already coincides with a steepest-descent contour, and the Laplace approximation applies directly. For more general kinematics, however, the integration contour must be `regularized', and the appropriate Lefschetz thimbles constructed carefully. We briefly review this procedure below.

The regularization of string worldsheet integrals was first discussed by Witten \cite{Witten:2013pra}, originally with the aim of resolving ambiguities in the causal and unitary propagation of strings in spacetime, which arise from the intrinsically Euclidean nature of the worldsheet formulation. Witten's $i \epsilon$-prescription amounts to Wick rotating from Euclidean to Lorentzian worldsheets near all the branch points of the integrand. This procedure has since then been refined in the string amplitude literature to describe general kinematics \cite{Yoda:2024pie, Eberhardt:2024twy} (see also \cite[App.~A]{Mizera:2019vvs}). In particular, this approach ties the characteristic oscillatory behaviour in the physical region---for example, the ratio of sines in \eqref{eq: Stirling physical}---to the topology of the regularized contour, namely through the contribution of an infinite family of complex saddles captured by thimble analysis, besides the single real saddle of \cite{Gross:1987kza}.

We first remind how the leading asymptotic behaviour of the four-point amplitude was historically determined. High-energy fixed-angle $s$-channel scattering corresponds to real $s \to - \infty$ while keeping the ratio $a < 0$ fixed. One may then rewrite \eqref{eq: integral form factor} as
\begin{equation}
    \label{eq: laplace form string form factor} 
    F(s,-as)  = -s \int_0^1 \deriv x \, f(x) e^{-s g(x)},
\end{equation} with $f(x) \coloneqq \tfrac{1}{x}$ and $g(x) \coloneqq \ln x - a \ln(1-x)$. To discuss the $\alpha' \to \infty$ limit, one solves the stationary point equation
\begin{equation}
    g'(x_0) \overset{!}{=} 0 \quad \Rightarrow \quad x_0 = \frac{1}{1-a}.
\end{equation} Given the restricted convergence of \eqref{eq: laplace form string form factor}, $s, a <0$ imply $x_0 \in (0,1)$ and $g''(x_0) < 0$, and this point is a global maximum of $g(x)$ and $(0,1)$ a steepest-descent cycle. The Laplace approximation therefore applies. Expanding to all orders around $x_0$ yields \eqref{eq: asymptotic series F} \cite{Kervyn:2025wsb}.

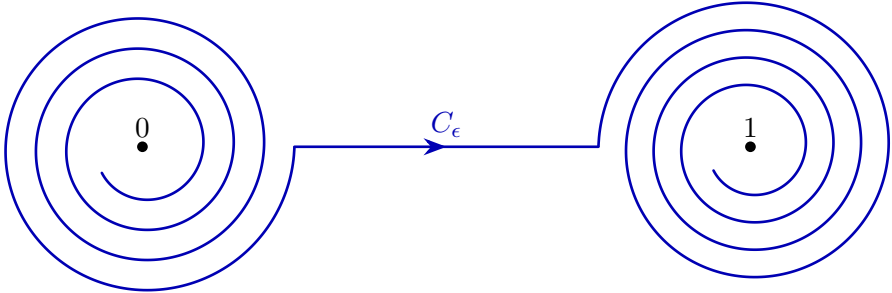
\begin{figure}[ht]
    \centering
    \input{tikzContour}
    \caption{Schematic regularized integration contour $C_\epsilon$ for the four-point amplitude.}
    \label{fig: new contour}
\end{figure}

For more general kinematics, we use Witten's $i\epsilon$-prescription and replace the original integral over $(0,1)$ by the complex contour integral
\begin{equation}
    \label{eq: regularized worldsheet integral}
    F_\epsilon(s,u)
    = -s \int_{C_\epsilon} \deriv z \, z^{-1 - s - i \epsilon} (1-z)^{-u - i \epsilon},
    \qquad \epsilon > 0,
\end{equation} with $C_\epsilon$ chosen to closely follow the boundaries of moduli space and wind infinitely many times around the branch points at $z=0$ and $z=1$ as depicted in Figure~\ref{fig: new contour}, thereby regulating the divergences and providing a well-defined analytic continuation. 

\noindent
In doing so, the path picks up infinitely many monodromies labelled by two winding numbers $(n,m) \in \mathbb{Z}^2$ such that 
\begin{equation}
    \ln(z_{n,m}) = \ln(z_{0,0}) + 2 \pi i n, \quad \ln(1 - z_{n,m}) = \ln(1 - z_{0,0}) + 2 \pi i m,
\end{equation} where $z_{0,0}$ is the trivial saddle on the principal $(0,0)$ sheet and $z_{n,m}$ the saddle on the $(n,m)$ sheet. The latter thus come with an extra phase $e^{- 2 \pi i n (s + 1 + i\epsilon)  - 2 \pi i m (i \epsilon +u)}$ compared to $z_{0,0}$. The next task is to identify which of these saddles contribute to the integral for a given choice of kinematics. This is achieved using thimble analysis (see, e.g., \cite{Witten:2010cx}), which gives a systematic way to deform $C_\epsilon$ into a sum over thimbles $\mathcal{J}_{n,m}$,
\begin{equation}
    \int_{C_\epsilon} \deriv z \, e^{- (s + 1 + i\epsilon) \ln z - (i \epsilon +u) \ln(1-z)} = \sum_{ \{\mathcal{J}_{n,m}\}} (-1)^{(n,m)} \int_{\mathcal{J}_{n,m}} \deriv z \, e^{- (s + 1 + i\epsilon) \ln z - (i \epsilon +u) \ln(1-z)},
\end{equation} where $(-1)^{(n,m)} = \pm 1$, called the \textit{intersection number}, gives the orientation of the thimble $\mathcal{J}_{n,m}$ with respect to $C_\epsilon$. Along each thimble, the integrand is non-oscillating and decays rapidly away from the saddle point, so the Laplace method may once again be applied. This yields
\begin{equation}
    \label{eq: F thimble expansion}
    F_\epsilon(s,u) = \sqrt{\frac{2\pi su}{t}}  (-s)^{-s} (-u)^{-u} t^{-t} \expterms(s,u) \sum_{\{\mathcal{J}_{n,m}\}} (-1)^{(n,m)} e^{- 2 \pi i n (s + 1 + i\epsilon)  - 2 \pi i m (i \epsilon +u)}.
\end{equation} The particular set of thimbles $\{\mathcal{J}_{n,m}\}$ depends on the kinematics considered, which in turn dictates the geometry of the gradient flow around the saddles. We refer to \cite{Witten:2010cx} and the Appendix of \cite{Yoda:2024pie} for details on thimbles, and simply borrow results from the latter below to illustrate the applicability of this method and its connection to resurgence later on.

In the unphysical regime, both $s,u < 0$. One cannot deform the original contour to pick up any complex saddle since the vector field of the thimble flow equation flows into the logarithmic branch points at $z=0$ and $z=1$. Therefore, only the trivial saddle $z_{0,0}$ on the principal sheet contributes, and \eqref{eq: F thimble expansion} yields exactly the form factor of \eqref{eq: Stirling open string}. 

Meanwhile, in the physical regime where $s > 0> t, u$, thimble analysis yields contributions from the saddles on the sheets $(-\infty, 0)$, $(-\infty, 1)$, \ldots, $(-1,0)$, $(-1,-1)$, and $(0,0)$ upon deforming the original contour $C(\epsilon)$ in \eqref{eq: regularized worldsheet integral}, before taking the limit $\epsilon \to 0$. Therefore, \eqref{eq: F thimble expansion} becomes
\begin{multline}
    F_\epsilon(s,u) = \sqrt{\frac{2\pi su}{t}}  (-s)^{-s} (-u)^{-u} t^{-t} \expterms(s,u) \\ \times \left[ 1 + \sum_{n \leq -1} e^{-2 \pi i n (s + 1 + i \epsilon)} -  e^{2 \pi i (i \epsilon +u)}\sum_{n \leq -1} e^{-2 \pi i n (s + 1 + i \epsilon)} \right].
\end{multline} The geometric series converges given $\abs{e^{2 \pi i n (s + 1 + i \epsilon)}} < 1$. Taking $\epsilon \to 0$, we land exactly on the form factor entering \eqref{eq: Stirling physical}. Thimble analysis thus gives a clear interpretation for the discontinuity noted between \eqref{eq: Stirling open string} and \eqref{eq: Stirling physical}, where the contributions of an infinity of complex saddles reproduce the appropriate structure of poles and zeros of the amplitude.

While this example shows that Lefschetz thimble analysis provides a powerful framework for extracting asymptotic information, it suffers from several important limitations. One must often contend with an infinite set of complex saddle points, and identifying which saddles contribute in a given kinematic regime---together with their behaviour across Stokes walls---is highly nontrivial, especially for higher multiplicity. The thimble decomposition itself is not uniform in parameter space, but undergoes discontinuous changes due to Stokes phenomena, complicating analytic continuation in kinematic space. Moreover, the explicit construction of Lefschetz thimbles requires solving gradient flow equations in complexified spaces, which is generally intractable beyond low-dimensional examples. Even when the relevant saddles are known, determining the associated intersection numbers that specify how the original contour decomposes into thimbles is a subtle global problem.

\section{Precision high-energy asymptotics from resurgence}
\label{sec: sec3}

Our discussion of the limitations of thimble analysis above motivates an alternative approach, that does not require carefully constructing steepest descent contours. 

\subsection{Multiplicity \texorpdfstring{$n=4$}{n=4}} 
\label{sec: transseries string}

In this section, we leverage the know resurgence properties of $\Gamma$ to systematically extract both the perturbative and non-perturbative asymptotic data at four points.

\subsubsection{Resurgence of the Euler Beta function}

In resurgence terms, the task at hand is to complete the asymptotic series \eqref{eq: Stirling open string} into a transseries. In this section, we achieve this using the known resurgent properties of the Gamma function. For notational convenience, let us introduce the scaled Gamma function 
\begin{equation}
    \Gamma^\ast(z) \coloneqq \sqrt{\frac{z}{2 \pi}} \left(\frac{z}{e}\right)^{-z} \Gamma(z), 
\end{equation} and similarly for its reciprocal, $1/\Gamma^\ast$. Transseries for $\Gamma^\ast$ and $1/\Gamma^\ast$ around $\abs{z}=\infty$ are given in \cite{Nemes:2022} and derived in Appendix~\ref{app: transseries Gamma and reciprocal} for completeness. One finds
\begin{equation}
    \label{eq: main transseries Gamma}
    \Gamma^\ast(z) = \exp[\sum_{k=1}^{\infty} \frac{B_{2k}}{2k(2k-1)} z^{1-2k} ] \left(1+ \sum_{k=1}^{\infty} \Stokes_k(\theta) \,  e^{\pm 2 \pi i k z} \right), \quad \abs{z} \to \infty,
\end{equation} where $\theta \coloneqq \arg(z) \in (- \pi, \pi)$ and we introduced the Stokes coefficients
\begin{equation}
    \label{eq: Stokes transseries Gamma}
    \Stokes_k(\theta) \coloneqq \begin{cases}
            0 & 0< \abs{\theta} < \frac{\pi}{2}, \\ 
            \frac{1}{k!} \left(\frac{1}{2}\right)_k & \theta = \pm \frac{\pi}{2}, \\ 
            1 & \frac{\pi}{2} < \abs{\theta} < \pi.
        \end{cases}
\end{equation} Likewise, the transseries for the reciprocal of Gamma writes 
 \begin{equation}
    \label{eq: main transseries reciprocal Gamma}
    \frac{1}{\Gamma^\ast(z)} = \exp[- \sum_{k=1}^{\infty} \frac{B_{2k}}{2k(2k-1)} z^{1-2k} ] \left(1 - \sum_{k=1}^{\infty} \tilde{\Stokes}_k(\theta) \, e^{\pm 2 \pi i k z} \right), \quad \abs{z} \to \infty,
\end{equation} only this time with different Stokes coefficients, namely
\begin{equation}
    \label{eq: Stokes transseries reciprocal Gamma}
    \tilde{\Stokes}_1(\theta) \coloneqq \begin{cases}
            0 & 0< \abs{\theta} < \frac{\pi}{2}, \\ 
            \frac{1}{2} & \theta = \pm \frac{\pi}{2}, \\ 
            1 & \frac{\pi}{2} < \abs{\theta} < \pi,
        \end{cases} \quad \text{and} \quad \tilde{\Stokes}_k(\theta) \coloneqq \begin{cases}
            0 & 0< \abs{\theta} < \frac{\pi}{2}, \\ 
            - \frac{1}{k!} \left( -\frac{1}{2} \right)_k & \theta = \pm \frac{\pi}{2}, \\ 
            0 & \frac{\pi}{2} < \abs{\theta} < \pi,
        \end{cases} \, k \geq 2.
\end{equation} The upper or lower sign in \eqref{eq: main transseries Gamma} and \eqref{eq: main transseries reciprocal Gamma} is taken according to whether $z$ lies in the upper or lower complex half-plane, respectively. Given \eqref{eq: string form factor}, these act as the building blocks for the full transseries of the four-point string form factor. We write
\begin{equation}
    \begin{aligned}
        \frac{\Gamma(-s)\Gamma(1-u)}{\Gamma(1+t)} &= \sqrt{-\frac{2 \pi u}{t}}(-s)^{-\frac{1}{2}} (-1)^{-s -u} s^{-s} u^{-u} t^{-t} \frac{\Gamma^\ast(-s) \Gamma^\ast(-u)}{\Gamma^\ast(t)}
    \end{aligned}
\end{equation} where $s = \abs{s} e^{i \theta_s}$, $u = \abs{u} e^{i \theta_u}$ and $t = \abs{t} e^{i \theta_t}$, with $\theta_i \in (-\pi, \pi)$. Using \eqref{eq: main transseries Gamma} and \eqref{eq: main transseries reciprocal Gamma}, the last term above is
\begin{equation}
    \frac{\Gamma^\ast(-s) \Gamma^\ast(-u)}{\Gamma^\ast(t)} = - \expterms(s,u) \npsaddles(s,u),
\end{equation} with $\expterms$ given in \eqref{eq: def exponential factor} and
\begin{equation}
    \label{eq: np terms}
    \npsaddles(s,u) \coloneqq \sum_{k=0}^{\infty} \Stokes_k(\theta_s \mp \pi) \, e^{\pm 2 \pi i k (-s)} \sum_{l=0}^{\infty} \Stokes_l(\theta_u \mp \pi) \, e^{\pm 2 \pi i l (-u)} \sum_{m=0}^{\infty} \tilde{\Stokes}_m(\theta_t) \, e^{\pm 2 \pi i m t}.
\end{equation} The upper or lower sign is taken according to whether the variable $-s$, $-u$ or $t$ lies in the upper or lower complex half-plane, respectively. The Stokes coefficients $\Stokes_k(\theta)$ and $\tilde{\Stokes}_k(\theta)$ are given by \eqref{eq: Stokes transseries Gamma}  and \eqref{eq: Stokes transseries reciprocal Gamma}, and we defined $\Stokes_0(\theta) = - \tilde{\Stokes}_0(\theta) = 1 \, \forall \theta$ for notational simplicity. The full transseries expansion of the four-point string form factor \eqref{eq: string form factor} then writes
\begin{equation}
    \label{eq: full transseries string form factor}
    F(s,u) = - \sqrt{\frac{2 \pi su}{t}}  (-1)^{-s -u} s^{-s} u^{-u} t^{-t} \expterms(s,u)  \npsaddles(s,u).
\end{equation} Taken together with \eqref{eq: def exponential factor}, \eqref{eq: np terms}, \eqref{eq: Stokes transseries Gamma} and \eqref{eq: Stokes transseries reciprocal Gamma}, this expression analytically continues \eqref{eq: asymptotic series F} across all Stokes sectors in the complex $(s,u)$-plane. Restricting to real invariants, this result correctly describes the high-energy asymptotics associated to any configuration of Table~\ref{tab: kinematics}, regardless of it being associated to physical or unphysical scattering processes.

As a sanity check, we now verify that \eqref{eq: full transseries string form factor} satisfies the Euler reflection property for $s,u \in \mathbb{R}$. In the unphysical domain $s, u < 0$ and $t > 0$, so only the perturbative series should survive in \eqref{eq: full transseries string form factor}. Clearly, we then have $\theta_s \mp \pi = \theta_u \mp \pi = \theta_t = 0$, so $\Stokes_k(\theta_s \mp \pi) = \Stokes_k(\theta_u \mp \pi) = \tilde{\Stokes}_k(\theta_t) = 0$ from \eqref{eq: Stokes transseries Gamma} and \eqref{eq: Stokes transseries reciprocal Gamma}, and we are left with only \eqref{eq: Stirling open string}.
On the other hand, in the physical domain one has $s > 0 > t, u$. Let $s = \abs{s}e^{i \epsilon}$, with $0 < \epsilon \ll 1$.
Then $\theta_s - \pi = \theta_t = \epsilon - \pi$, $\theta_u + \pi = \epsilon$, meaning that
\begin{equation}
    \begin{gathered}
        \lim_{\epsilon \to 0 } \tilde{\Stokes}_1(\epsilon - \pi) =  \lim_{\epsilon \to 0} \Stokes_k(\epsilon - \pi) = 1 \quad \forall k \geq 1, \\ 
        \lim_{\epsilon \to 0} \Stokes_l(\epsilon) =  \lim_{\epsilon \to 0} \tilde{\Stokes}_m(\epsilon - \pi) = 0 \quad \forall l \geq 1, \, m \geq 2.
    \end{gathered}
\end{equation} As a result, taking into account that $s$ lies in the upper complex half-plane so both $-s$ and $t$ are in the lower complex half-plane (i.e., we take the lower sign in \eqref{eq: full transseries string form factor}), we have
\begin{equation}
    \lim_{\epsilon \to 0} \npsaddles(s,u) = \sum_{k=0}^{\infty} e^{+ 2 \pi i k s}
        \left(1 - e^{-2 \pi i t} \right).
\end{equation} Given $\Im(s) > 0$, we have $\abs{e^{2\pi i k s}} < 1$ and the geometric series converges, yielding
\begin{equation}
    \label{eq: sanity check string transseries}
    \lim_{\epsilon \to 0} \npsaddles(s,u) = - (-1)^{-s-t} \frac{\sin(\pi t)}{\sin(\pi s)},
\end{equation} so we recover \eqref{eq: Stirling physical} as expected. The argument naturally extends away from $s,u \in \mathbb{R}$. Ultimately, $\npsaddles(s,u)$ captures exactly the contributions of the complex saddles uncovered in \S \ref{eq: steepest descent methods Yoda}. Resurgence theory therefore provides a precise algebraic understanding of the geometric phenomena underlying asymptotic expansions in kinematic space.

\subsubsection{Difference equations and Bernoulli numbers}
\label{sec: difference N=4}

We now seek to recover the transseries \eqref{eq: full transseries string form factor} for $F$ solely from a difference relation satisfied by its integral representation \eqref{eq: integral form factor}, \textit{before} explicit evaluation of the integral to a Beta function, \eqref{eq: string form factor}. We first assume $\Re(s) < 0, \Re(u) < 1$ and $\Re(t) > 0$ in the convergent domain. Integration by parts (IBP) yields the rank-1 difference system
\begin{subequations} \label{eq: functional indentity F2}
    \begin{align}
        \Delta_s \ln F(s,u) &= \ln \frac{s+u}{s}, \label{eq: IBP N=4 s} \\ 
        \Delta_u \ln F(s,u) &= \ln \frac{s+u}{u}, \label{eq: IBP N=4 u}
    \end{align}
\end{subequations} where we introduced the single-variable difference operator $\Delta_x f(x) \coloneqq f(x+1) - f(x)$. We now consider the solution to this system of difference equations in the asymptotic regime $\Re(s), \Re(u) \to - \infty$. As both $\Re(s), \Re(u) = -\infty$ are irregular singular points of \eqref{eq: functional indentity F2}, the full solution is naturally expected to take the form of a transseries \cite{Bender:1999box}.

Before doing so, we readily remark that the Bernoulli numbers in \eqref{eq: full transseries string form factor} can be generally understood to arise from the general structure of the difference equations we are dealing with. Indeed, writing $\Delta_x f(x) = (e^{\partial_x} -1) f(x)$---with $\sigma_x f(x) \coloneqq f(x+1) = e^{\partial_x} f(x)$ the forward shift operator---means that formally solving an inhomogeneous difference equation of the form 
\begin{equation}
    \label{eq: difference equation general}
    \Delta_x f(x) = g(x)
\end{equation} thus amounts to finding $f(x) = \Delta_x^{-1} g(x)$. Bernoulli numbers automatically enter through
\begin{equation}
    \label{eq: inverse difference Bernoulli}
    \Delta_x^{-1} = \frac{1}{e^{\partial_x}-1} = \sum_{n=0}^{\infty} \frac{B_n}{n!} \partial_x^{n-1} + \ker(\Delta_x), \quad \text{where } \partial_x^{-1} f(x) = \int^x \deriv x' \, f(x'), 
\end{equation} having used \eqref{eq: generating series Bernoulli}. With $g(x) = \ln(x)$, this procedure yields the asymptotic series for $\ln \Gamma(x)$ around $\abs{x} = \infty$ \cite{Dominici:2006}. In Appendix~\ref{app: example difference log Gamma}, we show that this method is equivalent to solving \eqref{eq: difference equation general} by method of controlling factors \cite{Bender:1999box}, with the advantage of making the appearance of the Bernoulli numbers $B_{2k}$ manifest, rather than requiring an a posteriori identification based on matching numerical coefficients. We thus take this approach below.
 
 Using \eqref{eq: inverse difference Bernoulli}, most general solution to \eqref{eq: IBP N=4 s} is
\begin{equation}
    \ln F(s,u) = \sum_{n=0}^{\infty} \frac{B_n}{n!} \partial_s^{n-1} \ln \frac{s+u}{s} + \phi(u) + \sum_{k \neq 0} \, c_k(u) e^{2\pi i k s},  
\end{equation} up to an additive constant $c_0$ in $s$ and $u$ to be fixed shortly. Imposing \eqref{eq: IBP N=4 u}, we find 
\begin{equation}
    \Delta_u \ln F(s,u) = \ln(u+s) + \phi(u+1) - \phi(u) + \sum_{k \neq 0} \left( c_k(u+1) - c_k(u)\right) e^{2\pi i k s},
\end{equation} so we must have $\phi(u+1)- \phi(u) = - \ln(u)$ and $c_k(u+1) - c_k(u) = 0$. The former is once again straightforwardly solved by 
\begin{equation}
    \phi(u) = - \sum_{n=0}^{\infty} \frac{B_n}{n!} \partial_u^{n-1} \ln u,
\end{equation} up to $1$-periodic terms in $u$ which we absorb in $c_k(u) = c_0 + \sum_{l \neq 0} c_{kl} \, e^{2\pi i l u}$. When the dust settles, using $B_0 = 1$ and $B_1 = - \tfrac{1}{2}$, we find 
\begin{multline}
    \ln F(s,u) = c_0 + \left(\frac{1}{2} -s \right)\ln s +  \left(\frac{1}{2} -u \right)\ln u + \left((u+s) - \frac{1}{2}\right) \ln(u+s) \\ + \sum_{k=1}^{\infty} \frac{B_{2k}}{2k(2k-1)} \left( (s+u)^{1-2k} - s^{1-2k} - u^{1-2k}\right) + \sum_{k, l\neq 0} c_{kl} \, e^{2\pi i (k s + lu)}.
\end{multline} The constant $c_0 = \tfrac{1}{2} \ln (2\pi)$ may be determined using, e.g., zeta-function regularization, see Appendix~\ref{app: example difference log Gamma}. Substituting $t = -s -u$ and isolating the $s,t,u$-contributions in the sum over exponentials, we identify the constants $c_k$ with the Stokes multipliers \eqref{eq: stokes constant log gamma} entering the transseries of $\ln \Gamma(z)$ around $\abs{z} = \infty$. Exponentiating, we then land precisely on \eqref{eq: full transseries string form factor}, without using any known asymptotic properties of the \textit{integrated} result \eqref{eq: string form factor}.

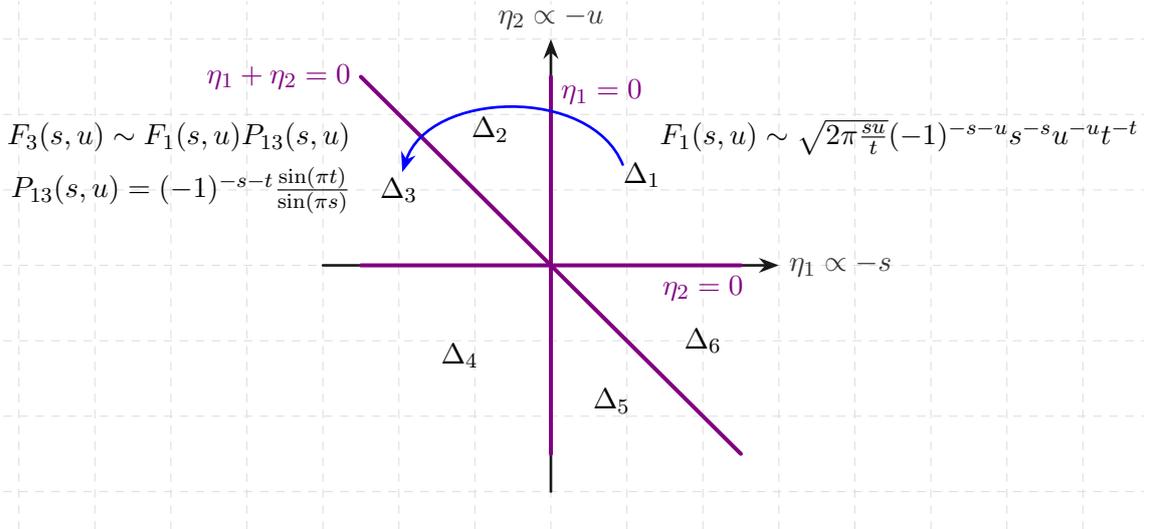
\begin{figure}[H]
    \centering
    \input{tikzRegular4.tex}
    \vspace{-0.5cm}
    \caption{Regular domains $\Delta_i$ of the $n=4$ difference system \eqref{eq: functional indentity F2} in the $\eta$-plane.}
    \label{fig: regularFan4}
\end{figure}

As a matter of fact, Stokes phenomena are already encoded at the level of difference equations, provided one consider a generic large-Mandelstam expansion in the two-dimensional kinematic lattice spanned by the standard basis $\{e_1,e_2\}$. For this, we consider again real Mandelstam invariants and introduce the two-vector $z \coloneqq (s,u) \in \mathbb{R}^2$. We then rewrite \eqref{eq: functional indentity F2} as $F(z - e_i) = \mathsf{M}_i(z) F(z)$ ($i=1,2$), with 
\begin{equation}
    \label{eq: shift matrices n=4}
    \mathsf{M}_1(z) = \frac{s+u}{s}, \quad \mathsf{M}_2(z) = \frac{s+u}{u}.
\end{equation} Next, consider a large-$z$ expansion along the generic lattice direction $\eta \coloneqq \eta_1 e_1 + \eta_2 e_2$ ($\eta_i \in \mathbb{Z}\backslash \{0\}$). Let $N \in \mathbb{Z}_{>0}$. To leading order as $N \to \infty$, we then have
\begin{equation}
    \mathsf{M}_1(z - N \eta) = \frac{\eta_1+\eta_2}{\eta_1} + \mathcal{O}(1/N), \quad \mathsf{M}_2(z - N \eta) = \frac{\eta_1+\eta_2}{\eta_2} + \mathcal{O}(1/N).
\end{equation} For the difference to remain well-defined, we thus require $(\eta_1 + \eta_2)\eta_1 \eta_2 \neq 0$. Such a choice of $\eta$ corresponds to a \textit{regular direction} of the difference system. The $\eta$-plane splits accordingly into six domains $\Delta_i$ ($i=1, \ldots, 6$), depicted in Figure~\ref{fig: regularFan4}.
\noindent
These reproduce exactly the six distinct (physical and unphysical) kinematic configurations identified in Table~\ref{tab: kinematics}; in particular, the physical $s$-channel region ($s > 0 > u,t$) corresponds to $\Delta_3$, while $s, u < 0 < t$ corresponds to $\Delta_1$, which is contained in the unphysical region where \eqref{eq: integral form factor} converges absolutely. 

For each regular direction $\eta$ belonging to a given domain, there exists a unique asymptotic expansion $F_\eta$ of $F$ for large values of $z$, built from iterating the difference system encoded by \eqref{eq: shift matrices n=4}; more on this shortly. Different domains are not independent, however: analytic continuation (crossing) from a direction $\eta \in \Delta$ to another $\eta' \in \Delta'$ (such that $\Delta \cap \Delta' = \emptyset$ and $\eta , \eta' \in \mathbb{Z}^2 \backslash \{0\}$) is accompanied by Stokes phenomena,\footnote{Note that, in the present context, Stokes phenomenon should be understood more broadly as the discontinuous change of the sectorial asymptotic realization under variation of the kinematic direction $\eta$, rather than solely in the usual one-variable sense associated with competing exponential scales in the asymptotic variable. In fact, there is only one single exponential scale at $n=4$.}
\begin{equation}\label{Stokesconnection}
    F_{\eta'}(z) \sim P_{\eta \eta'}(z) F_\eta(z) ,
\end{equation}
where $P_{\eta \eta'}(z)$ is a $1$-periodic function in the lattice directions, i.e., a meromorphic function of $e^{2\pi i s}$ and $e^{2\pi i u}$ in our case. This is precisely what we observed when passing from \eqref{eq: Stirling open string} to \eqref{eq: Stirling physical}, where the `non-perturbative' exponential contributions resum to produce the monodromy factor \eqref{eq: sanity check string transseries} via resurgence; in this new notation, $\npsaddles(s,u) = - P_{13}(s,u)$. 
Determining $P_{\eta \eta'}$ in general is known as a \textit{connection problem} in the literature, and thus naturally ties to the task of completing a given asymptotic series into a transseries using resurgence to fully characterise its discontinuities across Stokes lines. As we will see in the next section, this observation generalizes to difference systems of rank $r > 1$, with $P_{\eta \eta'}$ a $r \times r$ \textit{connection matrix}. We return to this problem in \S\ref{sec: physical kinematics and Stokes}, where we discuss the connection problem for higher-rank systems and Stokes phenomena in higher dimensions.

\subsection{Multiplicity \texorpdfstring{$n \geq 5$}{n geq 5}}

We now extend our analysis to the fixed-angle high-energy scattering of $n \geq 5$ open superstrings, fully exploiting the difference equations in the Mandelstam variables satisfied by the string worldsheet integrals to derive their asymptotic expansion as $\alpha' \to \infty$. 

\subsubsection{Asymptotics from a holonomic difference system}
\label{fivediff}

For $n\!=\!5$ we have a basis of four Euler integrals \req{Zint}
with $\pi,\sigma \in S_2$. Assuming canonical colour ordering $\pi\!=\!1$
these are given by~\cite{Mafra:2011nv}: 
\begin{equation} \label{eq: minimal functions N=5}
\begin{aligned}
    F^{(23)} &=\sum_{\rho \in S_2}Z_{1\rho}\;S[\rho|23] \\ &= s_{12} s_{34} \int_{0}^{1} \deriv z_3 \int_{0}^{z_3} \deriv z_2 \, z_2^{s_{12}-1} \ z_3^{s_{13}}\ (z_3 - z_2)^{s_{23}}\ (1-z_2)^{s_{24}} (1-z_3)^{s_{34}-1}, \\
    F^{(32)} &=\sum_{\rho\in S_2} Z_{1\rho}\;S[\rho|32] \\ &= s_{13} s_{24} \int_{0}^{1} \deriv z_3 \int_{0}^{z_3} \deriv z_2 \, z_2^{s_{12}} \ z_3^{s_{13}-1}\ (z_3 - z_2)^{s_{23}}\ (1-z_2)^{s_{24}-1} (1-z_3)^{s_{34}}.
\end{aligned} 
\end{equation}
There are $\tfrac{5}{2}(5-3) = 5$ independent Mandelstam invariants \req{Mandelstam}. 
An early account on high-energy single and double Regge limit of five-point function can be found in \cite{Bialas:1969jz}. Without 
 loss of generality, we pick the adjacent Mandelstam invariants $s_i \coloneqq \alpha'(k_i + k_{i+1})^2$ (with $k_{6} \equiv k_1$) as independent variables. We furthermore restrict to real Mandelstam invariants for now, construct the vector $s \coloneqq (s_i)_{i=1}^5 \in \mathbb{R}^5$, introduce the backward shift operator $\shift_i : s \mapsto s - e_i$ in $\mathbb{R}^5$ and the vector of functions $\Phi(s) \coloneqq (F^{(23)}(s), F^{(32)}(s))$. Using IBP, we find the system
\begin{equation}
    \label{eq: shift equation}
    \shift_i \Phi(s) = \mathsf{M}_i(s) \Phi(s), \quad i = 1, \ldots, 5,
\end{equation} with $\mathsf{M}_i$ some $2 \times 2$ contiguity matrices depending on the kinematic invariants $s$. These may be viewed as a `discrete connection' in kinematic space and are reported in Appendix~\ref{app: contiguity matrices}.

One may easily check that the connection is \textit{flat}, that is to say $(\sigma_i \mathsf{M}_j) \mathsf{M}_i - (\sigma_j \mathsf{M}_i) \mathsf{M}_j = 0$ for all $i,j = 1, \ldots, 5$. This follows from the compatibility of IBP relations of the same integral basis, or the statement that the order of shifts doesn't matter, 
\begin{equation}
    \label{eq: flatness}
    \Phi(s - e_i - e_j) = \mathsf{M}_i(s - e_j) \cdot \mathsf{M}_j(s) \Phi(s) \overset{!}{=} \mathsf{M}_j(s - e_i) \cdot \mathsf{M}_i(s) \Phi(s).
\end{equation} As such, $\mathsf{M}_i$ can be viewed as the discrete analogue of a flat Gauss-Manin connection. The system \eqref{eq: shift equation} satisfying \eqref{eq: flatness} is known as a \textit{rational holonomic system of first-order difference equations} \cite{Aomoto:2011ggg}, and amenable to Birkhoff-Trjitzinsky (BT) theory \cite{Birkhoff:1933}. We leverage this structure to derive asymptotic expansions around $\alpha' = \infty$ solely from \eqref{eq: shift equation}. The challenge amounts to reducing \eqref{eq: shift equation} to a system of scalar difference equations, so we first seek a formal gauge transformation that diagonalizes the full flat connection order by order. 

Consider the limit $\alpha' \to \infty$. There are infinitely many ways to approach infinity in $\mathbb{R}^5$. To describe this freedom, we introduce the directional vector $\eta \coloneqq (\eta_i)_{i=1}^5 \in \mathbb{Z}^5 \backslash \{0\}$ in the rank-$5$ lattice $\{e_1, \ldots, e_5\}$ underlying \eqref{eq: shift equation} (so $\hat s \propto \eta$), as well as a positive integer $N \in \mathbb{Z}_{>0}$. Each $\mathsf{M}_i$ then admits a Laurent expansion of the form
\begin{equation}
    \label{eq: Laurent expansion M}
    \mathsf{M}_i(s - N \eta) = \mathsf{M}_{i, \eta}^{(0)} \, N^{\mu_i} + \mathcal{O}\left(1/N\right) \quad \text{as } N \to \infty,
\end{equation} where $\mu_i \in \mathbb{Z}$. We find $\mu_i = 0$, $[\mathsf{M}_{i, \eta}^{(0)}, \mathsf{M}_{j, \eta}^{(0)}] = 0$ $\forall i,j=1, \ldots, 5$, and the determinants 
\begin{equation}
    \label{eq: determinants leading}
    \begin{aligned}
        \det \mathsf{M}_{1,\eta}^{(0)} &= \frac{\left(\eta _1+\eta _2-\eta _4\right) \left(\eta _1-\eta _3+\eta _5\right)}{\eta _1 \left(\eta _1-\eta _3-\eta _4\right)}, \\ 
        \det \mathsf{M}_{2,\eta}^{(0)} &= \frac{\left(\eta _1+\eta _2-\eta _4\right) \left(\eta _2+\eta _3-\eta _5\right)}{\eta _2 \left(\eta _2-\eta _4-\eta _5\right)}, \\ 
        \det \mathsf{M}_{3,\eta}^{(0)} &= \frac{\left(-\eta _1+\eta _3+\eta _4\right) \left(\eta _2+\eta _3-\eta _5\right)}{\eta _3 \left(-\eta _1+\eta _3-\eta _5\right)}, \\ 
        \det \mathsf{M}_{4,\eta}^{(0)} &= \frac{\left(\eta _1-\eta _3-\eta _4\right) \left(-\eta _2+\eta _4+\eta _5\right)}{\left(\eta _1+\eta _2-\eta _4\right) \eta _4}, \\ 
        \det \mathsf{M}_{5,\eta}^{(0)} &= \frac{\left(\eta _1-\eta _3+\eta _5\right) \left(-\eta _2+\eta _4+\eta _5\right)}{\left(\eta _5 - \eta _2 - \eta _3\right) \eta _5}.
    \end{aligned}
\end{equation} By construction, $\mathsf{M}_{i,\eta}^{(0)}$ doesn't depend on $s$. Any $\eta \in \mathbb{Z}^5\backslash\{0\}$ such that $\det \mathsf{M}_{i,\eta}^{(0)} \neq 0, \infty$ $\forall i$ defines a regular direction of \eqref{eq: shift equation}, along which the difference is well-defined and admits an irregular asymptotic (transseries) solution,\footnote{This is satisfied by, e.g., $\eta = (1,1,1,1,1)$.} just like in the rank-1 case in \S\ref{sec: difference N=4}.\footnote{Setting one of the $s_i$'s to zero amounts to reducing the dimension (not rank) of the system, suppressing one direction as $\alpha' \to \infty$. E.g.~for $n=5$ one would then have four $2\times 2$ contiguity matrices instead of five.}

Moreover, if $\eta, \eta' \in \mathbb{Z}^5 \backslash \{0\}$ are mutually proportional, i.e., $\eta' = \sigma \eta$ ($\sigma \in \mathbb{Q}_{>0}$) then $\eta$ and $\eta'$ clearly give the same direction. Therefore, we may restrict ourselves to \textit{primitive} regular directions in the lattice, i.e., to those vectors that cannot be written as an integer multiple of another integer vector: if $\eta = k \eta'$ ($k \in \mathbb{Z}$), then $\eta' = \pm \eta$. Assuming that $\eta$ is primitive, one may then find a unimodular change of variables $s = W_\eta \xi$ with $W_\eta \in \mathrm{GL}(5;\mathbb{Z})$ satisfying
\begin{equation}
    s_i = \sum_{j=1}^{5} w_{ij} \xi_j, \quad W_\eta := (w_{ij})_{i,j=1}^5, \quad \det W_\eta = \pm 1,  
\end{equation} such that $w_{i1} = \eta_i$. Note that the choice of $W_\eta$ is not unique. Once the first column is fixed to $\eta$, the remaining columns may be chosen arbitrarily as long as $W_\eta \in \mathrm{GL}(5,\mathbb{Z})$ and $\det W_\eta = \pm 1$, i.e., provided they complete $\eta$ to a basis of the lattice. Any two such matrices differ by a unimodular transformation acting only on the transverse coordinates $\xi_2,\ldots,\xi_5$, leaving the coordinate $\xi_1$ along the direction $\eta$ unchanged. Consequently, the asymptotic expansion obtained in the limit $N \to \infty$ depends only on the direction $\eta$, and not on the particular choice of $W_\eta$.

The upshot of this manipulation is that if $\eta$ is a regular direction with respect to \eqref{eq: shift equation} in the lattice spanned by $\{e_i\}$, then $e^\ast_1 = (1,0,0,0,0)$ is now a regular direction for
\begin{equation}
    \sigma_i^{\ast} \Psi(\xi) = \mathsf{A}_{i} (\xi) \Psi(\xi)
\end{equation} in the lattice spanned by $\{e_i^\ast\}$, given $\eta = W_\eta e_1^\ast$ and
\begin{equation}
    \begin{aligned}
        \sigma_\eta \Phi(W_\eta \xi) = \Phi(W_\eta \xi - \eta) = \Psi( \xi - W_\eta^{-1}\eta) =  \Psi(\xi - e_1^\ast) = \sigma^\ast_1 \Psi(\xi).
    \end{aligned}
\end{equation}
Above, $\sigma_i^{\ast}: \xi_i \mapsto \xi_i - 1$ are the backward shift operators of the variables $\xi = (\xi_i)_{i=1}^5 \in \mathbb{R}^5$, and $\mathsf{A}_i(\xi)$ are the new contiguity matrices in this basis. In particular, the only relevant contiguity matrix is given by the ordered product $\mathsf{A}_1 (\xi) \coloneqq \mathsf{P}_5(\xi) \cdot \mathsf{P}_{4}(\xi) \cdots \mathsf{P}_1(\xi)$, with
\begin{subequations}
    \label{eq: ordered product shifts}
    \begin{equation}
        \mathsf{P}_j(\xi) \coloneqq \mathsf{M}_j\left( s - (\eta_j -1) e_j - \sum_{k=1}^{j-1} \eta_k e_k \right) \cdots \mathsf{M}_j \left(s - \sum_{k=1}^{j-1} \eta_k e_k \right) \Bigg|_{s = W \xi},
    \end{equation} provided $\eta_j \geq 0$. If $\eta_j < 0$, one has instead 
    \begin{equation}
        \mathsf{P}_j(\xi) \coloneqq \mathsf{M}_j\left( s - \eta_j e_j - \sum_{k=1}^{j-1} \eta_k e_k \right)^{-1} \cdots \mathsf{M}_j \left(s + e_j - \sum_{k=1}^{j-1} \eta_k e_k \right)^{-1} \Bigg|_{s = W \xi}.
    \end{equation}
\end{subequations} Along this regular direction, $\mathsf{A}_1(\xi)$ admits a Laurent expansion with respect to $\xi_1$ only,
\begin{equation}
    \mathsf{A}_1(\xi) = \mathsf{A}_1^{(0)}  + \frac{1}{\xi_1} \mathsf{A}_1^{(1)}(\xi') + \ldots, \quad \xi' := (\xi_2, \ldots, \xi_5),
\end{equation} with $\mathsf{A}_1^{(0)}$ independent of $\xi$, $\det \mathsf{A}_1^{(0)} \neq 0$, and where $\mathsf{A}_1^{(l)}(\xi') \in \mathbb{C}[\xi']$ is a polynomial of degree at most $l$. Moreover, from \eqref{eq: ordered product shifts} we have 
\begin{equation}
    \label{eq: LO matrix relation}
    \mathsf{A}_1^{(0)} = \left( \mathsf{M}^{(0)}_{1,\eta} \right)^{\eta_1} \cdots \left( \mathsf{M}^{(0)}_{5,\eta} \right)^{\eta_5}
\end{equation} in terms of the original leading-order contiguity matrices of \eqref{eq: Laurent expansion M}.

Provided the eigenvalues $\lambda^\ast_{\alpha}$ ($\alpha=1,2$) of $\mathsf{A}_1^{(0)}$ are mutually different, one may construct a gauge transformation $\Psi(\xi) = \mathsf{G}(\xi) \tilde \Psi(\xi)$ as the formal Laurent series
\begin{equation}
    \mathsf{G}(\xi) := \mathsf{G}_0 + \frac{1}{\xi_1} \mathsf{G}_1(\xi') + \frac{1}{\xi_1^2} \mathsf{G}_2(\xi') + \ldots \in \text{GL} \left(2; \mathbb{C}[\xi'] \otimes \mathbb{C} \left( \left( \tfrac{1}{\xi_1} \right) \right) \right)
\end{equation} such that the solution 
\begin{equation}
    \label{eq: condition diag N=5}
    \mathsf{B}(\xi) := \mathsf{G}(\xi-e^\ast_i)^{-1} \mathsf{A}_1(\xi) \mathsf{G}(\xi)
\end{equation} of $\tilde \Psi(\xi - e^\ast_1) = \mathsf{B}(\xi) \tilde \Psi(\xi)$ takes the form \cite[Theorem 4.3]{Aomoto:2011ggg}
\begin{equation}
    \label{eq: solution B}
    \mathsf{B}(\xi) = \mathsf{\Lambda}_0 + \frac{1}{\xi_1} \mathsf{\Lambda}_1,
\end{equation} with $\mathsf{\Lambda}_0 \coloneqq \text{diag}(\lambda^\ast_{1}, \lambda^\ast_{2})$ and $\mathsf{\Lambda}_1 \coloneqq \text{diag}(\rho^\ast_{1}, \rho^\ast_{2})$. The proof hinges on the fact that the Laurent expansion of \eqref{eq: condition diag N=5} around large $\xi_1$, schematically of the form
\begin{equation}
    \label{eq: expansion gauge transf}
    \left(\mathsf{G}_0 + \frac{\mathsf{G}_1}{\xi_1 - 1}  + \ldots \right) \left(\mathsf{\Lambda}_0 + \frac{\mathsf{\Lambda}_1}{\xi_1}  + \ldots \right) = \left(\mathsf{A}_1^{(0)} + \frac{\mathsf{A}_1^{(1)}}{\xi_1}  + \ldots \right) \left(\mathsf{G}_0 + \frac{\mathsf{G}_1}{\xi_1}  + \ldots \right),
\end{equation} yields a recursive system of relations satisfied by the $\mathsf{G}_k$'s by using 
\begin{equation}
    (z - 1)^{-n} = z^{-n} \sum_{l=0}^{\infty} \frac{n(n+1) \cdots (n+l-1)}{l!} z^{-l}
\end{equation} and comparing coefficients on both sides of the equality, order by order in $1/\xi_1$. At leading order, one finds 
\begin{equation}
    \label{eq: leading order gauge}
    \mathsf{G}_0 \mathsf{\Lambda}_0 = \mathsf{A}_1^{(0)} \mathsf{G}_0,
\end{equation} which fixes $\mathsf{G}_0$ to be the matrix whose column vectors are the eigenvectors of $\mathsf{A}_1^{(0)}$ associated to $\lambda^\ast_{\alpha}$. At subleading order, \eqref{eq: expansion gauge transf} yields
\begin{equation}
    \mathsf{\Lambda}_1 = \mathsf{X} + [\mathsf{\Lambda}_0, \mathsf{G}_0^{-1} \mathsf{G}_1], \quad \text{where } \mathsf{X} := \mathsf{G}_0^{-1} \mathsf{A}_1^{(1)} \mathsf{G}_0.
\end{equation} The diagonal part of the commutator above vanishes, so we may absorb the off-diagonal entries of $\mathsf{X}$ in the definition of $\mathsf{G}_1$, and, by construction, $\mathsf{\Lambda}_1$ is the diagonal part of $\mathsf{X}$, whose entries we denote by $\rho_\alpha^\ast$ ($\alpha=1,2$). The construction carries on recursively. A second gauge transformation of the same form may then be applied to set all subleading matrices $\mathsf{\Lambda}_n$ ($n \geq 2$) to zero by induction, and one obtains \eqref{eq: solution B}. When the dust settles, this procedure yields a \textit{unique} gauge transformation $\mathsf{G}$ which diagonalizes $\mathsf{A}_1$ into $\mathsf{B}$.

With \eqref{eq: solution B} at hand, the rank-2 difference system $\sigma^\ast_1 \tilde \Psi(\xi) = \mathsf{B}(\xi) \tilde \Psi(\xi)$ decouples into one scalar equation per component $\alpha$ of $\tilde \Psi$
\begin{equation}
    \left(e^{-\partial_1} -1\right) \ln \tilde \Psi_\alpha(\xi) = \ln \lambda^\ast_{\alpha} + \ln(1+ \frac{r_\alpha}{\xi_1}), \quad \alpha = 1,2. \label{eq: Psi i=1}
\end{equation} Here, we introduced $r_\alpha := \rho^\ast_{\alpha}/\lambda^\ast_{\alpha}$, which is constant with respect to $\xi_1$ by construction. The Ansatz $\tilde \Psi_\alpha(\xi) = C_\alpha(\xi') \left(\lambda^\ast_{\alpha}\right)^{-\xi_1} \psi_\alpha(\xi_1)$ gives
\begin{equation}
    \label{eq: solution intermediaire}
     \frac{\psi_\alpha(\xi_1-1)}{\psi_\alpha(\xi_1)} = \frac{\xi_1 + r_\alpha}{\xi_1} \quad \Rightarrow \quad \psi_\alpha(\xi_1) = \frac{\Gamma(1+\xi_1)}{\Gamma(1+r_\alpha + \xi_1)} P_\alpha(\xi_1),
\end{equation} with $P_\alpha(\xi_1)$ a $1$-periodic function of $\xi_1$, i.e., a meromorphic function of $e^{2\pi i \xi_1}$, and an overall normalization $C_\alpha(\xi')$ depending only on the transversal directions. The solution $\tilde{\Psi}(\xi)$ is sometimes called the BT \textit{normal form}, and is the difference-equation analogue of WKB-type expansions for differential equations.

While the difference system alone doesn't resolve the $1$-periodic ambiguity $P_\alpha (\xi_1)$, \eqref{eq: solution intermediaire} still exposes the number-theoretic content of the asymptotic expansion of $\tilde \Psi$. Approaching infinity along $\eta$ corresponds to $\xi_1 \to + \infty$ in each regular sector. Using
\begin{equation}
    \ln \Gamma(z + h) \sim \left( z + h - \frac{1}{2} \right) \ln z - z + \frac{1}{2} \ln 2\pi + \sum_{k=2}^\infty \frac{(-1)^k B_k(h)}{k(k-1)} z^{1-k}, \quad \abs{z} \to \infty,
\end{equation}
where $\abs{\arg(z)} < \pi$ and $h \in \mathbb{C}$ is fixed, we find
\begin{equation}
    \label{eq: asymp series psi tilde}
    \tilde \Psi_\alpha(\xi) \sim C_\alpha(\xi') \left(\lambda^\ast_{\alpha}\right)^{- \xi_1} \frac{\xi_1^{1-r_\alpha} }{r_\alpha + \xi_1} \exp[\sum_{k=1}^\infty \frac{c_{k,\alpha}}{\xi_1^{k}}] P_\alpha(\xi_1), \quad \xi_1 \to \infty,
\end{equation} having introduced the coefficients 
\begin{equation}
    \label{eq: coefs general}
    c_{k, \alpha} \coloneqq  \frac{B_{k+1}- B_{k+1}(1-r_\alpha)}{k(k+1)}.
\end{equation} For completeness, we note that this result could also have been derived from \eqref{eq: inverse difference Bernoulli} by inverting the difference operator in \eqref{eq: Psi i=1}, writing $e^{-\partial_1}-1 = - e^{-\partial_1}(e^{\partial_1} - 1) = - e^{-\partial_1} \Delta_1$, and using the known expansion
\begin{equation}
    \label{eq: expansion bernoulli poly into num}
    B_n(x) = \sum_{k=0}^n \binom{n}{k} B_k \, x^{n-k}
\end{equation} of Bernoulli polynomials in terms of the eponym numbers. Owing to \eqref{eq: expansion bernoulli poly into num}, the coefficients  \eqref{eq: coefs general} contains only Bernoulli numbers and polynomials in the ratio $r_\alpha$ and no positive-weight MZVs nor mixed Tate periods, as expected from our discussion of \S\ref{sec: periods low high amplitudes}.

Undoing the gauge transformation, we have
\begin{equation}
    \Psi_\alpha(\xi) = \sum_{\beta=1}^{2} \left( \left(\mathsf{G}_0\right)_{\alpha \beta} + \frac{1}{\xi_1} \left(\mathsf{G}_1\right)_{\alpha \beta}(\xi') + \ldots \right) \tilde \Psi_{\beta}(\xi),
\end{equation} so $\Phi_\alpha(s) \sim \sum_\beta \left(\mathsf{G}_0\right)_{\alpha \beta} \tilde \Psi_\beta(\xi|_s)$ to leading order in $1/\alpha'$ in the limit $s \propto - \alpha' \eta$ ($\alpha' \to \infty$), with $\tilde \Psi_\beta$ given in \eqref{eq: asymp series psi tilde} and where we denote $\xi_i|_s \coloneqq (W^{-1}s)_i$ for notational conciseness. 

This structure generalizes straightforwardly to higher multiplicity $n > 5$. As explained in \S \ref{sec: periods low high amplitudes}, the full $n$-point tree-level open superstring amplitude is described by a minimal set of $(n-3)!$ independent Euler integrals \cite{Mafra:2011nv,Broedel:2013tta}
\be
F_\Pi{}^\si = (-1)^{n-3} \sum_{\rho \in S_{n-3}} S[\, \rho(2,\ldots,n-2) \, | \, \si(2,\ldots,n-2) \, ] \, Z_{\Pi}(1,\rho(2,3,\ldots,n-2),n,n-1),
\label{01,5}
\ee themselves depending on the $\tfrac{n}{2}(n-3)$ independent kinematic invariants \req{Mandelstam}. Integration by parts may again be used to derive a holonomic difference system of rank $(n-3)!$ of the form \eqref{eq: shift equation}, where the contiguity matrices $\mathsf{M}_i(s)$ ($i =1, \ldots, \tfrac{n}{2}(n-3)$) now have size $(n-3)! \times (n-3)!$. Given a suitable regular direction
\begin{equation}
    \eta \in \mathbb{Z}^{\frac{n}{2}(n-3)} \backslash \left\{ \eta \, : \, \det \mathsf{M}_{i,\eta}^{(0)} \neq 0, \infty, \, \Big|  \,i = 1, \ldots, \frac{n}{2}(n-3) \right\}
\end{equation} at infinity in the kinematic lattice, the leading contiguity matrices in a Laurent expansion of the form \eqref{eq: Laurent expansion M} mutually commute, and thus yield a set of $(n-3)!$ joint eigenvectors, which we assume mutually distinct. Moreover, one may always find a suitable unimodular matrix $W_\eta \in \mathrm{GL}(\tfrac{n}{2}(n-3), \mathbb{Z})$ allowing to project the asymptotic problem onto a unique direction in a new basis $\{e_i^\ast\}$ such that $s = W_\eta \xi$ and $\eta = W e_1^\ast$. In this new frame, the only relevant contiguity matrix is given by the ordered product
\begin{equation}
    \mathsf{A}_1(\xi) = \mathsf{P}_{\frac{n}{2}(n-3)}(\xi) \cdots P_1(\xi),
\end{equation} with $\mathsf{P}_i$ as in \eqref{eq: ordered product shifts}. The formal diagonalization of the difference system carries through as for $n=5$ and yields an asymptotic series of the form \eqref{eq: asymp series psi tilde} for $\tilde \Psi_\alpha(\xi)$, with $\lambda_\alpha^\ast$ the $(n-3)!$ distinct eigenvalues of 
\begin{equation}
    \label{eq: general A}
    \mathsf{A}_1^{(0)} = \prod_{i=1}^{\frac{n}{2}(n-3)} \left( \mathsf{M}_{i,\eta}^{(0)} \right)^{\eta_i}.
\end{equation} The ratios $r_\alpha = \rho^\ast_\alpha /\lambda_\alpha^\ast$ are obtained similarly as for $n=5$. Undoing the gauge transformation yields $\Phi_\alpha(s) \sim \sum_\beta \left(\mathsf{G}_0\right)_{\alpha \beta} \tilde \Psi_\beta(\xi|_s)$ again, and no new periods appear beyond \eqref{eq: coefs general}, which confirms our preliminary analysis of \S\ref{sec: periods low high amplitudes}. 

\subsubsection{From difference equations to scattering equations} 
\label{sec: diffeq scatteq}

Let us now briefly comment on the leading exponential growth of $\Phi_\alpha(s)$. We fix $n =5$ to be specific and generalize to $n > 5$ at the end. 

Given that $\mathsf{M}_{i,\eta}^{(0)}$ mutually commute, \eqref{eq: LO matrix relation} implies $\lambda^\ast_\alpha = \prod_{i=1}^5 \lambda_{i,\alpha}^{\eta_i}$, where $\lambda_{i,\alpha}$ are the (assumed mutually distinct) eigenvalues of the latter. This factorization enables rewriting the leading exponential growth of \eqref{eq: asymp series psi tilde} as
\begin{equation}
    \left( \lambda_{\alpha}^\ast \right)^{-\xi_1} = \exp[ \left(\sum_{j=2}^5 w_{ij} \xi_j -s_i \right) \ln \lambda_{i,\alpha}]
    \sim \prod_{i=1}^5 \lambda_{i, \alpha}^{-s_i}.
\end{equation} The last step follows from the fact that the transversal directions $\xi_j$ ($j \geq 2$) do not contribute to the asymptotic growth as $\abs{\xi_1} \to \infty$. More precisely, these terms remains bounded and independent of $\xi_1$, so they affect only the prefactor but not the leading exponential scaling in $\xi_1$, and may therefore be absorbed in $C_\alpha(\xi')$.

Clearly, the eigenvalues $\lambda_{i,\beta}$ determine the exponential sectors of the transseries. We now show that the branch $\beta$ directly corresponds to classical solutions of the worldsheet integrals \eqref{eq: minimal functions N=5}. Spelling \eqref{eq: leading order gauge} out in components and using \eqref{eq: LO matrix relation}, we have
\begin{equation}
    \label{eq: LO eigenvalue equation}
    \mathsf{A}_1^{(0)} v_\alpha = \lambda_\alpha^\ast v_\alpha \quad \Leftrightarrow \quad \mathsf{M}_{i,\eta}^{(0)} v_\alpha = \lambda_{i, \alpha} v_\alpha, \quad i = 1, \ldots, 5,
\end{equation} for $v_\alpha$ ($\alpha = 1,2$) the two independent eigenvectors common to all $\mathsf{M}_{i,\eta}^{(0)}$ with eigenvalues $\lambda_{i, \alpha}$. We find
\begin{equation}
    v_\pm = \left( 
       \frac{\eta _2 \eta _3-\eta _4 \eta _3+\eta _1 \left(\eta _2+2 \eta _3-\eta _5\right)+\eta _4 \eta _5 \pm\sqrt{\Delta}}{2 \left(\eta _1+\eta _2-\eta _4\right) \left(\eta _2+\eta _3-\eta _5\right)},  1
   \right)
\end{equation} where we now denote $\alpha \in \{1,2\} \eqqcolon \{+, -\}$ and the quadratic discriminant is
\begin{multline}
    \label{eq: quadratic discriminant}
    \Delta(\eta) \coloneqq \left(\eta _1 \eta _2+\eta _3 \left(\eta _4-\eta _2\right)\right){}^2+\left(\eta _1-\eta _4\right){}^2 \eta _5^2 \\ +2 \left(-\eta _2 \eta _1^2+\eta _2 \eta _3 \eta
   _1+\left(\eta _2+\eta _3\right) \eta _4 \eta _1+\eta _3 \left(\eta _2-\eta _4\right) \eta _4\right) \eta _5.
\end{multline} Provided $\Delta \neq 0$, the two independent vectors $v_\pm$ form the columns of $\mathsf{G}_0$. The associated eigenvalues are
\begin{equation}
    \label{eq: eigenvalues}
    \begin{aligned}
        \lambda_{1,\pm} &= \frac{-2 \eta _1^2-\left(\eta _2-2 \left(\eta _3+\eta _4\right)+\eta _5\right) \eta _1+\eta _2 \eta _3-\eta _3 \eta _4+\eta _4 \eta _5 \pm \sqrt{\Delta }}{2 \eta _1 \left(\eta _3+\eta _4 - \eta_1\right)}, \\ 
        \lambda_{2,\pm} &= \frac{-\left(2 \eta_2+\eta_3\right) \left(\eta_2-\eta_4\right)-\eta_1 \left(\eta_2-\eta_5\right) -\left(\eta_4-2 \eta_2\right) \eta_5 \pm \sqrt{\Delta}}{2 \eta_2 \left(\eta_4+ \eta_5- \eta_2\right)}, \\
        \lambda_{3,\pm} &= \frac{- \eta_2 \eta_3 - \left(2 \eta_3+ \eta_4\right) \left(\eta_3- \eta_5\right) - \eta_1 \left(-\eta_2-2 \eta_3+\eta_5\right) \pm \sqrt{\Delta }}{2 \eta_3 \left(\eta_1 + \eta_5 - \eta_3\right)}, \\ 
        \lambda_{4,\pm} &= \frac{\eta_2 \left(\eta_3+2 \eta_4\right)+\eta_1 \left(-\eta_2+2 \eta_4+\eta_5\right)-\eta_4 \left(\eta_3+2 \eta_4+\eta_5\right) \pm \sqrt{\Delta }}{2 \left(\eta_1+\eta_2-\eta_4\right) \eta_4}, \\ 
        \lambda_{5,\pm} &= \frac{-\eta_2 \left(\eta_3-2 \eta_5\right)+\eta_1 \left(\eta_2-\eta_5\right)+\left(\eta_3-\eta_5\right) \left(\eta_4+2 \eta_5\right) \pm \sqrt{\Delta }}{2 \left(\eta_2+\eta_3-\eta_5\right) \eta_5}.
    \end{aligned}
\end{equation} These are homogeneous of degree $0$ in $\eta$ and thus invariant under the rescaling $\eta \mapsto s$. 

Meanwhile, solving the scattering equations \eqref{SQE} for $n = 5$ in terms of the leftover (unfixed) punctures $z_2$ and $z_3$ yields the two solutions 
\begin{equation}
    \label{eq: saddles n=5}
    z_\pm(s) = \left\{ \begin{aligned}
        z_3 &= \frac{z_2 \left(s_1 z_2-s_3 z_2+s_5 z_2-s_1-s_2\right)}{s_1 z_2-s_2 z_2-s_3 z_2+s_5 z_2-s_1} \quad \text{and} \\ 
    z_2 &= \frac{-s_1 s_2+s_3 s_2+2 s_1 s_4-s_3 s_4+s_1 s_5+s_4 s_5 \pm \sqrt{\Delta_\text{SE}}}{2 \left(s_1-s_3+s_5\right) \left(-s_2+s_4+s_5\right)},
    \end{aligned} \right.
\end{equation} having introduced the discriminant
\begin{equation}
    \Delta_\text{SE} \coloneqq \left(s_2 s_3+s_4 \left(s_5-s_3\right)+s_1 \left(-s_2+2 s_4+s_5\right)\right){}^2-4 s_1 s_4 \left(s_1-s_3+s_5\right) \left(-s_2+s_4+s_5\right).
\end{equation} One can easily check that $\Delta_\text{SE} = \Delta(s)$, meaning that the two solutions of the scattering equations are in 1:1 correspondence with the two eigenspaces spanned by the solutions to \eqref{eq: LO eigenvalue equation} whenever $\Delta \neq 0$. As a matter of fact, we find
\begin{equation}
    \left(\lambda_{1, \pm}, \lambda_{2, \pm}, \lambda_{3, \pm}, \lambda_{4, \pm}, \lambda_{5, \pm}\right) \Bigg|_{\eta \mapsto s} =  \left(\frac{z_3}{z_2}, \frac{(1-z_2)z_3}{z_3-z_2}, \frac{1-z_2}{1-z_3}, \frac{1}{z_3}, \frac{1}{1-z_2}\right) \Bigg|_{z_\mp(s)},
\end{equation} which proves that
\begin{equation}
    \Phi_\alpha(s) \propto \sum_{\pm} (\mathsf{G}_0)_{\alpha \pm}  \exp(- \sum_{i=1}^{5} s_i \ln \lambda_{i,\pm}) = \sum_{\pm} (\mathsf{G}_0)_{\alpha \pm}  \exp(- \alpha' S(z_\mp(s);\hat s)),
\end{equation} with $S(z_2,z_3;\hat s)$ the Morse action \eqref{eq: morse action}. This makes the connection to the string worldsheet integral completely transparent: the two eigenvalue branches correspond to the two classical solutions of the scattering equations \eqref{SQE} for $n=5$, and the leading asymptotics of the holonomic difference system \eqref{eq: shift equation} reproduce the expected leading exponential scaling outlined in \S \ref{sec: periods low high amplitudes}. Put differently, the algebraic spectral curve of the difference connection is precisely the scattering equation curve. This observation also clarifies that the Stirling-type structure observed at four points persists at five points, now separately on each saddle. This structure generalizes straightforwardly to any $n > 5$, where the $(n-3)!$ distinct eigenvalues of \eqref{eq: general A} are in one-to-one correspondence with the $(n-3)!$ solutions to \eqref{SQE}.

\subsubsection{Connection problem and Stokes multipliers}
\label{sec: physical kinematics and Stokes}

Having constructed sectorial asymptotic solutions to the difference system satisfied by the Euler integrals, we now turn to the associated connection problem and its relation to Stokes phenomena. While we do not solve it completely in the present work, we outline its expected general structure and highlight the main difficulties. To our knowledge, this question remains largely unexplored from the perspective of resurgence theory for higher-dimensional difference systems, and we hope to return to it in future work.

The reduced asymptotic analysis of \S\ref{fivediff} associates to each regular direction $\eta$ at infinity a one-dimensional reduced difference system in the variable $\xi_1$, together with a formal asymptotic basis determined by the leading matrix $\mathsf{A}_1^{(0)}$. This leading system has rank $(n-3)!$ at $n$ points, and its eigenvalue branches are in one-to-one correspondence with the $(n-3)!$ branches of the scattering equations discussed in \S\ref{sec: diffeq scatteq}. The natural connection problem is therefore to compare the corresponding reduced asymptotic realizations attached to two distinct regular directions $\eta$ and $\eta'$. This should allow one to resolve the leftover 1-periodic ambiguity $P_\alpha(\xi_1)$ in \eqref{eq: solution intermediaire} by imposing compatibility between the sectorial asymptotic solutions across adjacent regular sectors, given an initial `seed' normalization in one regular sector, e.g., where the integral converges absolutely. 

Given three mutually distinct regular directions $\eta, \eta'$ and $\eta'' \in \mathbb{Z}^{\frac{n}{2}(n-3)} \backslash \{0\}$, the connection matrices are $1$-periodic in $\{s_i\}$ and satisfy 
\begin{equation}
    P_{\eta \eta'}(s) P_{\eta' \eta''}(s) = P_{\eta \eta''}(s), \quad P_{\eta \eta'}(s) P_{\eta' \eta}(s) = 1.
\end{equation} With this notation, we emphasize that the connection multipliers belong to the original $\tfrac{n}{2}(n-3)$-dimensional difference system in the kinematic $s$-variables, rather than to the reduced one-dimensional asymptotic variable $\xi_1$. Accordingly, we expect that solving the connection problem from the reduced one-dimensional sectorial asymptotic solutions only yields the pullback of the full higher-dimensional connection matrix along the ray.

A second important point is that this problem should not be confused with ordinary analytic continuation or Borel resummation in the reduced variable $\xi_1$. The coefficients \eqref{eq: coefs general} define a divergent series with Borel transform 
\begin{equation}
    \label{eq: borel transform psi tilde}
    \mathcal{B}\left[ \sum_{k=1}^\infty \frac{c_{k,\alpha}}{\xi_1^k} \right](\zeta) = \frac{1}{\zeta} \left[ \frac{1-e^{(1-r_\alpha)\zeta}}{e^\zeta-1} + 1 - r_\alpha \right],
\end{equation} which is meromorphic in the $\zeta$-plane with simple poles at $\zeta_k = 2\pi i k$ ($k \in \mathbb{Z}^\ast$) around which
\begin{equation}
    \mathcal{B}\left[ \sum_{k=1}^\infty \frac{c_{k,\alpha}}{\xi_1^k} \right](\zeta_k + \zeta) = \frac{1}{\zeta} \frac{1-e^{2\pi i k r_\alpha}}{2\pi i k} + \mathcal{O}(\zeta^0).
\end{equation} It is, however, incorrect to construct $P_\alpha(\xi_1)$ in \eqref{eq: solution intermediaire} by laterally Borel resumming \eqref{eq: borel transform psi tilde} across the Stokes ray at $\arg(\xi_1) = \pm \tfrac{\pi}{2}$. Indeed, the reduced connection $\mathsf{A}_1$ depends on the choice of regular direction through \eqref{eq: ordered product shifts}, so that varying $\eta$ modifies the reduced asymptotic problem itself, rather than merely analytically continuing a fixed one. In particular, $\lambda^\ast_\alpha$ and $r_\alpha$ may differ from one regular direction to another. 

The above obstruction is again tied to the genuinely multi-dimensional nature of the difference systems under consideration, which requires a reduction along $\xi_1$. In the simpler one-variable setting, the variable tending to infinity coincides with the formal asymptotic expansion parameter, and the precise relation between connection matrices and Stokes multipliers was established by Immink \cite{Immink:1988}. Stokes phenomena in higher dimensions were investigated by Sabbah \cite{Sabbah:1997, Sabbah:2012} within the broader language of meromorphic connections and sheaves; see also \cite{Kontsevich:2024esg} and \cite{Angius:2025drr} for a recent application of these ideas in physics. While this should provide the natural abstract framework for higher-dimensional difference systems as well, it remains unclear how to extract from it the explicit connection data relevant in the sense of Aomoto-Kita \cite{Aomoto:2011ggg} for the string amplitudes considered in the present work. In a number of special cases, Aomoto-Kita solve such connection problems using twisted de Rham homology; we discuss this perspective in \S\ref{sec: sec5}.

Finally, from five points onward, the connection problem is enriched by the coexistence of $(n-3)!$ distinct exponential scales and it is unclear whether the difference system alone can consistently prescribe their possible change of dominance from one regular sector to the other. Given a regular direction, let $\tilde{\Psi}(\xi)$ denote a reduced asymptotic basis in which the leading matrix $\mathsf{A}_1^{(0)}$ is diagonal. In this basis, the pullback of the connection matrix is diagonal too, $\mathsf{S}_{\mathrm{diag}}(\xi_1) = \mathrm{diag}[P_1(\xi_1), \ldots, P_{(n-3)!}(\xi_1)]$. However, returning to the original basis $\Psi(\xi)= \mathsf{G} (\xi)\,\tilde{\Psi}(\tilde{\xi})$, one obtains
\begin{equation}
    \Psi(\xi)\ \mapsto\ \mathsf{G}(\xi) \cdot \mathsf{S}_{\mathrm{diag}}(\xi_1)\,\tilde{\Psi}(\xi) = \mathsf{G}(\xi)\cdot \mathsf{S}_{\mathrm{diag}}(\xi_1)\cdot \mathsf{G}(\xi)^{-1}\Psi(\xi),
\end{equation} and the exponential scales effectively mix. The matrix $\mathsf{G}$ is defined as a Laurent series in $\xi_1$, so the non-perturbative content of $\mathsf{S}_\mathrm{diag}$ is \textit{a priori} dressed with factors of $1/\xi_1$ to construct the full asymptotic solution in the non-diagonal basis. 

\begin{figure}[ht]
    \centering
    \begin{subfigure}{0.49\linewidth}
        \centering
        \includegraphics[width=\linewidth]{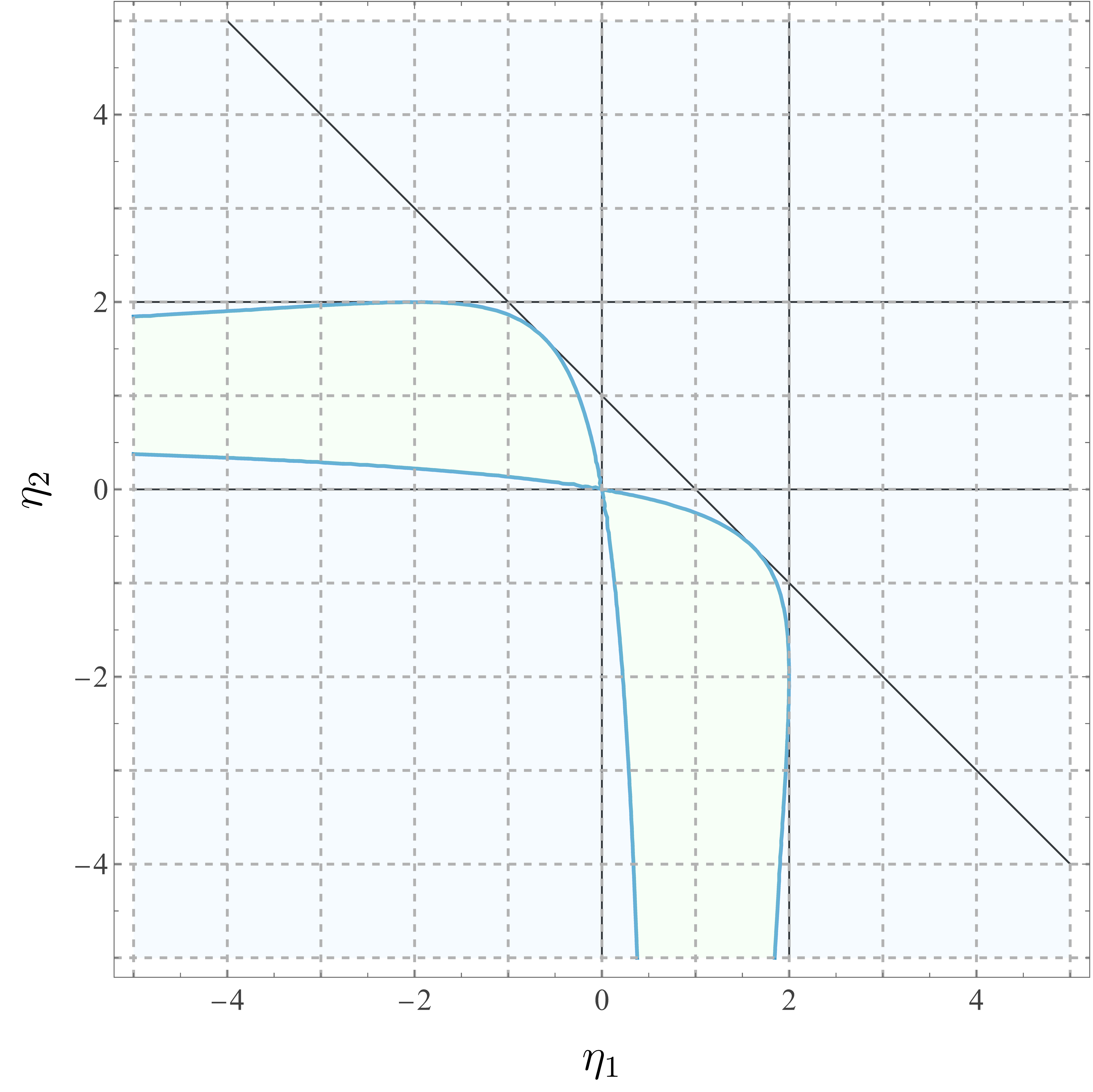}
        \caption{\label{fig: regular example}}
    \end{subfigure}
    \begin{subfigure}{0.49\linewidth}
        \centering
        \includegraphics[width=\linewidth]{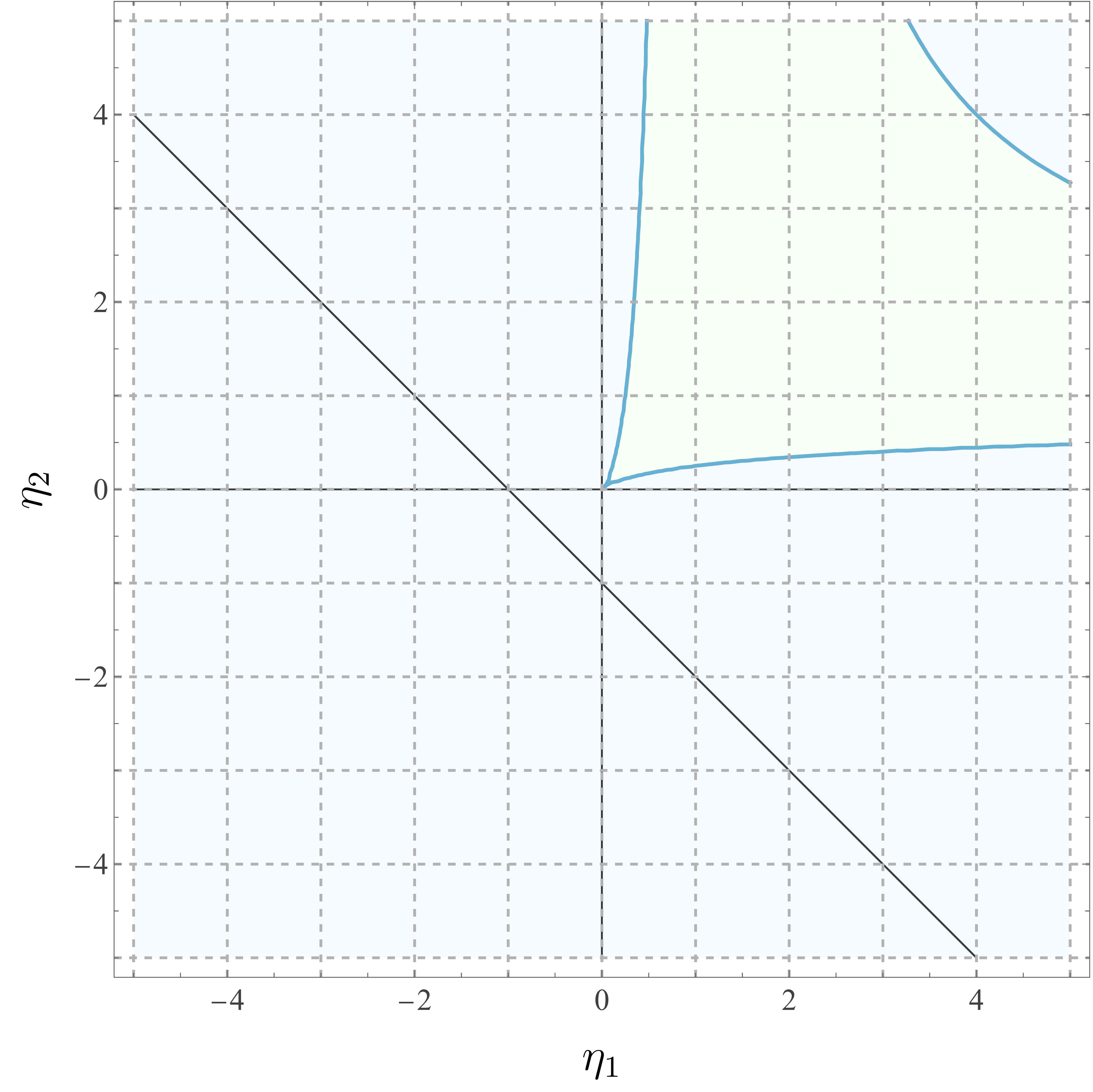}
        \caption{\label{fig: 5d stokes sector}}
    \end{subfigure}
    \caption{Regular sectors in the $(\eta_1, \eta_2)$ lattice for $n=5$ and \textbf{(a)} $\eta_3 = \eta_4 = \eta_5 =1$ or \textbf{(b)} $\eta_3 = - \eta_4 = \eta_5 = 1$. We depict with the shaded regions the regular directions where either $\Delta(\eta) > 0$ (blue region) or $\Delta(\eta) < 0$ (green region). The solid lines delineate the boundaries of the regular sectors as per \eqref{eq: determinants leading}. This is the (2D slice of the) 5D analogue of Figure~\ref{fig: regularFan4}.}
    \label{fig: eta slice}
\end{figure}

We illustrate the interplay between exponential dominance and regular sectors in the $n =5$ case in Figure~\ref{fig: eta slice}, where we display the regular sectors obtained from \eqref{eq: determinants leading} in the $(\eta_1,\eta_2)$ lattice, for fixed $\eta_3, \eta_4$ and $\eta_5$. Given $\lambda_\alpha^\ast = \prod_{i=1}^5 \lambda_{i, \alpha}^{\eta_i}$, the (complex/real) nature of the two roots $\lambda_\alpha^\ast$ and their relative dominance follows from the sign of the discriminant $\Delta(\eta)$ of \eqref{eq: quadratic discriminant} common to all $\lambda_{i,\alpha}$, which we indicate by the shaded regions in the plot. For $\Delta(\eta) > 0$, $\lambda_\pm^\ast$ are generically distinct and real, so one term is usually exponentially dominant over the other as $\xi_1 \to \infty$. For $\Delta(\eta) \leq 0$, the two roots are equal or complex conjugate, and there is no relative exponential dominance. 

In particular, Figure~\ref{fig: eta slice} shows that although the two eigenvalues may transition from being real to forming a complex-conjugate pair within a fixed regular sector (e.g., in Fig.~\ref{fig: 5d stokes sector}), their relative dominance cannot switch within the same regular sector. Throughout any such sector, either one eigenvalue has strictly larger modulus than the other, or the two have equal modulus, so that the relative dominance is either fixed or zero. Given \eqref{eq: LO matrix relation}, a change in this pattern can occur only across the boundaries of regular sectors, where the discriminant of the characteristic polynomial $\chi(t) = \prod_{\alpha=\pm} \left(t - \prod_{i=1}^5 \lambda_{i,\alpha}^{\eta_i} \right)$ becomes singular. The relative dominance may thus only change between regular sectors.

\begin{figure}[t]
    \centering
    \begin{subfigure}{0.49\linewidth}
        \centering
        \includegraphics[width=\linewidth]{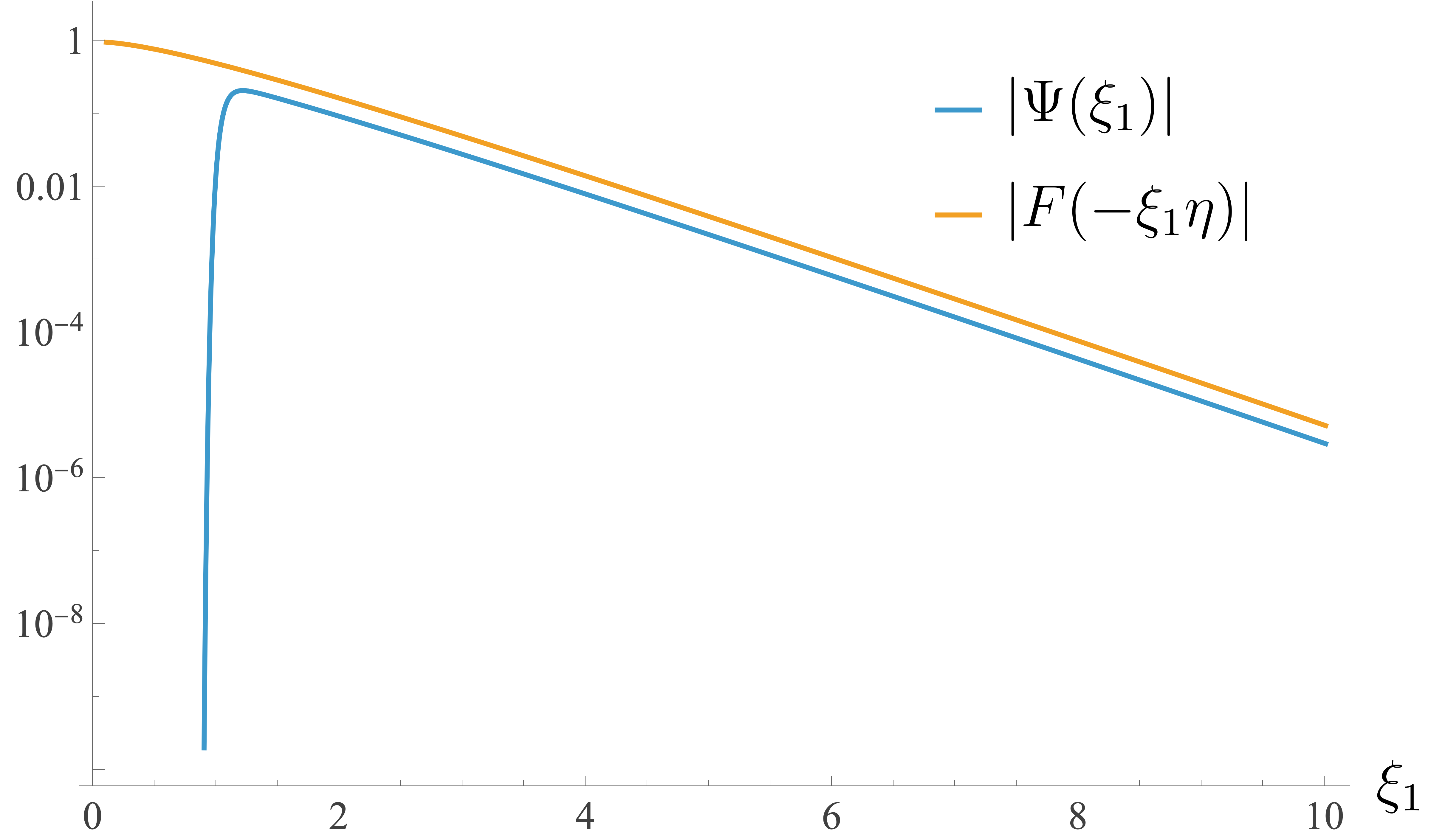}
        \caption{}
    \end{subfigure}
    \begin{subfigure}{0.49\linewidth}
        \centering
        \includegraphics[width=\linewidth]{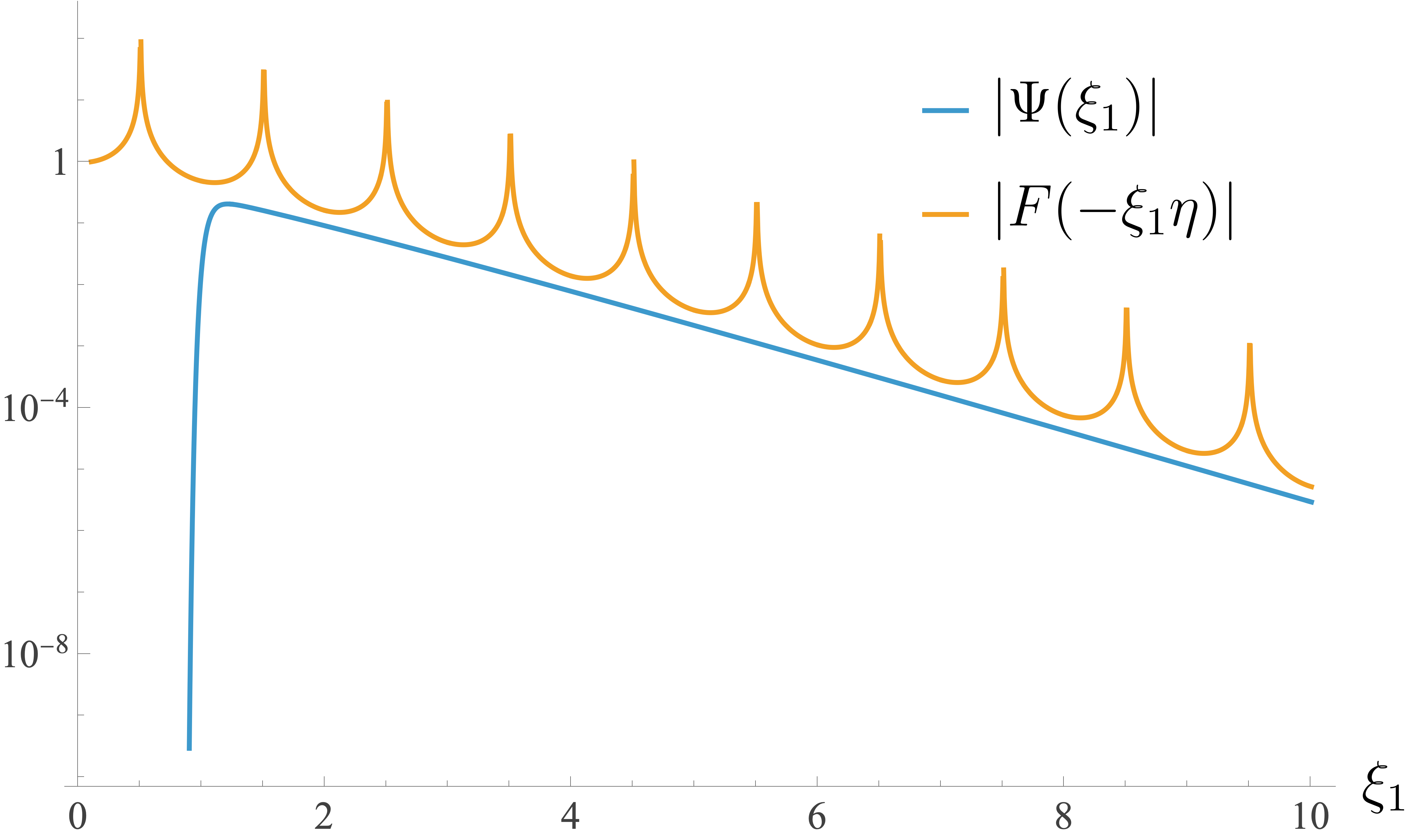}
        \caption{}
    \end{subfigure}
    \caption{Four-point asymptotic solution \eqref{eq: asymp series psi tilde} with $P(\xi_1) \equiv 1$ against the exact solution \eqref{eq: string form factor} for \textbf{(a)} $\eta = (1,1)$ and \textbf{(b)} $\eta = (-2,1)$, respectively in the unphysical and physical kinematic region. The asymptotic series in $\Psi(\xi_1)$ was truncated at $k_\text{max}=20$.}
    \label{fig: asymp check n=4}
\end{figure}

To assess what is already captured by the asymptotic solution alone, we plot $\Psi(\xi_1)$ without the connection function against the exact solution \eqref{eq: string form factor} for $n=4$ in Figure~\ref{fig: asymp check n=4}, for two choices of regular direction at infinity, respectively in the unphysical and physical kinematic region. The comparison (on a logarithmic scale) shows that the asymptotic expression correctly reproduces the overall exponential envelope of the exact result, up to a constant overall factor. On the other hand, in the physical region the sectorial expression does not recover the pole structure, which therefore lies beyond the information contained in the local sectorial form alone. This is mismatch is precisely what $\npsaddles(s,u)$ captures.

%% file: tikzContour.tex
\begin{tikzpicture}[>=Stealth, line join=round, line cap=round]

\newcommand\spiral{}
\def\spiral[#1](#2)(#3:#4:#5){
    \pgfmathsetmacro{\domain}{pi*#3/180+#4*2*pi}
    \draw [#1,shift={(#2)}, domain=10:\domain,variable=\t,smooth,samples=int(\domain/0.08)] 
    plot ({\t r}: {#5*\t/\domain})
}

\draw[blue!70!black, line width=1pt, postaction={decorate}, decoration={markings, mark=at position 0.5 with {\arrow{Stealth[length=3mm]}}}] (-2,0) -- (2,0);
\node[blue!70!black,above] at (0,0) {$C_\epsilon$};

\spiral[blue!70!black, line width=1pt](-4,0)(0:5:2);
\spiral[blue!70!black, line width=1pt](4,0)(180:5:2);

\fill (-4,0) circle (2pt);
\fill (4,0) circle (2pt);

\node[black,above] at (-4,0) {$0$};
\node[black,above] at (4,0) {$1$};

\end{tikzpicture}

%% file: tikzRegular4.tex
\begin{tikzpicture}[>=Stealth, line join=round, line cap=round] 
\draw[step=1cm, gray!25, dashed] (-7.2,-3.5) grid (7.8,3.5);

\draw[black!90, line width=1pt,->] (0,-3) -- (0,3);
\draw[black!90, line width=1pt,->] (-3,0) -- (3,0);
\node[black!80,above] at (0,3) {$\eta_2 \propto -u$};
\node[black!80,right] at (3,0) {$\eta_1 \propto -s$};

\draw[violet, line width=1.5pt] (-2.5,2.5) -- (2.5,-2.5);
\node[violet,left] at (-2.5,2.5) {$\eta_1+\eta_2=0$};
\draw[violet, line width=1.5pt] (-2.5,0) -- (2.5,0);
\node[violet,below] at (2,0) {$\eta_2=0$};
\draw[violet, line width=1.5pt] (0,-2.5) -- (0,2.5);
\node[violet,right] at (0,2.3) {$\eta_1=0$};

\node[black] at (1.2,1.2) {$\Delta_1$};
\node[black] at (-0.8,1.8) {$\Delta_2$};
\node[black] at (-2,1) {$\Delta_3$};
\node[black] at (-1.2,-1.2) {$\Delta_4$};
\node[black] at (0.8,-1.8) {$\Delta_5$};
\node[black] at (2,-1) {$\Delta_6$};

\node[black, right] at (1.3,1.7) {$F_1(s,u) \sim \sqrt{2\pi \frac{su}{t}} (-1)^{-s-u} s^{-s} u^{-u} t^{-t}$};
\node[black, left] at (-2.5,1.7) {$F_3(s,u) \sim F_1(s,u) P_{13}(s,u)$};
\node[black, left] at (-2.5,1) {$P_{13}(s,u) = (-1)^{-s-t} \frac{\sin(\pi t)}{\sin(\pi s)}$};

\node (A) at (1,1.2) {};
\node (B) at (-2,1.1) {};

\draw[blue, line width=1pt,->] (A) to[bend right=70] (B);

\end{tikzpicture}

%% file: complexMellinSpace.tex
\section{Unifying low- and high-energy expansions}
\label{MellinStokes}

Euler integrals \req{Zint} satisfy systems of linear differential equations,
such as Gauss-Manin or GKZ systems with respect to auxiliary parameters \cite{GKZ1994,Aomoto1987,Matsubara-Heo:2023ylc}, as well
as KZ-type equations with respect to extra marked points \cite{Broedel:2013aza}.
In addition, they obey difference equations arising from discrete shifts of
kinematic invariants.
While these structures probe the dependence of the integrals along specific
directions in parameter space, in this section we instead focus on their
variation in the $\ap$-space \req{Spacealpha}. In particular, we compare the
canonical local descriptions in the limits $\ap \to 0$ and $\ap \to \infty$
and develop a unified perspective that relates these two regimes.

\subsection{A Aomoto-Gauss-Manin connection}

We now define a connection constant relating the two local descriptions at $\ap\to0$ and $\ap\to\infty$ and discuss how  Stokes/resurgence enter for the irregular singularity at infinity.

The Euler-Beta function \req{eq: euler integral beta}
\be\label{Start}
I(\ap):=\frac{\Gamma(\ap \mu_0)\;\Gamma(\ap \mu_1)}{\Gamma(\ap \mu_0+\ap \mu_1)}
=B(\mu_0,\mu_1)
\ee
entering the string form factor \req{eq: string form factor} as $F(\ap)=\tfrac{\Gamma(1-\ap \hat{s})\;\Gamma(1-\ap \hat{u})}{\Gamma(1-\ap \hat{s}-\ap \hat{u})}=\ap \tfrac{su}{t}\;B(-s,-u)$
fulfills the following first order linear differential equation with respect to $\ap$:
\be\label{DGL}
\fc{\p I}{\p\ap}=\Omega\ I,
\ee
with the connection
\be\label{connection}
\Omega=\mu_0\;\psi(\ap \mu_0)+\mu_1\;\psi(\ap \mu_1)+\mu_\infty\;\psi(-\ap \mu_\infty),\quad \mu_0+\mu_1+\mu_\infty=0.
\ee
Differentiating Euler integrals \req{Zint} with respect to $\ap$ induces logarithmic insertions in the twisted 
$(n-3)$-forms underlying  \req{Zint}. In the present case these insertions are  expressible through Euler-type integrals  \req{eq: euler integral beta}, so that  the $\alpha'$-flow of \req{Start} is controlled  by the Aomoto-Gauss-Manin connection \req{connection} in the sense of \req{DGL}. Such logarithmic  (or  polylogarithmic) structures also arise naturally in AdS open string amplitudes \cite{Alday:2025bjp,Baune:2025hfu}.

With $\psi(\ap\mu)=-\tfrac{1}{\ap\mu}-\gamma_E + \zeta(2)\; \ap\mu- \zeta(3)\; (\ap\mu)^2+\ldots,$ and 
\be
\Omega=-\fc{1}{\ap}-2\;\mu_0\mu_1\;\zeta(2)\;\ap-3\;\mu_0\mu_1\mu_\infty\;\zeta(3)\;\ap^2+\Oc(\ap^3),
\ee
it is straightforward to verify that a perturbative series Ansatz
\be\label{Ansatz1}
I=\fc{c_0}{\ap}+c_1+c_2\;\zeta(2)\; \ap+c_3\;\zeta(3)\; \ap^2+\ldots,
\ee
 solves the equation \req{DGL} with $c_1=0,\; c_2=-\mu_0\mu_1c_0,\; c_3=-\mu_0\mu_1\mu_\infty c_0$, etc. Proceeding along this way with the integration constant $c_0=\tfrac{\mu_0+\mu_1}{\mu_0\mu_1}$ yields
\be\label{lowseries}
I=\fc{1}{\ap\mu_0}+\fc{1}{\ap\mu_1}-\zeta(2)\;\ap\;(\mu_0+\mu_1)+\ap^2\;(\mu_0+\mu_1)^2\;\zeta(3)+\ldots
\ee
On the other hand, we can also make an asymptotic transseries or WKB Ansatz as:
\be\label{Ansatz2}
I=\ap^{-1/2}\;\exp\{\phi(\ap)\}\; \sum_{n\geq0}\fc{c_n}{\ap^{n}}.
\ee
In this way, with $\psi(\ap\mu)=\ln(\ap\mu)+\sum\limits_{n=1}^{\infty} \frac{\zeta(1-n)}{(\ap\mu)^n}, \, \abs{\arg(\ap\mu)} < \pi$ and likewise
\begin{align}
    \Omega&=\mu_0\ln\mu_0+\mu_1\ln\mu_1+\mu_\infty\ln(-\mu_\infty)+\zeta(0)\;\ap^{-1}\\ \nonumber &+\zeta(-1)\;\ap^{-2}\;\lf(\fc{1}{\mu_0}+\fc{1}{\mu_1}+\fc{1}{\mu_\infty}\ri)+\Oc(\ap^{-3})
\end{align}
subject to $|\arg(\alpha' \mu_0)|,|\arg(\alpha' \mu_1)| < \pi$, we obtain 
\be\label{WKB}
\phi(\ap)=\ap[\mu_0\ln\mu_0+\mu_1\ln\mu_1+\mu_\infty\ln(-\mu_\infty)], 
\ee
$c_1=-c_0\zeta(-1)\lf(\fc{1}{\mu_0}+\fc{1}{\mu_1}+\fc{1}{\mu_\infty}\ri),$ etc., which together with  the integration constant $c_0\!=\!\sqrt{\tfrac{-2\pi\mu_\infty}{\mu_0\mu_1}}$ gives
\begin{align}\label{highseries}
I=\sqrt{\fc{2\pi(\mu_0+\mu_1)}{\mu_0\mu_1}}\;&\ap^{-1/2}\; \exp\lf\{\ap\lf[\mu_0\ln\mu_0+\mu_1\ln\mu_1+\mu_\infty\ln(-\mu_\infty)\ri]\ri\}\nonumber\\ 
&\times\lf[\ 1-\zeta(-1)\lf(\fc{1}{\ap\mu_0}+\fc{1}{\ap\mu_1}+\fc{1}{\ap\mu_\infty}\ri)+\Oc(\ap^{-3})\ \ri],
\end{align}
in agreement with Stirling formula for the Euler Beta function \req{Start}.
Because $\ap=\infty$ is irregular, the resulting formal series \req{highseries} is generally divergent but Borel-Laplace summable within Stokes sectors to a canonical analytic solution. 
Our solution \req{highseries} assumes $\Re \mu_0,\Re\mu_1>0$ and therefore corresponds to region $\Delta_1$ in  Figure~\ref{fig: regularFan4}.
The other five representations can be obtained by adjusting the Ansatz
\req{Ansatz2} accordingly.

\subsection{Universal description in complex Mellin space} 

We now derive a contour integral representation in complex Mellin space that simultaneously encodes the low- and high-energy expansions \req{lowseries} and \req{highseries} of the Euler-Beta integral \req{eq: euler integral beta}. From this perspective, the two asymptotic regimes are just different residue expansions of the same meromorphic complex integral.


Let $I(\alpha')$ as defined in \req{Start}
with $\Re\mu_i >0,\ \alpha'\in\mathbb{C},\ |\arg(\alpha' \mu_0)|,|\arg(\alpha' \mu_1)|<\pi$.
We find a single representation that yields \emph{both} the small-\(\alpha'\) and large-\(\alpha'\) asymptotics and explains the appearance of Riemann zeta values \(\zeta(m{+}1)\) vs.\ \(\zeta(1-2k)\).
We shall see that the two asymptotics can be obtained from the same contour.

Binet's first formula for the log gamma function \(\ln\Gamma(z)\) gives a Laplace-representation \cite[p.~258]{WhittakerWatson1990} (cf.~also Appendix~\ref{app: transseries log Gamma} and eq.~\req{logGamma})
\be
\ln\Gamma(z)=\lf(z-\fc{1}{2}\ri)\ln z-z+\fc12\ln(2\pi)+\int_0^\infty \deriv t \, K(t)\; \fc{e^{-tz}}{t}, \quad \Re(z)>0,
\ee
with
\begin{equation}\label{BernoulliK}
K(t):=\frac{1}{e^{t}-1}-\frac{1}{t}+\frac{1}{2}.
\end{equation}
After taking the combinations for
\(z\in\{\alpha' \mu_0,\alpha'\mu_1,\alpha'(\mu_0{+}\mu_1)\}\) we obtain
\begin{align}
\ln I(\ap)
&=\Phi(\mu_0,\mu_1;\ap)\;+\;\int_{0}^{\infty}\deriv t \, \frac{K(t)}{t}\,\Big[e^{-\ap \mu_0 t}+e^{-\ap \mu_1 t}-e^{-\ap(\mu_0+\mu_1)t}\Big], \label{eq:LB-mother}
\end{align}
with
\begin{align}
\Phi(\mu_0,\mu_1;\ap)&:=\Big(\alpha' \mu_0-\tfrac12\Big)\ln(\alpha' \mu_0)+\Big(\alpha' \mu_1-\tfrac12\Big)\ln(\alpha' \mu_1)\nonumber\\
&-\Big(\alpha'(\mu_0{+}\mu_1)-\tfrac12\Big)\ln(\alpha'(\mu_0{+}\mu_1))+\tfrac12\ln(2\pi)\nonumber\\
&\equiv \ap\;\phi(\ap)+\frac12\ln\frac{2\pi(\mu_0+\mu_1)}{\ap\mu_0\mu_1}.\label{Phi}
\end{align}
This \emph{single} Laplace transform \req{eq:LB-mother} drives both asymptotic regimes and provides both types of Riemann zeta values.  Analytic continuation in $\ap$ corresponds to rotating the Laplace contour (or taking lateral Borel sums of the large $\ap$ series); Stokes jumps are tied to the singular lattice \(t=2\pi i\mathbb{Z}\) of the kernel.

For small $\alpha'$ we expand \(e^{-\alpha' c t}=\sum_{m\ge0}\frac{(-\alpha' c t)^m}{m!}\) and use
\(\int_0^\infty \deriv t \, t^{m-1}\;K(t) =\zeta(m)\Gamma(m)$ following from \cite[Eq.~(2.8.1)]{Titchmarsh}
\begin{equation}\label{Titma}
\int_0^\infty \deriv t \, t^{s-1}\; K(t) =\zeta(s)\Gamma(s), \quad -1<\Re(s)<0,
\end{equation}
subject to analytic continuation to the region $\Re(s)>1$.
Eventually, one obtains the series
\begin{equation}
\ln I(\ap)
=\ln\frac{\mu_0+\mu_1}{\ap\mu_0\mu_1}
+\sum_{m\ge2}\frac{(-1)^{m}}{m}\,\zeta(m)\,
\big[\mu_0^{m}+\mu_1^{m}-(\mu_0{+}\mu_1)^{m}\big]\,\ap^{m},
\label{eq:small-alpha}
\end{equation}
matching (\ref{lowseries}). Thus the coefficients involve \(\zeta(2),\zeta(3),\dots\).

Let us comment on the intermediate steps leading to \req{eq:small-alpha}. 
Firstly, the identity \req{Titma} holds in the strip  $-1 < \Re s<0$ as a convergent integral and extends meromorphically to all 
$s\in\mathbb{C}$ by Zagier's generalized Mellin transform \cite{ZagierMellinQFT}.  For $\Re s>0$ one interprets the integral \req{Titma} in the sense of analytic continuation. 
Note that the subtractions in $K$ given in \req{BernoulliK} render the Mellin integration in  (\ref{Titma}) convergent through the generalized Mellin transform at $\Re s>1$ \cite{ZagierMellinQFT}. For  some function $f$ we have  
\begin{align}
\int_{0}^{\infty} \deriv x \, x^{s-1}\, f(x) &=
\int_{0}^{1} \deriv x \, x^{s-1}\!\left(f(x)-\frac{1}{x}\right) +\frac{1}{s-1}+\int_{1}^{\infty} \deriv x \,x^{s-1}\, f(x)\nonumber\\
&\simeq\int_{0}^{\infty} \deriv x \, x^{s-1}\!\left(f(x)-\frac{1}{x}\right) ,
\end{align}
 subject to the interpretation $\frac{1}{s-1}\simeq-\int_1^\infty \deriv x \, x^{s-2}$ for $\Re s<1$. That is to say, formally
we assume \cite{ZagierMellinQFT} 
$$\int_0^\infty x^{s-2}\,\deriv x=\int_0^1 x^{s-2}\,\deriv x+\int_1^\infty x^{s-2}\,\deriv x=\frac{1}{s-1}-\frac{1}{s-1}\simeq 0,$$ 
with the two integrals evaluated in their  domains of convergence 
\begin{align}
\int_0^1 \deriv x \, x^{s-2}&=\frac{1}{s-1},\quad \Re s>1,\\
\int_1^\infty \deriv x \, x^{s-2} &=-\frac{1}{s-1},\quad \Re s<1,
\end{align}
and analytically continued to $\mathbb{C}\slash\{1\}$.
Secondly,  the Mellin transform (\ref{Titma}) has simple poles at $s=0$ and $s=1$, so the naive Taylor expansion of the exponentials in the integrand 
\req{eq:LB-mother} requires a  regularization at $m=0$ and $m=1$, respectively. 
In \req{eq:small-alpha}
these poles are compensated by the factor $\big[\mu_0^{m}+\mu_1^{m}-(\mu_0{+}\mu_1)^{m}\big]\tfrac{(-\alpha')^{m}}{m!}$.
This way, the term $m=1$ yields $-\ap\phi(\ap)$ cancelling the same term in \req{Phi}. Furthermore, the $m=0$ term yields $-\tfrac12\ln(2\pi)+\tfrac12\ln\frac{(\mu_0+\mu_1)}{\ap\mu_0\mu_1}$ of \req{Phi} conspiring with the second term of \req{Phi} to give the first term $\ln\frac{(\mu_0+\mu_1)}{\ap\mu_0\mu_1}$ of \req{eq:small-alpha}. After these two `exceptional' cases are accounted for, the remaining terms $m\geq2$   generate the convergent $\zeta$-sum in \req{eq:small-alpha}.

On the other hand, for large $\alpha'$ we use the  Bernoulli expansion of \(K(t)\) at \(t=0\) yielding 
\(
K(t)=\sum_{k\ge1}\frac{B_{2k}}{(2k)!}\,t^{\,2k-1}
\). Termwise Laplace integration of \(
\frac{K(t)}{t}=\sum_{k\ge1}\frac{B_{2k}}{(2k)!}\,t^{\,2k-2}
\)
provides
\(\int_0^\infty \deriv t \, t^{2k-2} e^{-\alpha' c t}
\!=\!\Gamma(2k-1)(\alpha' c)^{1-2k},\; k\!>\!\tfrac12,\;\Re(\ap c)\!>\!0.\) 
Using \(B_{2k}\!=\!-2k\,\zeta(1-2k),\;k\geq1\) gives
\begin{align}
\ln I(\ap)
&=\Phi(\mu_0,\mu_1;\ap)\nonumber\\
&-\sum_{k\ge1}\frac{\zeta(1-2k)}{2k-1}\,
\left(\frac{1}{(\alpha' \mu_0)^{2k-1}}+\frac{1}{(\alpha' \mu_1)^{2k-1}}-\frac{1}{(\alpha'(\mu_0+\mu_1))^{2k-1}}\right)\ ,
\label{eq:large-alpha}
\end{align}
in agreement with \req{highseries}. Hence, the subleading terms are governed by zeta values \(\zeta(1-2k)\) with negative odd argument.

Note, that both \eqref{eq:small-alpha} and \eqref{eq:large-alpha} come from the \emph{same} kernel \(K(t)/t\) introduced in \eqref{BernoulliK}.
To derive a single-line MB bridge for the  Laplace-Binet expression \eqref{eq:LB-mother}
we insert  the Mellin inversion 
$$e^{-x}=\frac{1}{2\pi i}\int\limits_{-i\infty+c}^{+i\infty+c}\deriv s \, \Gamma(s)\,x^{-s}, \quad c>0$$ 
with vertical contour 
$\Re s = c$ in the strip $c_1 < \Re s < c_2$ and apply (\ref{Titma}) 
to get the \emph{one-line MB bridge}
\begin{align}
\ln I(\alpha')-\Phi(\mu_0,\mu_1;\ap)
&=\frac{1}{2\pi i}\int\limits_{-i\infty+c}^{+i\infty+c} \deriv s \, 
\Gamma(s)\,\Gamma(-s)\,\zeta(-s)\,
\big[\mu_0^{-s}+\mu_1^{-s}-(\mu_0{+}\mu_1)^{-s}\big]\,\alpha'^{-s} \label{eq:MB-bridge}\nonumber\\
&=\frac{1}{4\pi i}\int\limits_{-i\infty+c}^{+i\infty+c} \deriv s \,
(2\pi)^{-s}\,\frac{\Gamma(s)\,\zeta(1+s)}{\cos\lf(\frac{\pi s}{2}\ri)}\,
\big[\mu_0^{-s}+\mu_1^{-s}-(\mu_0{+}\mu_1)^{-s}\big]\,\alpha'^{-s}.
\end{align}
 The second equation follows from the identity $\Gamma(-s)\zeta(-s)=(2\pi)^{-s}\tfrac{\sin\lf(\tfrac{\pi s}{2}\ri)}{\sin(\pi s)}\zeta(1+s)=\tfrac12(2\pi)^{-s}\cos\lf(\tfrac{\pi s}{2}\ri)^{-1}\zeta(1+s)$.
Shifting the contour to the right and  collecting the residua at $s=1,2,\ldots$  by applying   ${\rm Res}_{s=n}\Gamma(-s)=-\tfrac{(-1)^n}{n!}$ 
 recovers \eqref{eq:large-alpha}.
 On the other hand, shifting the contour to the left and computing all residua at $s=0,-1,\ldots$  produces  \eqref{eq:small-alpha}. More precisely, for the residua at $s=-2,-3,\ldots$ we use ${\rm Res}_{s=-n}\Gamma(s)=\tfrac{(-1)^n}{n!}$ to yield the sum in \eqref{eq:small-alpha}. 
 Furthermore,  the residuum at $s=0$ arises from taking into account the double pole of
$\Gamma(s)\Gamma(-s)=-\tfrac{1}{s^2}-\zeta(2)+\Oc(s^2)$, expanding  $\zeta(-s)=-\tfrac12+\tfrac{s}{2}\ln(2\pi)+\Oc(s^2)$ and $\big[\mu_0^{-s}+\mu_1^{-s}-(\mu_0{+}\mu_1)^{-s}\big]\,\alpha'^{-s}=1+s\ln\tfrac{\mu_0+\mu_1}{\ap\mu_0\mu_1}+\Oc(s^2)$. Altogether this gives the contribution $-\tfrac12\ln(2\pi)+\tfrac12\ln\tfrac{\mu_0+\mu_1}{\ap\mu_0\mu_1}$ at $s\!=\!0$. In addition, for the residuum at $s=-1$ we have: $\Gamma(s)=-\tfrac{1}{s+1}+\Oc((s+1)^0)$, $\zeta(-s)=-\tfrac{1}{s+1}+\Oc((s+1)^0)$ and $\big[\mu_0^{-s}+\mu_1^{-s}-(\mu_0{+}\mu_1)^{-s}\big]\,\alpha'^{-s}=-\ap\phi(\ap)(s+1)+\Oc((s+1)^2)$, which in total gives the residuum $-\ap\phi(\ap)$ at $s=-1$. 

Let us comment on potential contributions from the infinite semicircles when closing the contours to the right or left. For $\Re\mu_0>0,\Re\mu_1>0$ the relevant part of the integrand \eqref{eq:MB-bridge} assumes the form $F(s)=\Gamma(s)\,\Gamma(-s)\,\zeta(-s)\alpha'^{-s}=-\tfrac{\pi}{\sin(\pi s)}\tfrac{\zeta(-s)}{s}\ap^{-s}$.
For $s=|s|e^{i\phi}$ away from the real axis ($\phi\neq0,\pi$) the sine factor provides exponential damping 
$e^{-\pi |s\sin\phi|}$, while the zeta factor $\zeta(-s)$ develops only polynomial growth in $|s|$. Altogether, with the factor $e^{-s\ln\ap}$, which behaves as $|e^{-s\ln\ap}|\sim e^{-|s|\cos\phi\ln\ap}$, i.e., the decay depends on the sign of $\ln\ap$, this ensures exponential decrease along the large arcs in the right (left) half-plane for $\alpha'\to\infty\ (\alpha'\to 0)$. Hence, in both cases the arc contributions vanish.

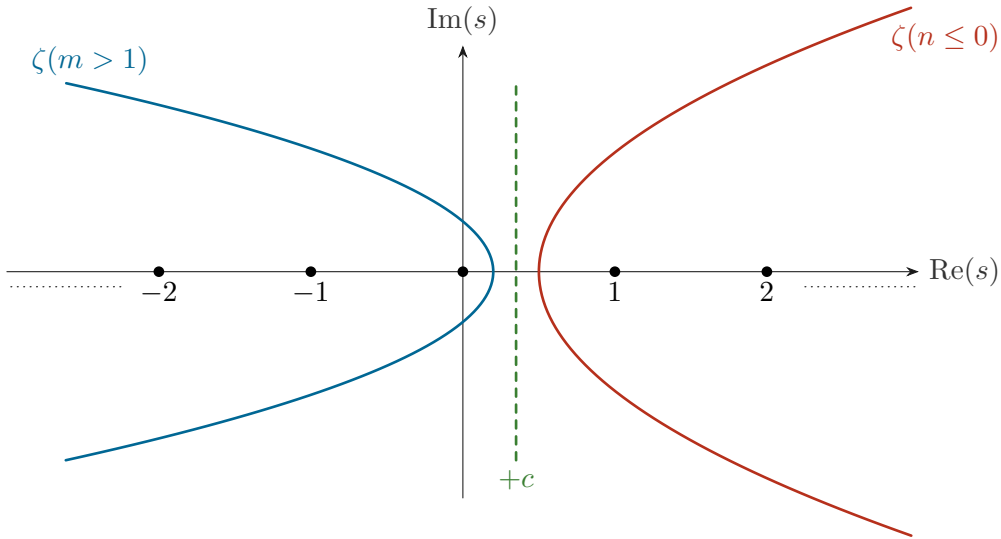
\begin{figure}[h]
    \centering
    \input{tikzMellin}
    \caption{Low-energy \eqref{eq:small-alpha} and high-energy \eqref{eq:large-alpha} expansions  in complex Mellin space $s$. The contour is closed to the left for $\ap\ra0$ and to the right for $\ap\ra\infty$.}
    \label{MellinBarnes}
\end{figure}

It is interesting to note that the WKB phase \req{WKB} entering \req{Phi} stems from the residuum at $s\!=\!-1$, while the low-energy field-theory part from $s\!=\!0$.
The Laplace-Binet representation \eqref{eq:LB-mother} and the single-line MB bridge \eqref{eq:MB-bridge} are \emph{transform-dual} descriptions of the same object; the two asymptotic regimes are just opposite contour shifts in the MB plane or, equivalently, two applications of Watson's lemma to the same Laplace kernel, cf. Fig.~\ref{MellinBarnes}.

The Mellin–Barnes representation \req{eq:MB-bridge} defines a single meromorphic function in the complex $s$-plane, whose pole structure simultaneously governs both asymptotic regimes. The small $\ap$-expansion is controlled by the rightmost boundary pole at $s=0$, while the leading large-$\ap$-behaviour---the WKB phase \req{WKB} entering \req{Phi}---originates from the next pole at $s=-1$. Thus the low- and high-energy limits are determined by different boundary contributions of the same Mellin transform.

\subsection{Two asymptotic regimes as Stokes sectors}

The differential equation \req{DGL} defines a meromorphic connection \req{connection} in the  complex $\ap$-plane with an irregular singularity at $\ap\!=\!\infty$. The Mellin-Barnes representation \req{eq:MB-bridge} lifts the amplitude to a single meromorphic function in the auxiliary complex Mellin variable $s$ whose pole structure governs both asymptotic regimes  in $\ap$.
In particular, it makes manifest that the fixed-angle high-energy limit \req{eq:large-alpha} and the low-energy  $\ap$-expansion \req{eq:small-alpha} arise as distinct asymptotic sectors of a single meromorphic object.
From this viewpoint, resurgence is not an additional structure imposed on the amplitude, but a direct consequence of the irregular singularity at $\ap=\infty$. 
The Stokes phenomenon is inherited from the irregular singularity of the differential equation \req{DGL} at $\ap=\infty$ and is reflected in the Mellin representation through the dominance properties of $(\ap)^{-s}$.

The contour deformation in the complex Mellin space $s$ is not arbitrary but depends on the sign of $\ln\ap$: for $\ap>1$ the integrand decays exponentially in the right half-plane and the contour may be closed to the right, whereas for $0<\ap<1$ it decays in the left half-plane and the contour must be closed to the left. In this way, the two asymptotic expansions arise from different regions of the same meromorphic integrand.
Upon analytic continuation in $\arg\ap$, the exponential factor $(\ap)^{-s}=e^{-s\ln\ap}$ changes its dominance properties.\footnote{The Mellin–Barnes parameter  $s$ controls multiplicative scaling in $\ap$ and is comparable to the Borel variable $\zeta$ of the large $\ap$-expansion. In fact, the exponential weight $\ap^{-s} = e^{-s\ln\ap}$ induces dominance regions in the $s$-plane analogous to those of a Laplace-Borel transform (with exponential factor $e^{-\zeta z}$, cf. \req{defTrans}), which explains the formal similarity of the contour deformations to standard Borel resummation pictures.} This leads to a Stokes phenomenon in parameter space: different exponential sectors become dominant in different angular regions of $\ap\in\mathbb{C}$.

From the viewpoint of twisted cohomology, the Mellin-Barnes integral \req{eq:MB-bridge} represents a twisted period. The low-energy (unipotent/KZ-type) expansion \req{eq:small-alpha} and the high-energy (WKB/Picard-Lefschetz-type) expansion \req{eq:large-alpha} correspond to different bases of twisted integration cycles (equivalently, different steepest–descent thimbles). The transition between them is governed by a Stokes matrix acting on the space of twisted cycles, cf. \cite[Sec. 4.3-4.5]{Aomoto:2011ggg} and \cite[Sec.~A.2.2]{Mizera:2019gea}. In this sense, the Mellin-Barnes bridge \req{eq:MB-bridge} makes manifest that the two asymptotic regimes are not independent structures, but rather two Stokes sectors of a single global twisted period. Twisted intersection theory provides a natural geometric framework to describe this relation, and is further discussed in Section~\ref{sec: sec5}.

Since both the low-energy expansion \req{lowseries} and the high-energy transseries \req{highseries} solve the same Aomoto-Gauss-Manin differential equation \req{DGL} in $\ap$, they must be related by a constant connection coefficient. In the present four-point case the twisted cohomology space is one-dimensional, so this reduces to a single scalar constant relating the two canonical local solutions. 
More explicitly, with \req{Start}
and normalizing the low-energy solution \req{lowseries} of \req{DGL} by $I_0(\alpha')=\tfrac{\mu_0\mu_1}{\mu_0+\mu_1}I(\ap)$ with $I_0(\alpha')\to 1$ as $\alpha'\to 0$ and extracting the canonical WKB factor at infinity as $I_\infty(\alpha')=\sqrt{\tfrac{\mu_0\mu_1}{2\pi(\mu_0+\mu_1)}}I(\alpha')$
one finds\footnote{I.e., $\cal S$ is simply the ratio of the two integration constants $c_0$ entering the Ans\"atze \req{Ansatz1} and \req{Ansatz2}.}
\begin{equation}\label{Associator}
\mathcal S=\fc{I_0(\alpha')}{I_\infty(\alpha')}=\sqrt{2\pi}\ 
\sqrt{\frac{\mu_0\mu_1}{\mu_0+\mu_1}}\ ,
\end{equation}
up to the conventional choice of asymptotic normalization at
$\alpha'=\infty$.
This constant encodes the transformation between the unipotent expansion at $\ap=0$, governed by positive zeta values $\zeta(n>1)$, and the irregular WKB sector at $\ap=\infty$, whose subleading corrections involve negative zeta values $\zeta(1-2k)$.
From the Mellin-Barnes perspective, this connection coefficient has a natural global origin: both asymptotic regimes arise as residue expansions of the same meromorphic integrand \req{eq:MB-bridge}. The constant \req{Associator} relating the two local solutions therefore reflects the global analytic continuation of this single Mellin transform rather than an additional dynamical input. In this sense, the arithmetic structures appearing in the two asymptotic sectors are not independent, but different boundary manifestations of one global twisted period.

In the case of regular singular differential equations, such connection constants are provided by Drinfeld’s associator, cf.~\cite{Schlotterer:2012ny}. Since $\ap=\infty$ is an irregular singular point in the present setting, following the same reasoning the corresponding object  should instead be interpreted as a Stokes-type connection coefficient  \req{Stokesconnection}---an irregular analogue of the connection coefficients familiar from the theory of associators acting on the space of twisted cycles. As an example comparing  the two regions $\Delta_1$ and $\Delta_3$ in Fig.~\ref{fig: regularFan4} yields \req{eq: sanity check string transseries}:
\be
P_{\Delta_1\Delta_3}=\fc{I_{\Delta_3}}{I_{\Delta_1}}=\fc{\sin(\pi \ap\mu_0)}{\sin[\pi\ap(\mu_0+\mu_1)]}\ e^{\pi i \ap\mu_1},
\ee
and similar for the other 14 ratios.

%% file: tikzMellin.tex
\begin{tikzpicture}[>=Stealth, line join=round, line cap=round] 

\draw[black!90,->] (0,-3) -- (0,3);
\draw[black!90,->] (-6,0) -- (6,0);
\node[black!80,above] at (0,3) {$\Im(s)$};
\node[black!80,right] at (6,0) {$\Re(s)$};

\draw[OliveGreen,dashed,line width=1pt] (0.7,-2.5) -- (0.7,2.5);
\node[OliveGreen, below] at (0.7,-2.5) {$+c$};

\foreach \y in {-2,-1,1,2}{ 
    \fill (2*\y,0) circle (2pt);
    \node[black,below] at (2*\y,0) {$\y$};
}
\fill (0,0) circle (2pt);

\draw[black,dotted] (4.5,-0.2) -- (6,-0.2);
\draw[black,dotted] (-4.5,-0.2) -- (-6,-0.2);

\node (A) at (6,3) {};
\node (C) at (1,0) {};
\node (B) at (6,-3) {};

\draw[MidnightBlue, line width=1pt, domain=-2.5:2.5, samples=100]
plot ({0.4 - 0.9*\x*\x}, {\x});
\node[MidnightBlue,left] at (-4,2.8) {$\zeta(m >1)$};

\draw[BrickRed, line width=1pt, domain=-3.5:3.5, samples=100]
plot ({1 + 0.4*\x*\x}, {\x});
\node[BrickRed,right] at (5.5,3.1) {$\zeta(n \leq 0)$};

\end{tikzpicture}

%% file: resurgenceClosedKLT.tex
\section{Lefschetz thimbles  and high-energy KLT relation}
\label{sec: sec5}
 
In this section, we reinterpret our results in terms of twisted intersection theory. In addition, we formulate a high-energy double-copy relation in terms of Lefschetz thimbles. 

\subsection{Stokes multipliers and (twisted) intersection numbers}

Tree-level $n$-point string amplitudes admit a natural formulation in terms of pairings in twisted de Rham (co)homology, which arises from endowing their moduli space $\mathcal{M}_{0,n}$ with a \textit{local system}. This perspective yields a purely topological reformulation of genus-zero string amplitudes, and supports the existence of the amplitude basis of integrals of the form \eqref{Zint} used throughout this paper. We briefly review this construction below, referring to \cite{Mizera:2019gea} for a self-contained introduction.

A local system is a representation of the fundamental group of a given space which encodes its monodromy data. In the case of the moduli space $\mathcal{M}_{0,n}$ of Riemann spheres with $n$ punctures, the monodromies are sourced by the complex Mandelstam invariants \eqref{Mandelstam} entering \eqref{Zint} so the local system furnishes a direct connection to the space of kinematics. The fundamental group $\pi_1(\mathcal{M}_{0,n})$ is generated by loops $\circlearrowleft_{ij}$ in which a puncture $z_j$ encircles another $z_i$ and returns to its initial position, so the local system $\mathcal{L}_\omega$ is prescribed by the closed one-form $\omega \coloneqq \deriv \ln(\KN)$ (the `twist') with $\KN$ the multi-valued Koba-Nielsen factor 
\be \label{KNfactors}
\KN \coloneqq \prod\limits_{i<j}^{n-1} z_{ij}^{-\ap\hat{s}_{ij}}=e^{-\ap S},
\ee
where $S$ is the Morse action given in \req{eq: morse action}, along with the map
\begin{equation}
    \mathcal{L}_\omega \, :  \, \gamma \mapsto \exp \int_\gamma \omega
\end{equation} assigning to every path $\gamma$ in $\mathcal{M}_{0,n}$ a \textit{local coefficient}. One can then consider paths with coefficients in $\mathcal{L}_\omega$, denoted by $(p,q) \otimes \KN$, which to every point $z \in \mathbb{CP}^{n-3}$ along the path $(p,q)$ in $\mathcal{M}_{0,n}$ associates the coefficient $\KN(z)$. Introducing the boundary operator
\begin{equation}
    \partial_\omega \left( (p,q) \otimes \KN(z) \right) \coloneqq - p \otimes \KN(z) + q \otimes \KN(z), \quad \partial_\omega^2 = 0,
\end{equation} one obtains a natural definition of \textit{twisted homology}: $H_k(\mathcal{M}_{0,n}, \mathcal{L}_\omega) \coloneqq \ker \partial_\omega / \mathrm{im} \,\partial_\omega$, i.e., homology groups of paths with coefficients in $\mathcal{L}_\omega$, whose elements
\be\label{TwCyc}
C_i\otimes \KN
\ee
are called \textit{twisted cycles}. The (Poincar\'e-dual) cohomological analogue is obtained by considering the twisted connection $\nabla_{\pm\omega} \coloneqq \deriv \pm \omega \wedge$. Since $\omega$ is closed, $\nabla_{\pm \omega}^2 =0$ and the \textit{twisted cohomology} groups are $H^k(\mathcal{M}_{0,n}, \nabla_{\pm \omega}) \coloneqq \ker \nabla_{\pm \omega} / \mathrm{im} \, \nabla_{\pm \omega}$, whose elements $\varphi_\pm$ are called \textit{twisted cocycles}. Aomoto \cite{Aomoto:1975} showed that the only non-vanishing twisted cohomology group has $k=n-3$; the natural dual basis of cycles of $H^k(\mathcal{M}_{0,n}, \mathcal{L}_\omega)$ is thus provided by the canonical $(n-3)$-dimensional integration cycles $C_\pi$ of \eqref{domain} on $\mathcal{M}_{0,n}$. In the following we identify $C_\pi\simeq C_i$ with $\pi\in S_{n-3}$ and $i=1,\ldots,(n-3)!$. 

These definitions naturally give rise to the pairing, or \textit{period},
\be\label{Period}
\langle C_i \otimes \KN |\varphi\rangle:=\int_{C_i}\KN \ \varphi, \quad \varphi\in H^{n-3}(\mathcal{M}_{0,n},\nabla_\omega).
\ee 
For real kinematics \req{Mandelstam} we have $\Lc_{\pm\om}\simeq \Lc_{\mp\ov\om}$.
A similar construction as \req{TwCyc} then applies for the dual cycles of the twisted homology group $H_{n-3}(\Mc_{0,n}, \mathcal{L}_{\ov \om})$ with $\ov \om \coloneqq \deriv \ln \overline{\KN}$, and gives rise to the period integrals
\be\label{Periods}
\langle \tilde C_j\otimes \overline{\KN}|\varphi\rangle:=\int_{\tilde C_j}\overline{\KN}\ \varphi, \quad \text{where} \quad \overline{\KN} \coloneqq \prod_{i<j}^{n-1} \ov z_{ij}^{-\ap\hat{s}_{ij}}.
\ee
These two sets of twisted cycles allow to consider pairings thereof since $\Lc_\om\otimes\Lc_{\overline\om}$ is monodromy invariant.
Given two bases $\{C_i\}_{i=1}^{(n-3)!}$ and $\{\tilde C_j\}_{j=1}^{(n-3)!}$
of canonical twisted cycles, one defines the \textit{intersection matrix} as
\be\label{interI}
\langle C_i\otimes \KN |\tilde C_j\otimes \overline{\KN }\rangle \eqqcolon H^{-1}_{ij}.
\ee  
This fundamental object relates different integration cycles $C_i$ and allows for the decomposition of both real and complex integrals. Its entries are called \textit{twisted intersection numbers}.
From this point of view, open string integrals \req{Zint} are periods in the sense of \req{Period} and \req{Periods} for a suitable choice of cocycle $\varphi$, integrated over the canonical Euler cycles \eqref{domain}. For instance, for $n=4$ and $C_1 = (0,1)$ we have $\langle (0,1) \otimes \KN | \deriv \ln z \rangle = -Z_{11}(\alpha')$ as in \eqref{eq: integral form factor}.

As mentioned above, the convergence (and thus well-definedness) of \eqref{Period} generically depends on the choice of kinematic variables \req{Mandelstam}, which enter here through $\KN$. The divergence of the worldsheet integrals can be handled within this framework by an appropriate \textit{regularization} of the twisted cycles \eqref{TwCyc}. The latter amounts to compactifying $C_i \otimes \KN$ by attaching infinitesimal loops to its endpoints \cite{Mizera:2019gea}. This method is equivalent in spirit to the regularization procedure of \cite{Witten:2013pra}, as reported in \S \ref{eq: steepest descent methods Yoda} for the four-point amplitude, which promoted $C_1$ to $C_\epsilon$. This construction was systematically extended and refined in \cite{Eberhardt:2024twy}, where a procedure for constructing generalized convergent and compact Pochhammer contours $\Gamma_n$ was provided. These arise upon the Deligne-Mumford compactifictation of $\mathcal{M}_{0,n}$ \cite{Deligne:1969}, i.e., by `blowing-up' points on the boundaries of $\mathcal{M}_{0,5}$ where three or more punctures collide, and then attaching `tubes' to each codimension-1 boundary. The contours $\Gamma_n$ are based on the combinatorics of associahedra, whose edges (facets) directly correspond to boundary components $s_{ij}\ra0$, or factorization channels of the amplitude.

\begin{figure}[ht]
    \centering
    \input{tikzPunctures.tex}
    \caption{Blow-up of the integration contour in the real part of $\mathcal{M}_{0,5}$ (left) to $A_4$ (right).}
    \label{fig: chambers n=5}
\end{figure}
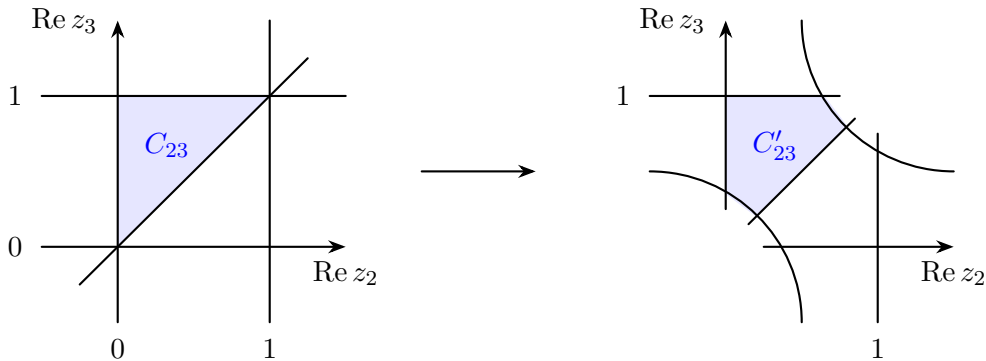

In Figure~\ref{fig: chambers n=5}, we display the Deligne-Mumford compactification of $\mathcal{M}_{0,5}$, which effectively promotes the contour $C_{23}$ in the real subspace of $\mathcal{M}_{0,5}$ to the new contour $\tilde C_{23}$, where the divisors at $(0,0)$ and $(1,1)$ have been blown up.
The connected components in the real part of $\mathcal{M}_{0,n}$ are sometimes called \textit{chambers}. The boundaries of $C'_{23}$ draw a pentagon, which corresponds to the associahedron $A_4$. 
The convergent compact contour $\Gamma_5$ is obtained by attaching tubes to the boundaries of $ C'_{23}$, see \cite{Eberhardt:2024twy} for details and $n > 5$. 

The compactification and regularization procedures discussed above provide a way of making string theory integrals well-defined beyond their region of absolute convergence by suitable alteration of the canonical twisted cocycles $C_i \otimes \KN$. This approach enabled the evaluation of $n$-point tree-level string amplitudes for finite $\alpha'$ beyond the domain of convergence \cite{Eberhardt:2024twy}. There exists, however, an alternative basis of twisted homology which is better suited for extracting their high-energy resurgent structure: the \textit{thimble basis}.

As discussed in \S\ref{sec: periods low high amplitudes}, the Morse function \req{eq: morse action}  has $(n-3)!$ saddles solving the equations \req{SQE}.
Each saddle $z^{(i)}$ has $n-3$ complex coordinates corresponding to $2(n-3)$ real directions. These split into $n-3$ downward (steepest-descent) and $n-3$ upward (steepest ascent) gradient-flow directions. At each saddle  these directions generate one descent thimble $\Jc_m$ and one ascent thimble $\Kc_n$. These cycles also form a basis of $H_{n-3}(\mathcal{M}_{0,n},\Lc_\omega)$ \cite{Witten:2010cx, Aomoto:2011ggg}. This means that any string integral of the form \req{Zint} can be decomposed into a sum over exactly these, by expanding the canonical cycle $C_i$ onto the thimble basis $\{\mathcal{J}_m\}$, with coefficients given by twisted intersection numbers between these two bases. The dual thimbles $\Kc_n$ (in ascent directions) refer to upward gradient flows and generically do not define convergent integration cycles, as they typically run into singular regions or infinity, and thus do not necessarily end on edges. They are mainly used to determine intersection numbers. Meanwhile, the descent thimbles $\Jc_m$ always flow to the boundary of  the moduli space corresponding to factorization channels. For $n=5$ the latter are the edges of the chambers in Figure~\ref{fig: chambers n=5}. 

The intersection matrix of Lefschetz thimbles is given by the topological intersection numbers between two sets of manifolds associated with the critical points of a holomorphic function (the action or the integrand's exponent) 
\be
I_{mn}=\langle\Jc_m\otimes \KN|\Kc_n\otimes \overline{\KN}\rangle,
\ee
with the stable thimbles $\Jc_m$ (`steepest descent' manifolds where the real part $\Re S$ of the action \req{eq: morse action} decreases, ensuring the integral converges) and unstable thimbles $\Kc_n$ also called `anti-thimbles' (manifolds of `steepest ascent' where the real part increases).
Thus the matrix element $I_{mn}$
represents the signed number of times the stable thimble 
$\Jc_m$ intersects the unstable thimble $\Kc_n$. Due to the way these are constructed from gradient flows, this matrix is typically the identity matrix $I_{mn}=\delta_{mn}$ after appropriate normalization and orientation choices, in a basis where each thimble is paired with its own unique anti-thimble, i.e., where they intersect only once at the corresponding saddle point.

Unlike canonical cycles, the configuration of thimbles (and anti-thimbles) depends explicitly on the choice of kinematics through the associated Morse function \eqref{Morse}. As the kinematic parameters vary, so do the critical points and their steepest descent and ascent contours, leading to different thimble configurations. Consequently, their overlap with the fixed canonical cycles can change discontinuously. These discontinuities occur when crossing Stokes walls, where the relative phases of saddle contributions align, and the corresponding jumps in intersection numbers thus provide the topological origin of Stokes phenomena in the asymptotic structure of the amplitude. 

\begin{figure}[H]
    \centering
    \begin{subfigure}{0.49\linewidth}
        \centering
        \includegraphics[width=\linewidth]{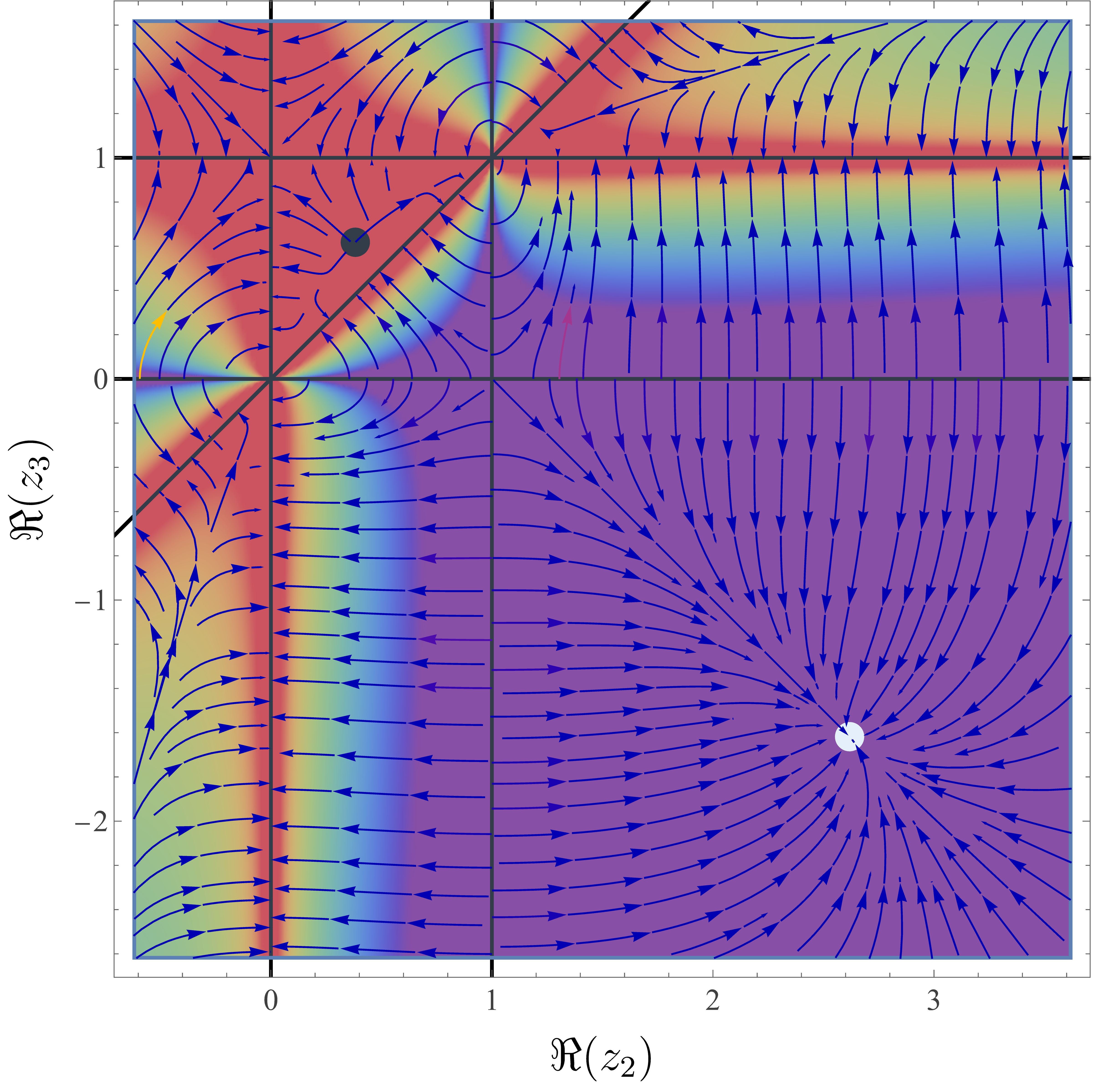}
        \caption{}
    \end{subfigure}
    \begin{subfigure}{0.49\linewidth}
        \centering
        \includegraphics[width=\linewidth]{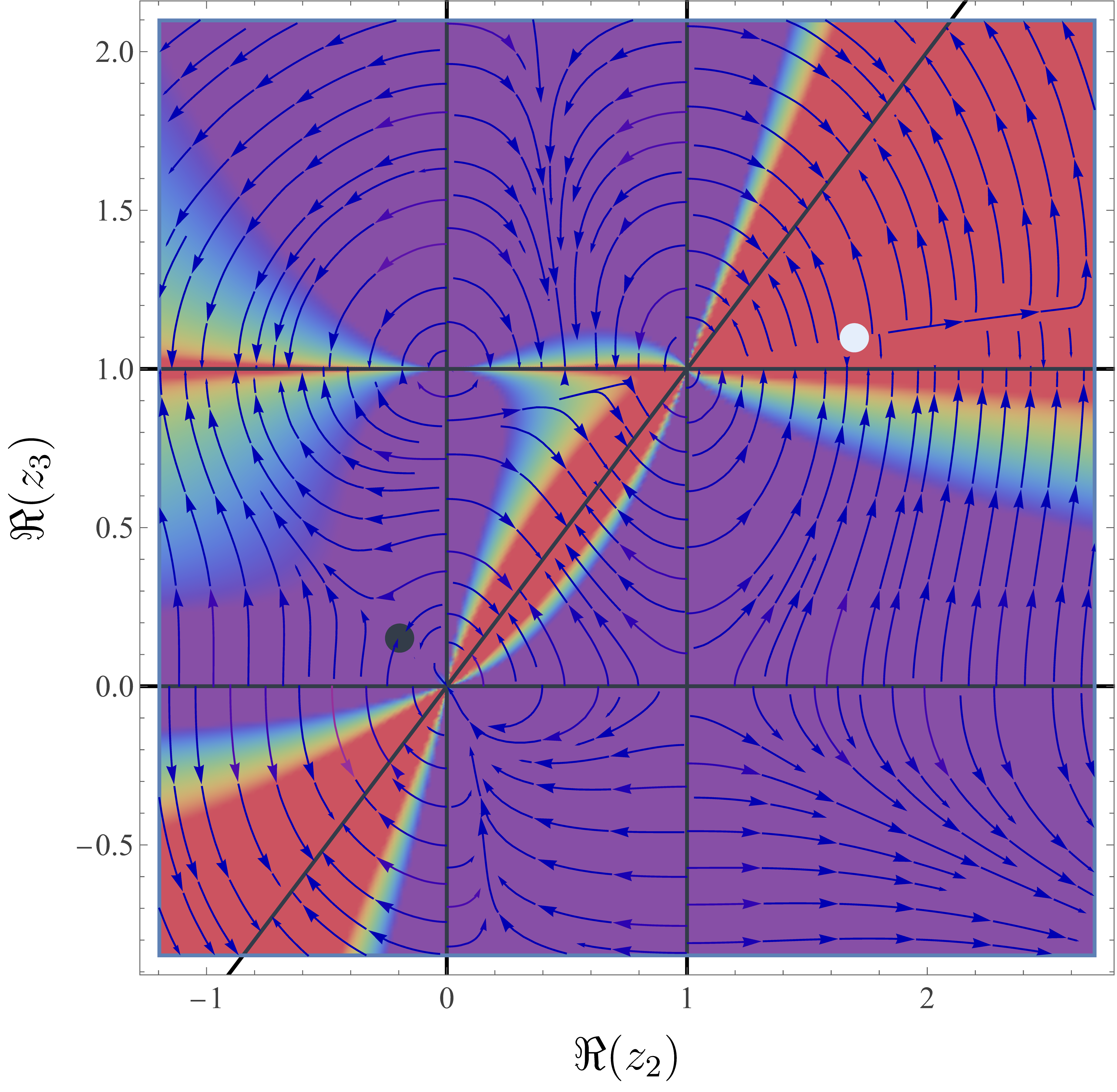}
        \caption{}
    \end{subfigure}
    \caption{Positions of the saddle points $z_\pm$ within the chamber decomposition (cf.~Figure~\ref{fig: chambers n=5}) for the $n=5$ amplitude and regular direction (a) $\eta=(1,1,1,1,1)$ and (b) $\eta' = (-2,5,1,1,1)$ at infinity, Note that $\eta$ and $\eta'$ lie in different sectors of Figure~\ref{fig: regular example}. The colour gradient depicts $-\Re S$ to show the relative dominance of the two saddles $z_+$ (white dot) and $z_-$ (black dot). We have $\Delta_\mathrm{SE} > 0$ in both cases, so the saddles are real. The arrows indicate the downward flow of $\Re S$.}
    \label{fig: thimbles saddles n=5}
\end{figure}

We illustrate this phenomenon for $n=5$ in Figure~\ref{fig: thimbles saddles n=5}, where we plot both the positions of the saddle points \eqref{eq: saddles n=5} within the chamber decomposition (see Figure~\ref{fig: chambers n=5}) and the corresponding values $\Re S$ of the leading contribution to the action \eqref{Morse} evaluated at these saddles, for two given fixed regular directions $\eta$ and $\eta'$ at infinity compatible with \eqref{eq: determinants leading}.
In the density plots\footnote{The actual thimbles are curves of steepest descent in the complexified space $z_2,z_3\in \mathbb{C}$, which are generally four-dimensional surfaces when decomposed into real and imaginary parts. Projecting the thimbles onto the real plane produces only a shadow of the full complex flow; as a consequence, features such as sinks or spirals in the real-plane gradient do not necessarily correspond to the directions along which the integral is dominated. In particular, a saddle with two negative eigenvalues of the Hessian of $\Re S$ appears as a sink in the real-plane gradient plot, since all real directions point toward the local maximum of $\Re S$. Conversely, a saddle with complex-conjugate Hessian eigenvalues produces a spiral pattern, reflecting that the local flow rotates around the stationary point. These patterns do not fully capture the contribution of the thimble in the complex integral.} of $\Re S$, the arrows indicate the downward flow of the action in the real $(z_2,z_3)$-plane, but they should not be interpreted as the true Lefschetz thimbles.

Figure~\ref{fig: thimbles saddles n=5} shows that a given regular direction $\eta$ is associated with a specific configuration of saddles distributed across the different chambers, together with a definite hierarchy (or dominance ordering) of the action evaluated on these saddles. In fact, as long as $\eta$ remains within a fixed regular region, the saddles stay in the same chambers and their relative dominance is unchanged. However, upon crossing into a different regular sector with regular direction $\eta'$, the configuration changes discontinuously: saddle points move from one chamber to another, and the relative dominance of the action can also change. 
\noindent
This was already inferred from Figure~\ref{fig: eta slice}. Since the construction of Lefschetz thimbles depends explicitly on the location of the saddles, the thimbles themselves also undergo a corresponding transformation when crossing between regular sectors. At the same time, this transition involves a change in the chamber decomposition of thimbles. Since chambers provide a basis of constant cycles in terms of which any integration cycle can be expressed, this reorganization of saddles and chambers manifests as a Stokes jumps.

Let us refine this statement. The thimbles $\Jc$ under consideration refer  to a certain asymptotic expansion \req{KLaplace} related to a specific Stokes sector $\eta$
(accounting for regions in the parameter space of $s_{ij}$), hence we should 
bookkeep them as $\Jc_m(\eta)$.
In a given sector $\eta$ any cycle $C_i$ can be decomposed into a basis of thimbles \cite{Aomoto:2011ggg}: 
\be\label{decTH}
C_i\otimes \KN=\sum_{m=1}^{(n-3)!} \langle C_i\otimes \KN|\Kc_m(\eta)\otimes \overline{\KN}\rangle\ \Jc_m(\eta)\otimes \KN\ .
\ee
The coefficients $\langle C_i|\Kc_m(\eta)\rangle$
are determined by the intersection of cycles $C_i$
with the dual anti-thimbles $\Kc_m(\eta)$.  In Picard-Lefschetz theory the Stokes matrix $\Pc_{mn}$  describes how the thimble basis jumps when crossing Stokes walls \cite{Aomoto:2011ggg}
\be
\Jc_m(\eta)=\sum_{n=1}^{(n-3)!}\Pc_{mn}(\eta,\eta')\;\Jc_n(\eta'),
\ee
where $\eta$ and $\eta'$ denote adjacent Stokes sectors separated by a Stokes wall. In this sense, the matrix
\be\label{decT}
\Pc_{im}\simeq \langle C_i\otimes \KN|\Kc_m(\eta)\otimes \overline{\KN}\rangle,
\ee 
encodes the Stokes data of the integral, as it transforms under changes of
Stokes sector via the Stokes matrices acting on the thimble basis, i.e.,~\req{decT} describes the projection of the cycle $C_i$ onto the thimble $\Jc_m(\eta)$. Because $C_i$ is a steepest descent cycle for kinematics within the region of convergence, \eqref{decT} can also be understood as the analytic continuation of thimbles to parameters outside of the convergent domain.

The explicit decomposition \req{decTH} was given by Aomoto in \cite{Aomoto:1987} for hypergeometric integrals associated with the value of ${}_3F_2$ at unity, which are closely related to the building blocks of the $n\!=\!5$ string amplitude \req{eq: minimal functions N=5}.
There, Aomoto also derives the difference system \eqref{eq: shift equation} from a homological perspective, considering a single cocycle integrated over two independent real chambers. In contrast, we adopted here a cohomological viewpoint in \S \ref{sec: sec3}, fixing a single cycle ($C_{23}$) and expressing the two integrals \eqref{eq: minimal functions N=5} in terms of two cocycles. The two approaches are of course equivalent. While the correspondence still holds at higher multiplicities, constructing the twisted cycles explicitly becomes much more intricate, while relating cocyles through IBP reduction remains feasible.

In Aomoto’s ${}_3F_2$ setting, one should distinguish two notions of `sectors'. The real chambers are the connected components of the complement of the real singular locus in the integration-variable space (punctures), see Figure~\ref{fig: chambers n=5}. Each chamber defines a real integration cycle, but these satisfy relations in twisted homology, leaving only two independent cycle classes. By contrast, the Stokes/asymptotic sectors lie in parameter (kinematic) space, where asymptotic limits are taken. These are angular regions in which a given asymptotic expansion---and corresponding canonical basis of solutions---is valid; crossing a Stokes ray changes this basis via a connection (Stokes) matrix. 

Ultimately, the link between the two is that chamber cycles provide global generators which, after analytic continuation in parameters and deformation into the complex domain, decompose into sectorial steepest-descent cycles governing asymptotics. When a Stokes ray is crossed, this decomposition jumps, producing the connection functions. Thus, these functions relate sectorial asymptotic bases, while the chamber description provides a concrete topological model of the change of basis.

\subsection{High-energy KLT relation and resurgent closed string}

By applying worldsheet monodromies, Kawai, Lewellen and Tye (KLT) derived a representation of closed string amplitudes in terms of open string amplitudes, valid at tree level and finite $\ap$ \cite{Kawai:1985xq}. Meanwhile, the number of critical points of the $n$-point Koba-Nielsen potential \req{eq: morse action} equals $(n-3)!$, matching the dimension of the basis of ordered open string integrals \req{Zint}. This correspondence underlies the compatibility between the thimble decomposition and the KLT representation.
In this section, we give the KLT relation for $\ap\ra\infty$ in terms of 
Lefschetz thimbles and their intersection numbers.

 At the mathematical level the KLT relation equates multi-dimensional complex integrals on $\Mc_{0,n}(\mathbb{C})$ to linear combinations of squares of real iterated integrals \req{Zint} on $\Mc_{0,n}(\mathbb{R})$.
In terms of the underlying worldsheet geometries a splitting of complex integration into holomorphic and anti-holomorphic
sectors is achieved by performing an analytic continuation of complex coordinates $z_i$ leading to the following
homological splitting \cite{Kawai:1985xq,Broedel:2013tta}
\begin{align}
\Mc_{\alpha\beta}&=\int_{\mathbb{C}^{n-3}}\deriv^2z_i\    \fc{\prod_{i<j}^{n-1} \ov z_{ij}^{-\ap\hat{s}_{ij}}}{\ov z_{1 2_\al} \ov z_{2_\al 3_\al} \cdots \ov z_{(n-3)_\al (n-2)_\al}}\ \fc{\prod_{i<j}^{n-1} \ov z_{ij}^{-\ap\hat{s}_{ij}}}{z_{1 2_\beta} z_{2_\beta 3_\beta} \cdots z_{(n-3)_\rho (n-2)_\beta}}\nonumber\\
&=(-1)^{n-3}\sum_{\pi,\rho\in S_{n-3}}^{(n-3)!} Z_{\pi\alpha}(\ap)\ \Sc[\pi|\rho]\ \widetilde{Z}_{\rho\beta}(\ap),\label{KLTZZ}
\end{align}
with  the string KLT kernel \cite{Kawai:1985xq,Bjerrum-Bohr:2010pnr} 
\be\label{SKernel}
\Sc[\rho|\si]:=\Sc[\, \rho(2,\ldots,n-2) \, | \, \si(2,\ldots,n-2) \, ]= \prod_{j=2}^{n-2} \sin\Big( \, s_{1,j_\rho} \ + \ \sum_{k=2}^{j-1} \theta(j_\rho,k_\rho) \, s_{j_\rho,k_\rho} \, \Big),
\ee
related to the field-theory kernel \req{KLTkernel} by taking appropriately the limit $\ap\ra0$. In \req{KLTZZ} the open string integral $\tilde{Z}_{\rho\beta}$ refers to \req{Zint} with the ordering $\rho\simeq(1,\rho(2,\ldots,n-2),n,n-1)$. This result was reformulated in terms of twisted cycles $C_i$ and within twisted intersection theory into \cite{aomoto1987complex,Mizera:2017cqs}
\be\label{KLT}
\Mc_{\alpha\beta}=\sum_{i,j=1}^{(n-3)!}\lf(\int_{C_i} \overline{\KN}\; \overline{\tilde\varphi}_\alpha\ri)\ H_{ij}\ \lf(\int_{C_j} \KN\; \varphi_\beta\ri),
\ee
with the Koba-Nielsen factors \req{KNfactors}, the intersection matrix (with orthogonal/dual cycles $C^\vee$)  \req{interI}:
\be
H_{ij}^{-1}=\langle C_i\otimes \KN|C_j\otimes \overline{\KN}\rangle\equiv(-1)^{n-3}\;\Sc[\pi|\rho]^{-1},\quad H_{ij}=\langle C^\vee_i\otimes \KN|C_j^\vee\otimes \overline{\KN}\rangle,
\ee
and the $n-3$-forms $\overline{\tilde\varphi}_\alpha, \varphi_\beta$ referring to the specific configurations $\al,\bet$ in \req{KLTZZ}. Note, that \req{decTH} gives rise to the decomposition:
\be\label{newDec}
\int_{C_j} \KN\; \varphi_\beta=\sum_{m=1}^{(n-3)!}\langle C_j\otimes \KN|\Kc_m\otimes\overline{\KN}\rangle\int_{\Jc_m} \KN\; \varphi_\beta.
\ee
With \req{newDec}  we can rewrite \req{KLT} in terms of a basis of Lefschetz thimbles
\be\label{KLTa}
\Mc_{\alpha\beta}=\sum_{m,n=1}^{(n-3)!}\lf(\int_{\Jc_m} \overline{\KN}\; \overline{\tilde\varphi}_\alpha\ri)\ \widetilde H_{mn}\ \lf(\int_{\Jc_n} \KN\; \varphi_\beta\ri),
\ee
with the (inverse) intersection matrix $\tilde H^{-1}_{mn}=\langle \Jc_m\otimes \KN | \Jc_n\otimes \overline{\KN}\rangle$ 
\be
\widetilde H_{mn}=\langle \Kc_m\otimes \KN | \Kc_n\otimes \overline{\KN}\rangle=\sum_{i,j=1}^{(n-3)!}\overline{\langle\Kc_m|C_i\rangle}\  H_{ij}\ \langle C_j|\Kc_n\rangle,
\ee
i.e.  for a  given Stokes sector:
\be\label{ThimbleKLT}
\widetilde H=\Pc^\dagger\ H\ \Pc.
\ee
For $n=4$ we have $H=2i\; \tfrac{\sin(\pi u)}{\sin(\pi t)}\sin(\pi s)$ and $\langle C|\Kc\rangle=e^{-i\pi u}\fc{\sin(\pi t)}{\sin(\pi s)} = \Pc(s,u)$ for physical kinematics. Thus \req{ThimbleKLT} gives \cite{Mizera:2019gea}:
\be
\widetilde H=2i\ \fc{\sin(\pi t)\;\sin(\pi u)}{\sin(\pi s)}=\lng \Kc|\Kc\rangle.
\ee

Finally, since in our KLT formula \req{KLTa} we integrate along thimbles $\Jc_n$ and the 
latter  furnish a canonical saddle-point basis in which the worldsheet integrals \req{Zint} admit asymptotic stationary-phase expansions \req{Morse}, we may replace the integrals by their corresponding individual stationary phase expansions \req{Morsei}:
\be
\lf(\int_{\Jc_n} \KN\; \varphi_\beta\ri)\rightarrow 
\lf(\fc{2\pi}{-\ap}\ri)^{\tfrac12 (n-3)}\ Z^{(n)}_{\beta}(\ap).
\ee
Note, that in the (high-energy) KLT representation \req{KLTa} 
 (in the thimble basis) each of the open string amplitudes is in one-to-one correspondence to {\it one} thimble and {\it one} expansion \req{Morsei} while in the original formula \req{KLT} each open string amplitude develops a full sum 
\req{Morse} over all $(n-3)!$ saddle points in its high-energy limit. 
To summarize, the high-energy limit of the KLT relation \req{KLTa} acquires a geometric
interpretation: closed string amplitudes arise from the intersection pairing of
Lefschetz thimbles associated with the saddle points of the Koba-Nielsen potential \req{eq: morse action}.

%% file: tikzPunctures.tex
\begin{tikzpicture}[>=Stealth, line join=round, line cap=round]

\begin{scope}[shift={(0,0)}]
\fill[color=blue!10!white, draw=none] (0,0) -- (2,2) -- (0,2);
\draw[] (0.65,1.35) node {\textcolor{blue}{$C_{23}$}};
\draw[thick,->] (-1,0) node[xshift=-10] {$0$} -- (3,0) node[yshift=-10] {$\Re z_2$};
\draw[thick] (-1,2) node[xshift=-10] {$1$} -- (3,2);
\draw[thick,->] (0,-1) node[yshift=-10] {$0$} -- (0,3) node[xshift=-20] {$\Re z_3$};
\draw[thick] (2,-1) node[yshift=-10] {$1$} -- (2,3);
\draw[thick] (-0.5,-0.5) -- (2.5,2.5);
\end{scope}

\begin{scope}[shift={(8,0)}]
\fill[color=blue!10!white, draw=none] (0.4,0.4) -- (1.6,1.6) -- (1.3,2) -- (0,2) -- (0,0.7);
\draw[] (0.65,1.35) node {\textcolor{blue}{$C'_{23}$}};
\draw[thick,->] (0.5,0) -- (3,0) node[yshift=-10] {$\Re z_2$};
\draw[thick] (-1,2) node[xshift=-10] {$1$} -- (1.5,2);
\draw[thick,->] (0,0.5) -- (0,3) node[xshift=-20] {$\Re z_3$};
\draw[thick] (2,-1) node[yshift=-10] {$1$} -- (2,1.5);
\draw[thick] (0.3,0.3) -- (1.7,1.7);
\draw[thick] (1,-1) arc (0:90:2);
\draw[thick] (3,1) arc (-90:-180:2);
\end{scope}

\draw[->, thick] (4,1) -- (5.5,1);

\end{tikzpicture}

%% file: ODEcoefs.tex
\section{Recurrence relation for the asymptotic series coefficients}
\label{app: ODEcoefs}

In this appendix, we provide details on the method used to study the resurgence of the asymptotic series \eqref{eq: asymptotic series F} for the tree-level form factor $F$. As explained in \S\ref{sec: four point tree string}, there is no closed-form expression for its coefficients $c_{n}(a)$. Here, we show that a recurrence relation obeyed by the latter may nonetheless be obtained, enabling fast and reliable numerics. 

We start from $F(s, -as) = -s B(-s, 1+as)$ and look for an ODE for
\begin{equation}
    B(-s,1+as) = \frac{\Gamma(-s)\Gamma(1+as)}{\Gamma(1+(a-1)s)}. 
\end{equation} Taking the logarithmic derivative in $s$ on both sides, we arrive at 
\begin{equation}
    \partial_s B(-s, 1+as) = B(-s, 1+as) \left[ - \psi(-s) + a \psi(1+as) - (a-1) \psi(1+(a-1)s) \right],
\end{equation} with $\psi(z) = \partial_z \ln \Gamma(z)$ the digamma function, and where all arguments are non-negative for our choice of kinematics. Using $\psi(1+z) = \psi(z) + \tfrac{1}{z}$, this further simplifies to 
\begin{equation}
    \label{eq: ODE beta function}
    \partial_s B(-s, 1+as) = B(-s, 1+as) \left[ - \psi(-s) + a \psi(as) - (a-1) \psi((a-1)s) \right].
\end{equation} Next, we ask that the perturbative series \eqref{eq: asymptotic series F} solve this equation, i.e.~we write the Ansatz
\begin{equation}
    B(-s,1+as) \sim \sqrt{-\frac{2 \pi a}{1-a}}(-s)^{-\frac{1}{2}} \left[B(a) \right]^s \sum_{n=0}^{\infty} \frac{c_{n}(a)}{(-s)^n},
\end{equation} for the Beta function as $s \to - \infty$, having introduced $c_0(a)= 1$ and $B(a) = (-a)^{a} (1-a)^{1-a}$ for ease of notation. The LHS of \eqref{eq: ODE beta function} is
\begin{equation}
    \partial_s B(-s,1+as) = \sqrt{-\frac{2 \pi a}{1-a}} [B(a)]^s \sum_{n=0}^{\infty} c_{n}(a) \left[\ln B(a)(-s)^{-n - \frac{1}{2}} + \left(n + \frac{1}{2}\right) (-s)^{-n - \frac{3}{2}}\right].
\end{equation} For the RHS, we use the known large-$z$ asymptotics \cite[5.15.8]{NIST:DLMF}
\begin{equation}
    \psi(z) \sim \ln z + \sum_{n=1}^{\infty} \frac{\zeta(1-n)}{z^n}, \quad \abs{\arg(z)} < \pi,
\end{equation} which hold in this kinematic regime where both $s,a < 0$ since the arguments of all $\psi$'s in \eqref{eq: ODE beta function} are then all strictly positive. Distributing and comparing the coefficients of $(-s)^m$ on both sides of \eqref{eq: ODE beta function}, we find the recurrence relation
\begin{equation}
    \label{eq: recurrence C_2m Bernouilli}
    c_{m}(a) = - \frac{1}{m} \sum_{k=1}^{\lfloor \frac{m+1}{2} \rfloor} c_{1+m-2k}(a) \frac{B_{2k}}{2k} \ell_{2k-1}(a), \quad m \geq 1, \quad c_0(a) = 1.
\end{equation} We verified that \eqref{eq: recurrence C_2m Bernouilli} reproduces the first few $c_n(a)$'s derived using saddle-point techniques in \cite{Kervyn:2025wsb}. This algorithm may be used to efficiently compute higher-order coefficients $c_{m}(a)$ recursively (up to $n_\text{max} \sim 200$ on standard machines with \verb|Mathematica|), as needed for resurgence methods, for instance to check that the series is Gevrey-1 and determine its parameters, or perform a Borel-Padé analysis like that displayed in Figure~\ref{fig: pade plot string amp}.

%% file: resurgenceGamma.tex
\section{Resurgence of the Gamma function and its logarithm}
\label{app: resurgence Gamma}

In this appendix, we provide details of the resurgence analysis of the Gamma function and its logarithm, which serve as fundamental building blocks of the four-point tree-level string amplitude \eqref{eq: string form factor}. Albeit this structure is well known in the asymptotic analysis literature \cite{WhittakerWatson1990, Nemes:2022} we review it here in detail, both for completeness and in order to introduce, in a concrete and familiar setting, several key concepts of resurgence theory used in the main text. We refer the reader to the reviews \cite{Dorigoni:2014hea,Aniceto:2018bis} for a more thorough introduction to resurgence and transseries in a broader context within physics.

\subsection{Transseries for \texorpdfstring{$\ln \Gamma$}{log Gamma}} 
\label{app: transseries log Gamma}

In order to understand the resurgent features of the Gamma function around large complex values of its argument, we first study its logarithm. Our starting point is the asymptotic expansion \cite{Gradshteyn:1702455}
\begin{equation}
    \label{eq: asymptotic log Gamma}
    \ln \Gamma(z) \sim \left(z - \frac{1}{2}\right) \ln z - z + \frac{1}{2} \ln 2\pi + \underbrace{\sum_{k=1}^{\infty} \frac{B_{2k}}{2k(2k-1)} z^{1-2k}}_{S_0(z)},
\end{equation} valid for $\abs{z} \to \infty$, $\abs{\arg(z)} < \pi$ (see, e.g., \cite{WhittakerWatson1990} for a derivation). The series $S_0(z)$ is asymptotic, owing to the factorial growth of Bernoulli numbers. Its Borel transform is given by
\begin{equation}
    \label{eq: derivation Borel transform}
    \mathcal{B}[S_0](\zeta) = \frac{1}{\zeta^2} \sum_{k=1}^{\infty} B_{2k} \frac{\zeta^{2k}}{(2k)!}.
\end{equation} Using the generating series \eqref{eq: generating series Bernoulli}, this becomes 
\begin{equation}
    \label{eq: derivation Borel transform pt2}
    \begin{aligned}
        \mathcal{B}[S_0](\zeta) &= \frac{1}{\zeta^2} \left( \sum_{n=0}^{\infty} B_n \frac{\zeta^n}{n!} - B_0 - B_1 \zeta \right) \\ 
        &= \frac{1}{\zeta} \left(\frac{1}{e^\zeta- 1} - \frac{1}{\zeta} + \frac{1}{2} \right), \quad \abs{\zeta} < 2\pi.
    \end{aligned}
\end{equation}
The original series was asymptotic, so $\mathcal{B}[S_0]$ has singularities, and the latter contain information about additional sectors of the theory that are `invisible' to the perturbative expansion. In the present case, \eqref{eq: derivation Borel transform} only has simple poles for every $\zeta_k$ such that $e^{\zeta_k} -1 = 0$, i.e.,~$\zeta_k \coloneqq 2 \pi i k$, $k \in \mathbb{Z}\backslash\{0\}$.\footnote{The assumption $\abs{\zeta} < 2\pi$ in \eqref{eq: derivation Borel transform pt2} is consistent with the location of the nearest poles of $\mathcal{B}[S_0]$.} Around these, $\mathcal{B}[S_0]$ admits the local expansion
\begin{equation}
    \label{eq: expansion Borel close to singularity}
    \begin{aligned}
        \mathcal{B}[S_0](\zeta_k + \zeta) 
        &= \frac{1}{\zeta_k} \frac{1}{\zeta} - \frac{2}{\zeta_k^2} + \mathcal{O}(\zeta), \quad k \in \mathbb{Z} \backslash \{0\}. 
    \end{aligned}
\end{equation}
Meanwhile, the naive singularity at the origin $\zeta = 0$ ($k=0$) is in fact removed, since
\begin{equation}
    \mathcal{B}[S_0](\zeta) \sim \frac{1}{\zeta} \left(\cancel{\frac{1}{\zeta} - \frac{1}{2} - \frac{1}{\zeta} + \frac{1}{2}} + \mathcal{O}(\zeta) \right) = \mathcal{O}(1), \quad \zeta \to 0.
\end{equation} The form of \eqref{eq: expansion Borel close to singularity}, where the singularity at $\zeta_k$ is a simple pole with residue $\tfrac{1}{\zeta_k}$ and there are no logarithmic singularity, tells us that the perturbative series around the associated instanton sector is not asymptotic itself, and carries through the entire resurgence analysis. The Stokes rays lie exactly on the imaginary axis.

Next, we turn to the Borel resummation of $S_0$. The Borel resummation of a given series $S$ is given by the directional Laplace transform of its Borel transform $\mathcal{B}[S]$ along a ray $\mathcal{C}^\theta = e^{i \theta} \mathbb{R}_+$ in the Borel $\zeta$-plane, 
\begin{equation}\label{defTrans}
        \mathcal{L}^\theta[S](z) \coloneqq \int_{\mathcal{C}^\theta} \deriv \zeta \, \mathcal{B}[S](\zeta) e^{-z\zeta}, \quad \text{where } \theta = \arg(z).
\end{equation} For $\theta =0$ and $S_0$, the integral runs along the positive real line,
\begin{equation}
    \mathcal{L}^0[S_0](z) = \int_{0}^{\infty} \deriv \zeta \, \frac{e^{-z\zeta}}{\zeta} \left(\frac{1}{e^\zeta- 1} - \frac{1}{\zeta} + \frac{1}{2} \right),
\end{equation} and we recover the exact integral formula for $\ln \Gamma(z)$   \cite[p.~258]{WhittakerWatson1990},
\begin{equation}\label{logGamma}
    \ln \Gamma(z) = \left(z - \frac{1}{2}\right) \ln z - z + \frac{1}{2} \ln(2\pi) + \mathcal{L}^0[S_0](z).
\end{equation} In fact, due to the presence of poles only on the imaginary axis, it is clear that this expression only holds for all $z$ such that $\abs{\arg(z)} < \tfrac{\pi}{2}$, i.e., as long as we do not cross any of the two Stokes lines. On a given Stokes line, one generally needs to resort to the \textit{lateral} Borel resummations
\begin{equation}
    \mathcal{L}^\theta_\pm[S](z) \coloneqq \int_{0}^{e^{i (\theta \pm \epsilon)}\infty} \deriv \zeta \, \mathcal{B}[S](\zeta) e^{-z\zeta}.
\end{equation} Deforming the contours, Cauchy's theorem yields
\begin{equation}
    \left(\mathcal{L}^\theta_+ - \mathcal{L}^\theta_-\right)[S](z) = \sum_{\zeta_k \in \mathcal{Z}} \oint_{\zeta_{k}} \deriv \zeta \, \mathcal{B}[S](\zeta) e^{-z\zeta},
\end{equation} where $\mathcal{Z} \coloneqq \{\zeta_k\}$ is the (possibly infinite) set of all singularities $\zeta_k$ of $\mathcal{B}[S](\zeta)$ lying on the Stokes ray $\mathcal{C}^\theta$. The lateral Borel resummations lead to distinct sectorial resummations of the original asymptotic series, but they are nonetheless connected via the so-called \textit{Stokes automorphism} $\underline{\mathfrak{S}}_\theta$, $\mathcal{L}^\theta_+ = \mathcal{L}^\theta_- \circ \underline{\mathfrak{S}}_\theta$. The operator $\underline{\mathfrak{S}}_\theta$ relates both resummations, and essentially encodes all the singular structure along the Stokes ray $\mathcal{C}^\theta$. Put differently, it quantifies how different asymptotic expansions hold, on diffent sides of a Stokes line.

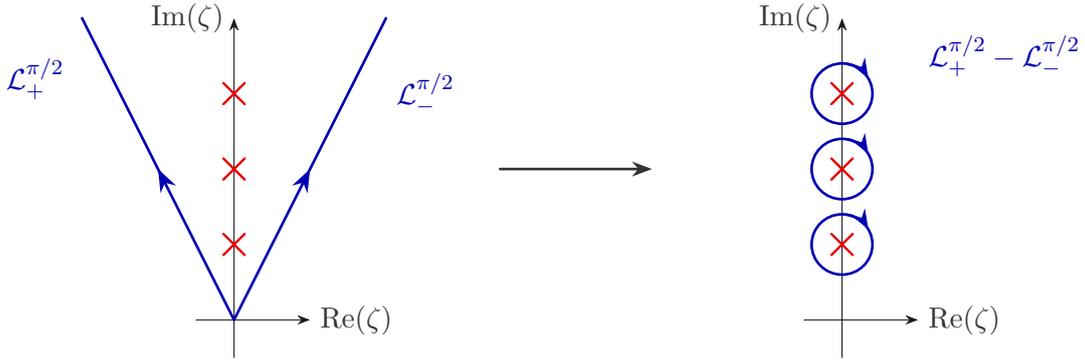
\begin{figure}[t]
    \centering
    \input{tikzBorel}
    \caption{Contour deformation for the lateral Borel resummation of $S_0$ around $\theta= \pi/2$.}
    \label{fig: laplace contour pi 2}
\end{figure}

Looking at $S_0$ and picking $\theta = \tfrac{\pi}{2}$ we have singularities at all $\zeta_k \coloneqq 2 \pi i k$, $k \in \mathbb{Z}_{>0}$. Enclosing all of them, we find
\begin{equation}
    \begin{aligned}
        \left(\mathcal{L}^{\pi/2}_+ - \mathcal{L}^{\pi/2}_- \right)[S_0](z) &= \sum_{k=1}^{\infty} \oint_{\zeta_k} \deriv \zeta \, \frac{e^{-z\zeta}}{\zeta} \left(\frac{1}{e^\zeta- 1} - \frac{1}{\zeta} + \frac{1}{2} \right) \\ 
        &= \sum_{k=1}^{\infty} e^{- z \zeta_k} \oint_{0} \deriv \zeta \, \mathcal{B}[S_0](\zeta_k + \zeta) \\
        &= - \sum_{k=1}^{\infty} \frac{e^{-2\pi i k z}}{k},
    \end{aligned}
\end{equation} where we used \eqref{eq: expansion Borel close to singularity}, and accounted for the fact that the contour runs clockwise around the poles, as depicted in Figure~\ref{fig: laplace contour pi 2}.
Recognizing the Mercator series for the logarithm,
\begin{equation}
    \label{eq: Taylor logarithm}
    \ln(1+x) = \sum_{k=1}^{\infty} \frac{(-1)^{k-1} x^k}{k}, \quad \abs{x} \leq 1, \, x \neq -1,
\end{equation} we arrive at
\begin{subequations}
\label{eq: lateral sums}
\begin{equation}
    \label{eq: lateral plus pi 2}
    \left(\mathcal{L}^{\pi/2}_+ - \mathcal{L}^{\pi/2}_- \right)[S_0](z) = + \ln(1- e^{-2 \pi i z})
\end{equation} for the Stokes jump across the ray $\mathcal{C}^{\pi/2}$. Likewise, one finds
\begin{equation}
    \label{eq: lateral minus pi 2}
    \left(\mathcal{L}^{-\pi/2}_+ - \mathcal{L}^{-\pi/2}_- \right)[S_0](z) = - \sum_{k=1}^{\infty} \frac{e^{+ 2\pi i k z}}{-k} = - \ln(1 - e^{+ 2\pi i z})
\end{equation}
\end{subequations} for the other Stokes ray $\mathcal{C}^{-\pi/2}$, where the integration contour also runs clockwise around the poles that lie on it. From the point of view of the resummed function (the result of carrying out the Laplace transform), the analytic continuation of $S_0$ has generated an infinite string of exponential terms.

As a sanity check, we note that \eqref{eq: lateral plus pi 2} and \eqref{eq: lateral minus pi 2} are consistent with the multi-valuedness of $\ln \Gamma(z)$ around the origin: starting at $\theta=0$ and performing a full rotation $z \mapsto ze^{2\pi i}$, we find
\begin{equation}
    \begin{aligned}
        \ln \Gamma(z) &\mapsto \left(z - \frac{1}{2}\right) (\ln z + 2\pi i) -z + \frac{1}{2} \ln 2 \pi + S(z) + \ln \left(\frac{1 - e^{-2 \pi i z}}{1 - e^{2 \pi i z}}\right) \\ 
        &= \ln \Gamma(z) - 2\pi i,
    \end{aligned}
\end{equation} as expected given the well-known Laurent series expansion of $\Gamma(z)$,
\begin{equation}
    \Gamma(z) = \frac{1}{z} - \gamma_E + \mathcal{O}(z) \quad \Rightarrow \quad \ln \Gamma(z) = - \ln (z) + \ln(1 - \gamma_E z + \mathcal{O}(z^2)).
\end{equation} These exponential corrections are also consistent with Euler's reflection formula for $\Gamma(z)$. Without loss of generality, we assume $\tfrac{\pi}{2} < \abs{\arg(z)} < \pi$. The formula writes
\begin{equation}
    \Gamma(z) \Gamma(ze^{ \mp i \pi}) = \frac{e^{\pm i \pi} \pi}{z \, \sin(\pi z)},
\end{equation} or equivalently
\begin{equation}
    \label{eq: constraint reflection formula}
    \begin{aligned}
        \ln \Gamma(z) + \ln \Gamma(z e^{\mp i \pi}) = \ln \pi - \ln z - \ln( e^{\mp i \pi} \sin(\pi z)).
    \end{aligned}
\end{equation} The LHS above is 
\begin{equation}
    \begin{aligned}
        \ln \Gamma(z) + \ln \Gamma(z e^{\mp i \pi})
        &= - \ln z  + \ln 2 \pi \pm \left(z + \frac{1}{2}\right) i \pi + S_0(z) + S_0(z e^{\mp i \pi}).
    \end{aligned}
\end{equation} Meanwhile, using
\begin{equation}
    \ln( e^{\mp i \pi} \sin(\pi z)) = \mp i \pi \left(z + \frac{1}{2}\right) + \ln(1 - e^{\pm 2 \pi i z}) - \ln 2,
\end{equation} the RHS becomes
\begin{equation}
    \begin{aligned}
        \ln \pi - \ln z - \ln( e^{\mp i \pi} \sin(\pi z)) 
        &= - \ln z + \ln 2 \pi \pm \left(z + \frac{1}{2}\right) i \pi - \ln(1 - e^{ \pm 2 \pi i z}),
    \end{aligned}
\end{equation} and \eqref{eq: constraint reflection formula} amounts to the functional relation
\begin{equation}
    \begin{aligned}
        S_0(z)
        &= - S_0(z e^{\mp i \pi}) + \sum_{k=1}^{\infty} \frac{e^{\pm 2 \pi i k z}}{k},
    \end{aligned}
\end{equation} with the choice of upper or lower sign depending on whether $z$ is in the upper or lower left half of the complex plane, respectively. This is consistent with \eqref{eq: lateral sums}.

This analysis of non-perturbative corrections arising upon crossing Stokes lines dictates the form of the \textit{transseries} associated to the problem, which yields an unambiguous and analytic continuation of the perturbative series across all Stokes sectors. We find
\begin{equation}
    \mathcal{S}(z) \coloneqq \underbrace{\sum_{k=1}^{\infty} \frac{B_{2k}}{2k (2k-1)} z^{1-2k}}_\text{perturbative} + \sum_{k=1}^{\infty} \frac{\stokes^\pm_{k}(\theta)}{k} e^{\pm 2 \pi i k z}, \quad \abs{\arg(z)} < \pi,
\end{equation} where $\theta \coloneqq \arg(z)$ and the Stokes multipliers $\stokes^\pm_{k}(\theta)$ are
\begin{equation}
    \label{eq: stokes constant log gamma}
    \stokes_{k}^+(\theta) \coloneqq \begin{cases}
        0 & -\pi < \theta < \frac{\pi}{2}, \\ 
        \frac{1}{2} & \theta = + \frac{\pi}{2}, \\ 
        1 & \frac{\pi}{2} < \theta < \pi,
    \end{cases} \quad \text{and} \quad
    \stokes_{k}^-(\theta) \coloneqq \begin{cases}
        0 & - \frac{\pi}{2} < \theta < \pi, \\ 
        \frac{1}{2} & \theta = - \frac{\pi}{2}, \\ 
        1 & - \pi < \theta < - \frac{\pi}{2}.
    \end{cases}
\end{equation} The value at $\theta = \pm \tfrac{\pi}{2}$ follows from Dingle's final rule \cite{Nemes:2022} and will not play any role in this work. When the dust settles, we read off the full transseries for $\ln \Gamma(z)$ at $\abs{z} = \infty$,
\begin{equation}
    \label{eq: transseries log Gamma}
    \ln \Gamma(z) = \left(z - \frac{1}{2}\right) \ln z - z + \frac{1}{2} \ln 2 \pi + \sum_{k=1}^{\infty} \frac{B_{2k}}{2k(2k-1)} z^{1-2k} + \sum_{k=1}^{\infty} \frac{\stokes^\pm_{k}(\theta)}{k} e^{\pm 2 i \pi k z}
\end{equation} for $\theta = \arg(z) \in (- \pi, \pi)$, interpreted through its Borel-resummation, along with \eqref{eq: stokes constant log gamma}.

\subsection{Transseries for \texorpdfstring{$\Gamma$}{Gamma} and its reciprocal} 
\label{app: transseries Gamma and reciprocal}

Let us now move on to the Gamma function, which is the building block of the four-point string amplitude of Section~\ref{sec: four point tree string}. For notational convenience, we introduce the scaled Gamma function 
\begin{equation}
    \Gamma^\ast(z) \coloneqq \sqrt{\frac{z}{2 \pi}} \left(\frac{z}{e}\right)^{-z} \Gamma(z). 
\end{equation} Exponentiation of the asymptotic series of $\ln \Gamma(z)$ around $\abs{z}=\infty$, \eqref{eq: asymptotic log Gamma}, produces the well-known asymptotic expansion
\begin{equation}
    \label{eq: series Gamma}
    \Gamma^\ast(z) \sim \exp(\sum_{k=1}^{\infty} \frac{B_{2k}}{2k(2k-1)} z^{1-2k}) \coloneqq \sum_{n=0}^{\infty} (-1)^n \frac{\gamma_n}{z^n}, \quad \text{as } \abs{z} \to \infty, \, \abs{\arg(z)} < \frac{\pi}{2},
\end{equation} where $\gamma_n$ are the Stirling coefficients. While a closed form for the latter remains unknown, their values can be generated numerically by means of the recurrence relation \cite[2.1.12]{Paris:2001}
\begin{equation}
    \gamma_k = (-2)^k \frac{\Gamma\left(k + \frac{1}{2}\right)}{\sqrt{\pi}} d_{2k}, \quad d_n = \frac{n+1}{n+2} \left( \frac{d_{n-1}}{n} - \sum_{j=1}^{n-1} \frac{d_j d_{n-j}}{j+1}\right), \quad n \geq 1, \, d_0 = 1. 
\end{equation} The asymptotic series for the reciprocal of the scaled Gamma function is similar,
\begin{equation}
    \label{eq: series reciprocal Gamma}
    \frac{1}{\Gamma^\ast(z)} \sim \exp(-\sum_{k=1}^{\infty} \frac{B_{2k}}{2k(2k-1)} z^{1-2k}) \coloneqq \sum_{k=0}^{\infty} \frac{\gamma_k}{z^k},
\end{equation} with $\gamma_k$ again the Stirling coefficients.

Next, we complete these asymptotic expansions into transseries. While the Borel transform of \eqref{eq: series Gamma} and \eqref{eq: series reciprocal Gamma} cannot be analytically computed, a transseries for these may still be obtained by exponentiating \eqref{eq: transseries log Gamma}, yielding
\begin{equation}
    \Gamma^\ast(z) = \exp[\sum_{k=1}^{\infty} \frac{B_{2k}}{2k(2k-1)} z^{1-2k} + \sum_{k=1}^{\infty} \frac{\stokes^\pm_{k}(\theta)}{k} e^{\pm 2 \pi i k z}].
\end{equation} The Mercator series \eqref{eq: Taylor logarithm} for the logarithm, the binomial expansion  in terms of Pochhammer symbols \cite[4.6.7]{NIST:DLMF},
\begin{equation}
    \label{eq: binomial Pochhammer}
    (1+x)^{-y} = \sum_{n=0}^{\infty} (-1)^n \frac{(y)_n}{n!} x^n, 
\end{equation} and the identity $(1)_k = k!$ yield \cite[(2.7)]{Nemes:2022}
\begin{equation}
    \label{eq: transseries Gamma}
    \Gamma^\ast(z) = \exp[\sum_{k=1}^{\infty} \frac{B_{2k}}{2k(2k-1)} z^{1-2k} ] \left(1+ \sum_{k=1}^{\infty} \Stokes_k(\theta) \, e^{\pm 2 \pi i k z} \right),
\end{equation} where we introduced $\theta = \arg(z)$ as well as the Stokes coefficients
\begin{equation}
    \Stokes_k(\theta) = \begin{cases}
            0 & 0< \abs{\theta} < \frac{\pi}{2}, \\ 
            \frac{1}{k!} \left(\frac{1}{2}\right)_k & \theta = \pm \frac{\pi}{2}, \\ 
            1 & \frac{\pi}{2} < \abs{\theta} < \pi.
        \end{cases}
\end{equation} The upper sign in \eqref{eq: transseries Gamma} is taken if $z$ is the the upper complex plane, and inversely. Likewise, one finds 
\begin{equation}
    \frac{1}{\Gamma^\ast(z)} = \exp[- \sum_{k=1}^{\infty} \frac{B_{2k}}{2k(2k-1)} z^{1-2k} ] \times \begin{cases}
        1 - e^{+ 2\pi i z} & + \frac{\pi}{2} < \theta < + \pi, \\
        1 & - \frac{\pi}{2} < \theta <+  \frac{\pi}{2}, \\ 
        1 - e^{- 2\pi i z} & - \pi < \theta < - \frac{\pi}{2}, \\ 
        (1 - e^{\pm 2\pi i z})^{\frac{1}{2}} & \theta = \pm \frac{\pi}{2},
    \end{cases}
\end{equation} which \eqref{eq: binomial Pochhammer} and $(-1)_k = 0$ $\forall k \geq 2$ simplify to
 \begin{equation}
    \frac{1}{\Gamma^\ast(z)} = \exp[- \sum_{k=1}^{\infty} \frac{B_{2k}}{2k(2k-1)} z^{1-2k} ] \left(1 - \sum_{k=1}^{\infty} \tilde{\Stokes}_k(\theta) \, e^{\pm 2 \pi i k z} \right),
\end{equation} with different Stokes coefficients, namely
\begin{equation}
    \tilde{\Stokes}_1(\theta) = \begin{cases}
            0 & 0< \abs{\theta} < \frac{\pi}{2}, \\ 
            \frac{1}{2} & \theta = \pm \frac{\pi}{2}, \\ 
            1 & \frac{\pi}{2} < \abs{\theta} < \pi,
        \end{cases} \quad \text{and} \quad \tilde{\Stokes}_k(\theta) = \begin{cases}
            0 & 0< \abs{\theta} < \frac{\pi}{2}, \\ 
            - \frac{1}{k!} \left( -\frac{1}{2} \right)_k & \theta = \pm \frac{\pi}{2}, \\ 
            0 & \frac{\pi}{2} < \abs{\theta} < \pi,
        \end{cases} \, k \geq 2.
\end{equation}

As a sanity check, one may verify that \eqref{eq: transseries Gamma} is consistent with the Euler reflection formula. Without loss of generality, let $z \in \mathbb{C}$, $\arg(z) \in \left(0, \tfrac{\pi}{2}\right)$, so $-z = e^{- i \pi} z$. Then
\begin{equation}
    \Gamma(z) \Gamma(e^{- i \pi} z) = \frac{ e^{i \pi} \pi}{z \, \sin(\pi z)} \quad \Leftrightarrow \quad \Gamma^\ast(z) \Gamma^\ast(e^{- i \pi} z) = \frac{e^{i \pi (1-z)}}{e^{i \pi z} - e^{- i \pi z}}. 
\end{equation} Using \eqref{eq: transseries Gamma} and the fact that $-z$ is in the lower half of the complex plane, the LHS of the expression above is
\begin{equation}
    \begin{aligned}
        \Gamma^\ast(z) \Gamma^\ast(e^{-i \pi} z) &= \exp(0) \times 1 \times \left(1 + \sum_{k=1}^{\infty} e^{- 2 \pi i k (-z)}\right) = \frac{1}{1 - e^{2 \pi i z}} = \frac{e^{- i \pi z}}{e^{- i \pi z} - e^{i \pi z}},
    \end{aligned}
\end{equation} and the two expression agree. Note that the second equality above relies on the fact that $\Im(z) > 0$ for $\arg(z) \in \left(0, \tfrac{\pi}{2}\right)$, so $\abs{e^{2\pi i kz }} = e^{-2\pi k \Im(z)} < 1$ and the geometric series converges.
Likewise, for $\arg(z) \in \left(- \tfrac{\pi}{2}, 0\right)$ we have $-z = e^{i \pi} z$ and the reflection formula
\begin{equation}
    \Gamma^\ast(z) \Gamma^\ast(e^{i \pi}z ) = \frac{e^{- i \pi z}}{e^{i \pi z} - e^{- i \pi z}},
\end{equation} which agrees again with the transseries \eqref{eq: transseries Gamma}, given
\begin{equation}
    \begin{aligned}
        \Gamma^\ast(z) \Gamma^\ast(e^{+i \pi} z) &= \exp(0) \times 1 \times \left(1 + \sum_{k=1}^{\infty} e^{+ 2 \pi i k (-z)}\right) = \frac{1}{1 - e^{-2 \pi i z}} = \frac{e^{- i \pi z}}{e^{i\pi z} - e^{- i \pi z}}.
    \end{aligned}
\end{equation} This time $\arg(z) \in \left(- \tfrac{\pi}{2}, 0\right)$ implies $\Im(z) < 0$ and $\abs{e^{-2\pi i kz}}< 1$, as required for the geometric series to converge. This concludes our resurgence analysis of the Gamma function.

\subsection{Revisiting \texorpdfstring{$\ln \Gamma$}{log Gamma}: controlling factors}
\label{app: example difference log Gamma}

We now revisit the transseries for $\ln \Gamma(z)$ around $\abs{z} = \infty$. This function and its resurgence properties are studied in detail in Appendix~\ref{app: transseries log Gamma}, where we derive the transseries
\begin{equation}
    \ln \Gamma(z) = \left(z - \frac{1}{2}\right) \ln z - z + \frac{1}{2} \ln(2 \pi) + \sum_{k=1}^{\infty} \frac{B_{2k}}{2k(2k-1)} z^{1-2k} + \sum_{k=1}^{\infty} \frac{\stokes^\pm_{k}(\theta)}{k} e^{\pm 2 i \pi k z},
\end{equation} see \eqref{eq: transseries log Gamma}, where $\theta = \arg(z)$ and the Stokes multipliers are given by
\eqref{eq: stokes constant log gamma}. Here, we aim to recover this result solely from the inhomogeneous linear finite difference equation $\ln \Gamma(z)$ obeys,
\begin{equation}
    \label{eq: difference equation log Gamma}
    F(z+1) - F(z) = \ln z,
\end{equation} which follows from the functional identity
\begin{equation}
    \label{eq: identity Gamma}
    \Gamma(z+1)= z \, \Gamma(z)
\end{equation} for the Gamma function, along with the definition $F(z) \coloneqq \ln \Gamma(z)$. We refer to, e.g., \cite{Bender:1999box,Costin:2008} for a comprehensive treatment of transseries solutions to finite difference equations. Since we want to be agnostic to any results for the Gamma function, we stress that \eqref{eq: difference equation log Gamma} and \eqref{eq: identity Gamma} could be deduced by direct partial integration on the integral representation
\begin{equation}
    \label{eq: integral rep Gamma}
    \Gamma(z) = \int_0^\infty \deriv t \, t^{z-1} e^{-t}, \quad \Re(z) > 0, 
\end{equation} without the need to evaluate the integral to $\Gamma(z)$.

To derive the solution to \eqref{eq: difference equation log Gamma} around $z = \infty$, we resort to the method of controlling factors (see, e.g.,  \cite{Bender:1999box}), and thus start by considering the corresponding differential equation. We have $F(z+1) - F(z) \leftrightarrow F'(z)$ for large $z$, so \eqref{eq: difference equation log Gamma} behaves like the first-order inhomogeneous ordinary differential equation
\begin{equation}
    \label{eq: ODE log Gamma}
    \partial_z F(z) = \ln z.
\end{equation} Clearly, $z=0$ is an irregular singular (branch) point of \eqref{eq: difference equation log Gamma}. Changing variables to $t = 1/z$, we find
$\partial_t F(t) = \ln(t)/t^2$, so $z= \infty$ is also an irregular singularity. We therefore expect an asymptotic series solution around this point. Integrating \eqref{eq: ODE log Gamma}, 
\begin{equation}
    F_0(z) \sim \int \deriv z \, \ln(z) = z \ln(z) - z + c_0,
\end{equation} with $c_0 \in \mathbb{C}$ an integration constant to be determined. We stress that this is forced by the leading behaviour of the difference equation. Inserting this back into \eqref{eq: difference equation log Gamma} yields
\begin{equation}
    \begin{aligned}
        F_0(z+1) - F_0(z) &\sim (z+1) \ln(z+1) - (z+1) - z\ln z + z \\ 
        &\sim \ln (z)+\frac{1}{2 z}-\frac{1}{6 z^2}+\frac{1}{12 z^3}-\frac{1}{20
   z^4}+...
    \end{aligned}
\end{equation} Comparing this with \eqref{eq: difference equation log Gamma}, we have extra unwanted terms in powers of $1/z$. Given 
\begin{equation}
    \ln(z+1) - \ln(z) = \frac{1}{z}-\frac{1}{2 z^2}+\frac{1}{3 z^3}-\frac{1}{4
   z^4}+ ..., 
\end{equation} we can cancel the $+\tfrac{1}{2z}$ term by adding $- \tfrac{1}{2} \ln z$ to $F_0$, i.e.,~by considering instead the solution $F_1(z) = F_0(z) - \tfrac{1}{2}\ln z$. This yields
\begin{equation}
    F_1(z+1) - F_1(z) \sim \ln (z)+\frac{1}{12 z^2}-\frac{1}{12 z^3}+\frac{3}{40
   z^4}+... 
\end{equation} There are still unwanted powers of $1/z$, starting from $1/z^2$. To fix these, we need to add corrections in inverse powers of $z$, i.e.~we posit a solution of the form
\begin{equation}
    \hat F(z) \sim \left(z - \frac{1}{2}\right) \ln z - z + c_0 + \sum_{k=1}^\infty \frac{a_k}{z^k}, 
\end{equation} with the coefficients $a_k$ to be determined. Inserting this Ansatz into \eqref{eq: difference equation log Gamma}, expanding for large $z$ and cancelling unwanted powers order by order, we find $a_1 = \tfrac{1}{12}$, $a_2 = 0$, $a_3 = - \tfrac{1}{360}$, $a_4 = 0$, $a_5 = \tfrac{1}{1260}$, ... and more generally
\begin{equation}
    a_{2k-1} = \frac{B_{2k}}{2k (2k-1)}, \, a_{2k} = 0, \quad k \geq 1.
\end{equation} Hence, the particular `perturbative' solution to the inhomogeneous equation \eqref{eq: difference equation log Gamma} is of the form
\begin{equation}
    \label{eq: pert log Gamma temp}
    \hat F(z) \sim \left(z - \frac{1}{2}\right) \ln z - z + c_0 + \sum_{k=1}^\infty \frac{B_{2k}}{2k (2k-1)} z^{1-2k}. 
\end{equation} Thus, by means of a local analysis and by `peeling off' the leading behaviour of the solution order by order, we recovered the asymptotic expansion of $\ln \Gamma(z)$ as $\abs{z} \to \infty$, up to an undetermined additive constant $c_0$. From either the knowledge of $\ln \Gamma$ or \eqref{eq: integral rep Gamma}, one may be tempted to enforce the initial condition $\hat F(1)=0$, but the nature of local analysis precludes the use of information from points far away from $z = \infty$ to determine $c_0$.\footnote{If we could analyse the difference equation simultaneously at $z=1$ and $z=\infty$, we would be doing global and not local analysis. In fact, \eqref{eq: pert log Gamma temp} simply doesn't exist at $z = 1$ in any literal sense, so plugging $z=1$ is not meaningful, neither analytically nor asymptotically, unless one regularizes the divergent series.} 

The typical way of solving this problem would be to match \eqref{eq: pert log Gamma temp} to the known large-$z$ asymptotics of the Gamma function \cite{Bender:1999box} or the Legendre duplication formula \cite{Namias:1986}, but this seems circular. We here showcase an alternative way to proceed, while remaining agnostic to existing results on the Gamma function. Given $F(1) = 0$, \eqref{eq: difference equation log Gamma} telescopes on integers and implies the exact relation
\begin{equation}
    \label{eq: difference log integers}
    F(n+1) = \sum_{k=1}^{n} \ln k. 
\end{equation} The finite part of \eqref{eq: difference log integers} as $n \to \infty$ may be computed by zeta regularization \cite[25.6.11]{NIST:DLMF}
\begin{equation}
    \sum_{k=1}^\infty \ln k \coloneqq - \zeta'(0) = \frac{1}{2} \ln (2\pi).
\end{equation} Above, we used the series representation of the Riemann zeta function 
\begin{equation}
    \zeta(s) = \sum_{k=1}^\infty \frac{1}{k^s} \quad \Rightarrow \quad \zeta'(s) = - \sum_{k=1}^\infty \frac{\ln k}{k^s},
\end{equation} which converges provided $\Re(s) > 1$ but may be analytically extended to a meromorphic function with simple pole at $s=1$. Our prescription is therefore $c_0 \coloneqq \tfrac{1}{2} \ln(2\pi)$. This fixes the particular solution to \eqref{eq: difference equation log Gamma} completely. An equivalent derivation of this result is presented in \cite{Dominici:2006}, where $c_0$ arises through optimal truncation of the divergent series
\begin{equation}
    \label{eq: truncation C0}
    c_0 \coloneqq \lim_{n\to \infty} \left(1 - \sum_{k=1}^{n} \frac{B_{2k}}{2k(2k-1)} \right)
\end{equation} around $n=7$, where $c_0 \approx 0.913912$. This is consistent with our analysis and the zeta-regularization. In fact, one may obtain \eqref{eq: truncation C0} by as the finite part coming from using the Euler-Maclaurin summation formula on \eqref{eq: difference log integers} as $n \to \infty$.

Next, we look at the homogeneous part of \eqref{eq: difference equation log Gamma}, $ F(z+1) - F(z) = 0$. This is trivially solved by $1$-periodic functions of the form $F(z) \propto e^{2\pi i k z}$, for $k \in \mathbb{Z}\backslash\{0\}$, where we account for the fact that the $k=0$ term is already absorbed in $c_0$. The full solution to \eqref{eq: difference equation log Gamma} must then feature a linear combination thereof, 
\begin{equation}
    \ln \Gamma(z) = \left(z - \frac{1}{2}\right) \ln z - z + \frac{1}{2} \ln(2\pi) + \sum_{k=1}^\infty \frac{B_{2k}}{2k (2k-1)} z^{1-2k} + \sum_{k=1}^\infty \sigma^\pm_k e^{\pm 2 \pi i k z}
\end{equation} for some constant coefficients $\sigma^\pm_k$. We stress that the difference equation by itself does not fix $c_0$, nor $\sigma^\pm_k$, and only gives rise to a resurgence \textit{class}. We fixed $c_0$ by regularizing the integer relation \eqref{eq: difference log integers}. To reproduce the correct asymptotic behaviour of $\ln \Gamma(z)$, $\sigma^\pm_k$ ought to be matched with the Stokes multipliers \eqref{eq: stokes constant log gamma} using resurgence on the perturbative expansion \eqref{eq: pert log Gamma temp}. This yields $\sigma_k^\pm = \stokes^\pm_k(\arg(z)) / k$. When the dust settles, we recover the transseries \eqref{eq: transseries log Gamma}. It is straightforward to exponentiate the latter to arrive to a transseries for the Gamma function and its reciprocal, as done above.

%% file: tikzBorel.tex
\begin{tikzpicture}[>=Stealth, line join=round, line cap=round] 

\draw[black!90, ->] (0,-2.5) -- (0,2);
\node[black!80,left] at (0,2) {$\Im(\zeta)$};
\draw[black!90, ->] (-0.5,-2) -- (1,-2);
\node[black!80,right] at (1,-2) {$\Re(\zeta)$};

\foreach \y in {-1,0,1}{ 
    \draw[red!90!black, line width=0.9pt] (-0.14,\y-0.14) -- (0.14,\y+0.14); 
    \draw[red!90!black, line width=0.9pt] (-0.14,\y+0.14) -- (0.14,\y-0.14); 
}

\draw[blue!70!black, line width=1pt, postaction={decorate}, decoration={markings, mark=at position 0.5 with {\arrow{Stealth[length=3mm]}}}] (0,-2) -- (-2.0,2.0); 
\draw[blue!70!black, line width=1pt, postaction={decorate}, decoration={markings, mark=at position 0.5 with {\arrow{Stealth[length=3mm]}}}] (0,-2) -- (2.0,2.0); 

\node[blue!70!black,left] at (-2.1,1.2) {$\mathcal{L}^{\pi/2}_+$};
\node[blue!70!black,right] at (2.0,1) {$\mathcal{L}^{\pi/2}_-$};

\draw[black!80, line width=1pt, ->] (3.5,0) -- (5.5,0);

\draw[black!90, ->] (8,-2.5) -- (8,2); 
\node[black!80,left] at (8,2) {$\Im(\zeta)$};
\draw[black!90, ->] (7.5,-2) -- (9,-2);
\node[black!80,right] at (9,-2) {$\Re(\zeta)$};

\foreach \y in {1,0,-1}{ 
    \draw[blue!70!black, line width=1pt] (8,\y) circle (0.4); 
    \draw[blue!70!black, line width=1pt, ->] ($ (8,\y) + (-280:0.4) $) arc (-280:-330:0.4); 
    \draw[red!90!black, line width=0.9pt] ($ (8,\y) + (-0.14,-0.14) $) -- ($ (8,\y) + (0.14,0.14) $); 
    \draw[red!90!black, line width=0.9pt] ($ (8,\y) + (-0.14,0.14) $) -- ($ (8,\y) + (0.14,-0.14) $); 
} 

\node[blue!70!black,right] at (9,1.5) {$\mathcal{L}^{\pi/2}_+ - \mathcal{L}^{\pi/2}_-$}; 

\end{tikzpicture}

%% file: contiguitymatrices.tex
\section{Five-point contiguity matrices}
\label{app: contiguity matrices}

In this appendix, we report the explicit contiguity matrices of \S\ref{fivediff}. The matrices entering \eqref{eq: shift equation} may be obtained using integration by parts on \eqref{eq: minimal functions N=5}, and are given by
\begin{equation}
    \label{eq: shift matrices N=5}
    \begin{aligned}
        \mathsf{M}_1(s) &= \begin{pmatrix}
            \frac{-s_1^2+s_1 (2 s_3+s_4-s_5+1)+s_3(s_2-s_4-1)+(s_4+1) s_5}{s_1(-s_1+s_3+s_4+1)} & -\frac{s_3}{ 1 + s_3+s_4 - s_1} \\ 
            \frac{ (s_1+s_2-s_4-1) (s_2+s_3-s_5)}{s_1 (-s_1+s_3+s_4+1)} & \frac{-s_1-s_2+s_4+1}{-s_1+s_3+s_4+1}
        \end{pmatrix}, \\
        \mathsf{M}_2(s) &= \begin{pmatrix}
            \frac{s_1 s_3-s_2^2+s_2 (s_4+s_5+1)}{s_2 (s_4+s_5+1-s_2)} & -\frac{s_1
       s_3}{s_2 (s_4+s_5+1-s_2)} \\ \frac{(s_1+s_2-s_4-1) (s_2+s_3-s_5-1)}{s_2 (s_4+s_5+1-s_2)} &
       -\frac{(s_1+s_2-s_4-1) (s_2+s_3-s_5-1)}{s_2 (s_4+s_5+1-s_2)}
        \end{pmatrix}, \\ 
        \mathsf{M}_3(s) &= \begin{pmatrix}
            \frac{s_1 (s_2+2s_3-s_5-1)-(s_3+s_4) (s_3-s_5-1)}{s_3 (s_1-s_3+s_5+1)} & -\frac{s_1}{s_1-s_3+s_5+1} \\ \frac{(s_1+s_2-s_4) (s_2+s_3-s_5-1)}{s_3 (s_1-s_3+s_5+1)} & \frac{s_5+1-s_2-s_3}{s_1-s_3+s_5+1}
        \end{pmatrix}, \\ 
        \mathsf{M}_4(s) &= \begin{pmatrix}
            \frac{s_3+s_4}{s_4} & -\frac{s_1 s_3}{s_4 (s_1+s_2-s_4)} \\ \frac{s_2+s_3-s_5}{s_4} & \frac{s_1 (s_2+s_3-s_4-s_5)+s_4 (s_4+s_5-s_2)}{s_4 (s_4-s_1-s_2)}
        \end{pmatrix}, \\ 
        \mathsf{M}_5(s) &= \begin{pmatrix}
            \frac{s_1+s_5}{s_5} & -\frac{s_1 s_3}{s_5 (s_2+s_3-s_5)} \\ \frac{s_1+s_2-s_4}{s_5} & \frac{(s_3-s_5) (s_4+s_5-s_2)- s_1 s_3}{s_5 (s_2+s_3-s_5)}
        \end{pmatrix}.
    \end{aligned}
\end{equation}